\documentclass[11pt,a4paper]{article}
\usepackage{jheppub}
\usepackage{graphicx}
\usepackage{url}
\usepackage{cancel}
\usepackage{caption}
\usepackage{placeins}
\usepackage{bm}
\usepackage{hyperref}
\usepackage{amsmath}
\usepackage{amssymb}
\usepackage{listings}
\usepackage{float}
\usepackage[toc, page]{appendix}
\usepackage{color}
\usepackage{booktabs}
\newcommand{\order}{{\cal O}}

\newcommand{\sherpa}{{\sc Sherpa}}
\newcommand{\powheg}{{\sc Powheg}}

\newcommand{\amcfast}{{\sc aMCfast}}
\newcommand{\applgrid}{{\sc APPLgrid}}
\newcommand{\apfelgrid}{{\sc APFELgrid}}

\newcommand{\amc}{{\sc MadGraph5\_aMC@NLO}}

\def\G1{{\bf \gamma^{(1)}_N}}

\title{Single Top Production in PDF fits}
\author[a]{Emanuele R.~Nocera,}
\author[b]{Maria Ubiali,}
\author[c]{and Cameron Voisey}
\affiliation[a]{Nikhef, Science Park 105, NL-1098 XG Amsterdam, The Netherlands}
\affiliation[b]{DAMTP, University of Cambridge, Wilberforce Road, Cambridge, CB3 0WA, United Kingdom}
\affiliation[c]{Cavendish Laboratory (HEP), JJ Thomson Avenue, Cambridge, CB3 0HE, United Kingdom}
\emailAdd{e.nocera@nikhef.nl}
\emailAdd{M.Ubiali@damtp.cam.ac.uk}
\emailAdd{voisey@hep.phy.cam.ac.uk}

\abstract{We study the impact of recent LHC $t$-channel single top-quark and
top-antiquark measurements at centre-of-mass energies of 7, 8 and 13 TeV on the
parton distribution functions (PDFs) of the proton. We consider, 
namely, total cross sections, top-antitop cross section ratios, and 
differential distributions. We present a critical appraisal of the data, 
studying in particular how their description is affected by the
theoretical details that enter the computation of the corresponding observables:
QCD and electroweak higher-order corrections, the flavour scheme, and the value
of the bottom-quark threshold. We perform a series of fits to the data within 
the NNPDF3.1 framework, whereby next-to-next-to-leading order QCD corrections 
are applied to single top measurements in a systematic way.
We find that there exists an optimal combination of data that maximises
consistency with the rest of the dataset, and efficiency in constraining the
up, down and, partially, gluon PDFs.}

\subheader{\small Cavendish-HEP-19/18, DAMTP-2019-28, Nikhef 2019-053}

\begin{document}

\maketitle
\flushbottom

\section{Introduction}

At hadron colliders, single top quarks and top antiquarks are produced 
in the Standard Model (SM) via the weak interaction involving, at leading 
order (LO), a $Wtb$ vertex. Three channels contribute to this process: 
the exchange of a space-like $W$ boson, or $t$-channel 
($q\bar{b}\to q^\prime\bar{t}$ and $\bar{q}b\to \bar{q}^\prime t$ at LO); 
the exchange of a time-like $W$ boson, or $s$-channel 
($q\bar{q}^\prime\to t\bar{b}, \bar{t}b$ at LO); and the associated production 
of a top quark or antiquark with an on-shell $W$ boson
($gb\to tW^-$ and $g\bar{b}\to \bar{t}W^+$ at LO). The $t$-channel leads to the 
largest cross section at the Large Hadron Collider (LHC).

Single top-quark and top-antiquark production is a signal to 
probe several aspects of the SM. For instance, it allows for the direct 
determination of the Cabibbo-Kobayashi-Maskawa (CKM) matrix element 
$|V_{tb}|$ without assuming unitarity~\cite{Alwall:2006bx,Cao:2015qta}, and
for the extraction of the top-quark mass~\cite{Sirunyan:2017huu,
Alekhin:2016jjz}.
It is also an important source of background events in searches for
physics beyond the SM that could modify the structure of the $Wtb$ vertex, 
or allow new gauge bosons or heavy quarks to interact via a
flavour-changing neutral current~\cite{Tait:2000sh,AguilarSaavedra:2008gt,
Hartland:2019bjb,Gao:2011fx,Cao:2015doa}. Furthermore, single top-quark and top-antiquark
production has also been suggested as a probe of parton distribution functions
(PDFs), in particular to 
constrain the ratio of up and down-quark distributions at large momentum
fractions $x$. The impact of measurements of inclusive total cross sections 
from the Tevatron and the LHC were specifically appraised in 
Refs.~\cite{Alekhin:2015cza,Alekhin:2017kpj}. 

In this work, we study the impact of the $t$-channel single top-quark and 
top-antiquark production data on the PDFs of the proton in a systematic way.
On the one hand, we will scrutinise the available experimental measurements,
in particular from the LHC. We will focus on total top-quark and 
top-antiquark cross sections (or their ratio) from 
ATLAS~\cite{Aad:2014fwa,Aaboud:2017pdi,Aaboud:2016ymp} and 
CMS~\cite{Chatrchyan:2012ep,Khachatryan:2014iya,Sirunyan:2016cdg} at 
centre-of-mass energies of 7, 8 and 13 TeV; and on differential cross sections,
both absolute and normalised to the total cross section, 
from ATLAS at 7 TeV~\cite{Aad:2014fwa} and 8
TeV~\cite{Aaboud:2017pdi}. On the other hand, we will investigate
how the description of the data is affected by the theoretical
details that enter the computation of single top observables.
Specifically, we will pay attention to the inclusion of higher-order 
quantum chromodynamics (QCD) corrections~\cite{Bordes:1994ki,Pittau:1996rp,
Stelzer:1997ns,Harris:2002md,Falgari:2010sf,Schwienhorst:2010je}, 
recently computed up to next-to-next-to-leading order (NNLO)
accuracy~\cite{Brucherseifer:2014ama,Berger:2016oht,Berger:2017zof}; to the
effect of next-to-leading order (NLO) electroweak (EW) 
corrections~\cite{Beccaria:2006ir,Mirabella:2008gj,Bardin:2010mz}; and to the
choice of the heavy flavour scheme. In this last respect we will
restrict our analysis to NLO QCD, and compare the four-flavour scheme
(4FS)~\cite{Campbell:2009ss}, in which the bottom quark is treated as a massive 
final state and does not contribute to the proton wave-function, with the
five-flavour scheme (5FS)~\cite{Bordes:1994ki}, in which the bottom quark is 
treated as a light quark, perturbatively generated within the proton 
above a certain matching scale (often taken to be of the order of the 
bottom-quark pole mass). The phenomenological dependence on the exact value of 
the bottom-quark threshold, in particular on a higher bottom-quark matching
scale as suggested in Ref.~\cite{Bertone:2017djs}, will be investigated as well.

The inclusion of single top-quark and top-antiquark data in a PDF fit will
be realised in the framework of the NNPDF3.1 global 
analysis~\cite{Ball:2017nwa}. We will investigate the internal consistency of 
the new data (and the consistency of the new data with the rest of the NNPDF3.1 
dataset); which theoretical details lead to its optimal description; whether 
it is advantageous to use absolute instead of normalised distributions; and 
which combination of LHC data maximises efficiency in constraining 
the PDFs. Such a combination will be suitable for inclusion in the next
generation of global PDF fits.

The outline of the paper is as follows. In Sect.~\ref{sec:exp_data} we present
the experimental measurements of single top-quark and top-antiquark production 
that we will be studying. In Sect.~\ref{sec:data-theory-comparison} we examine 
how this data is described upon various assumptions in the underlying theory. 
Specifically, we provide a quantitative assessment of the NNLO QCD corrections, 
of the NLO QCD$\times$EW corrections (in both cases computed within the 5FS), 
and of NLO QCD corrections computed either in the 4FS or in the 5FS, possibly 
with a higher bottom-quark matching scale. In Sect.~\ref{sec:fits} we assess 
the impact of the data in a PDF fit: we study the correlations between the data 
and the various PDF flavours; we discuss the details of our fits; we formulate 
a recommendation on the data that, if included in a fit, maximises consistency 
with the rest of the dataset and efficiency in constraining the PDFs; we 
analyse the correlations between single top data and data that is already 
included in NNPDF fits as standard; and we assess the phenomenological
implications of our optimal fit. Finally, we draw our conclusions in 
Sect.~\ref{conclusions}. An additional investigation of the differential 
distributions from ATLAS at 8 TeV is presented in Appendix~\ref{app:ATLAS8TeVdiff}.
These data are not included in the fits because complete information on the
experimental correlations within the data is not currently available.

\section{Experimental data} 
\label{sec:exp_data}

In this section we present the $t$-channel single top-quark and top-antiquark 
production data from ATLAS and CMS that we study in this work. 
We first discuss measurements of top-quark and top-antiquark total cross 
sections and of their ratios, and we then discuss the absolute and normalised 
differential cross sections available as a function of various kinematic 
variables.

\subsection{Total top-quark and top-antiquark cross sections and their ratios}
\label{sec:totxsecs}

The total cross section for $t$-channel single top production was 
first measured in proton-antiproton collisions at a centre-of-mass energy of 
1.96 TeV by the D0~\cite{Abazov:2009pa} and CDF~\cite{Aaltonen:2010jr} 
experiments at the Tevatron. These pioneering measurements established
electroweak top-quark production in hadronic collisions, gave insight into the 
properties of the $Wtb$ vertex, and allowed for a first extraction of the CKM 
matrix element $|V_{tb}|$. However, because of the rather large statistical
uncertainty in each case --- a consequence of the low luminosities achieved --- 
these measurements are not suitable for inclusion in modern PDF sets.

In this work, we therefore exclusively use the more precise measurements
from the LHC. Such measurements were first made at a centre-of-mass 
energy of 7 TeV by the ATLAS~\cite{Aad:2012ux} and CMS~\cite{Chatrchyan:2011vp} 
experiments, and have been subsequently complemented at a higher 
luminosity~\cite{Aad:2014fwa,Chatrchyan:2012ep}, and at centre-of-mass energies 
of 8 TeV~\cite{Aaboud:2017pdi,Khachatryan:2014iya} 
and 13 TeV~\cite{Aaboud:2016ymp,Sirunyan:2016cdg}.
In such analyses the top quark and top antiquark are reconstructed 
from a selection of their leptonic decay channels, 
$t\to e\nu b$, $t\to \mu\nu b$, and $t\to \tau\nu b$, with leptonic $\tau$ 
decays.

While total cross sections, $\sigma_{t+\bar{t}}$, are provided for all of the 
above measurements, individual cross sections for single top-quark and 
top-antiquark production, $\sigma_t$ and $\sigma_{\bar{t}}$, and their ratio, 
$\sigma_t/\sigma_{\bar{t}}$, are available only for the
ATLAS measurements~\cite{Aad:2014fwa,Aaboud:2017pdi,Aaboud:2016ymp}, 
and for the CMS measurements at 8 TeV~\cite{Khachatryan:2014iya} 
and 13 TeV~\cite{Sirunyan:2016cdg}. In these cases, we include the ratio 
$\sigma_t/\sigma_{\bar{t}}$ because its experimental and theoretical analysis 
largely benefits from uncertainty cancellations between the numerator and the 
denominator. For the CMS measurement at 7 TeV~\cite{Chatrchyan:2012ep}, 
we include the total cross section $\sigma_{t+\bar{t}}$; we discard earlier 
measurements from both ATLAS and CMS 
at 7 TeV~\cite{Aad:2012ux,Chatrchyan:2011vp}, 
which are based on reduced luminosities, and are therefore affected by rather
large statistical uncertainties. 
The data points for the total top-quark and top-antiquark cross sections, and 
their ratio, that are included in this work are summarised in 
Table~\ref{tab:tot_measurements}, where we list their centre-of-mass energy, 
$\sqrt{s}$, their integrated luminosity, $\mathcal{L}$, the corresponding 
observable, $\mathcal{O}$, their measured value, and their reference.

\begin{table}[!t]
  \centering
  \scriptsize
\begin{tabular}{lcccccc}
\toprule
  Experiment   
& $\sqrt{s}$ [TeV] 
& $\mathcal{L}$ [fb$^{-1}$] 
& $\mathcal{O}$
& Measurement
& Ref. \\
\midrule
ATLAS       
&  \, 7 
&  4.59 
& $\sigma_t/\sigma_{\bar{t}}$ 
& 2.04 $\pm$ 0.13 (stat.) $\pm$ 0.12 (syst.) 
& \cite{Aad:2014fwa}\\
&  \, 8 
& 20.2  
& $\sigma_t/\sigma_{\bar{t}}$ 
& 1.72 $\pm$ 0.05 (stat.) $\pm$ 0.07 (syst.) 
& \cite{Aaboud:2017pdi}\\
&    13 
&  3.2  
& $\sigma_t/\sigma_{\bar{t}}$ 
& 1.72 $\pm$ 0.09 (stat.) $\pm$ 0.18 (syst.) 
& \cite{Aaboud:2016ymp}\\
\midrule
CMS 
&  \, 7 
&  2.73 
& $\sigma_{t+\bar{t}}$ & 67.2 $\pm$ 6.1 pb                         
& \cite{Chatrchyan:2012ep}\\
&  \, 8 
& 19.7  
& $\sigma_t/\sigma_{\bar{t}}$             
& 1.95 $\pm$ 0.10 (stat.) $\pm$ 0.19 (syst.) 
& \cite{Khachatryan:2014iya}\\
&    13 
&  2.2  
& $\sigma_t/\sigma_{\bar{t}}$             
& 1.81 $\pm$ 0.18 (stat.) $\pm$ 0.15 (syst.) 
& \cite{Sirunyan:2016cdg}\\
\bottomrule
\end{tabular}

  \caption{Measurements of total $t$-channel single top-quark and top-antiquark 
  cross sections and their ratio analysed in this work. For each of them we 
  indicate the experiment, the centre-of-mass energy, $\sqrt{s}$, the 
  integrated luminosity, $\mathcal{L}$, the measured observable, $\mathcal{O}$,
  its value (stat.~and syst.~denote statistical and systematic 
  uncertainties), and its reference.}
\label{tab:tot_measurements}
\end{table}

\subsection{Differential cross sections}
\label{sec:diffdistrs} 

We also study differential cross section measurements for $t$-channel single
top-quark and top-antiquark production from ATLAS at a centre-of mass energy of 
7 TeV~\cite{Aad:2014fwa}. These measurements are provided as a function of the
transverse momentum, $p_T(t)$ ($p_T(\bar{t})$), and the absolute value of the 
rapidity, $|y(t)|$ ($|y(\bar{t})|$), of the top quark (top antiquark), and are
provided in two forms: one in which they are normalised to the corresponding 
total cross section evaluated by integrating over all kinematic bins, and one
in which they are not. In the former case, we always explicitly remove the last 
bin of the distribution, because its value is fixed by the normalisation
condition. In principle, the inclusion of the distributions in either form 
should have the same statistical impact in a fit of PDFs. In this respect,
to avoid loss of information on the overall normalisation, we always accompany
normalised distributions with the total cross sections corresponding to the 
same experimental analysis. In contrast, the 
simultaneous inclusion of total cross sections and absolute differential
distributions would result in double counting, because these are both 
determined from the same data sample.

In practice, however, the inclusion of absolute or normalised distributions
in a fit might make a difference, because some approximations have to be made 
in the analysis of the data. First, the experimental correlation between the 
differential distributions and the total cross section --- which should be 
taken into account --- is not available. We therefore assume that, because the 
total cross section amounts to a single data point, and its exclusion from the
fit would make no difference, the neglect of such correlation will 
{\it a fortiori} make little difference. Second, the theoretical 
description of normalised distributions benefits from a partial cancellation
of theoretical uncertainties that occurs in the ratio. Similarly to what happens
for experimental uncertainties, such a cancellation would be compensated by the 
theoretical uncertainty on the total cross section and the associated 
correlations. While theoretical uncertainties could be included in a 
fit~\cite{AbdulKhalek:2019bux,AbdulKhalek:2019ihb}, they are not in this study.

We account for all of the available information on systematic and statistical
correlations. Systematic uncertainties are correlated, for each separate 
source, bin-by-bin within each distribution. Likewise, we also include 
bin-by-bin correlations of statistical uncertainties within each distribution.
Systematic and statistical correlations across different distributions, and 
between normalised distributions and the total cross sections, are not 
determined experimentally, and are therefore not considered here. The only 
exception is the luminosity uncertainty (for absolute distributions and the
total cross section), which is 100\% correlated across all 
of the ATLAS measurements at the same centre-of-mass energy. Because of this 
partial knowledge of correlations, and to avoid 
double counting, only one of top-quark distributions and one of top-antiquark 
distributions can be included in a fit at a time.
In Table~\ref{tab:distributions} we summarise the features of each kinematic
distribution, indicating whether it is an absolute or a normalised distribution,
the number of data points, $N_{\rm dat}$, and its kinematic coverage.

\begin{table}[!t]
  \centering
  \scriptsize
\begin{tabular}{lcc}
\toprule
  Distribution 
& $N_{\rm dat}$ 
& Kinematic range 
\\
\midrule
  $d\sigma/dp_T(t)$ 
& 5 
& $0 < p_T(t) < 500$ GeV 
\\
  $d\sigma/dp_T(\overline{t})$ 
& 5 
& $0 < p_T(\overline{t}) < 500$ GeV 
\\
  $d\sigma/d |y(t)|$ 
& 4 
& $0 < |y(t)| < 3$ 
\\
  $d\sigma/d|y(\overline{t})|$ 
& 4 
& $0 < |y(\overline{t})| < 3$ 
\\
  $(1/\sigma)d\sigma/dp_T(t)$ 
& 4 
& $0 < p_T(t) < 500$ GeV 
\\
  $(1/\sigma)d\sigma/dp_T(\overline{t})$ 
& 4 
& $0 < p_T(\overline{t}) < 500$ GeV 
\\
  $(1/\sigma)d\sigma/d|y(t)|$ 
& 3 
& $0 < |y(t)| < 3$ 
\\
  $(1/\sigma)d\sigma/d|y(\overline{t})|$ 
& 3 
& $0 < |y(\overline{t})| < 3$ 
\\
\bottomrule
\end{tabular}

  \caption{The ATLAS measurements~\cite{Aad:2014fwa} of $t$-channel single 
  top-quark and top-antiquark differential cross sections at a centre-of-mass 
  energy of 7 TeV studied here. Distributions are 
  a function of the top-quark (top-antiquark) transverse momentum, $p_T(t)$ 
  ($p_T(\bar{t})$), and absolute rapidity, $|y(t)|$ ($|y(\bar{t})|$).
  The bin edges for the transverse momentum distributions are 0, 45, 75, 110, 
  150 and 500~GeV, while for the rapidity distributions they are 0, 0.2, 0.6, 
  1.1 and 3.0. For each distribution, we indicate the number of bins, 
  $N_{\rm dat}$, and their overall kinematic range.}
  \label{tab:distributions}
\end{table}

Finally, we shall note that other measurements of differential distributions
for $t$-channel single top-quark and top-antiquark production have become available 
recently. Namely, the ATLAS measurements at a centre-of-mass energy of 
8 TeV~\cite{Aaboud:2017pdi}, and the CMS measurements at a centre-of-mass
energy of 8 TeV~\cite{Khachatryan:2015dzz} and 13 TeV~\cite{Sirunyan:2019hqb}.
However, because no experimental information is provided on the individual
sources of systematic uncertainty and their correlations for these measurements,
they are not currently suitable for use in precision PDF determinations.
Therefore, they are not included in the fits presented in 
Sect.~\ref{sec:fits}. Nevertheless, we illustrate the 8 TeV ATLAS data of 
Ref.~\cite{Aaboud:2017pdi} in Appendix~\ref{app:ATLAS8TeVdiff}.

\section{Data-theory comparison} 
\label{sec:data-theory-comparison}

In this section we scrutinise the description of the datasets discussed in 
Sect.~\ref{sec:exp_data} upon various assumptions in the underlying theory.
Specifically, we investigate the impact of QCD corrections, EW corrections, 
and the flavour scheme adopted, especially in relationship
with the choice of the factorisation scale and, in the case of the 5FS,
the bottom-quark threshold. Our study aims at identifying the optimal 
theoretical settings to include the data in a PDF fit.

\subsection{QCD corrections}
\label{sec:QCD}

Computations of NLO QCD corrections to $t$-channel single top-quark and
top-antiquark production total cross sections were performed for the first time 
more than twenty years ago~\cite{Bordes:1994ki}, and have been extended to 
fully differential cross sections~\cite{Harris:2002md}, including 
decays~\cite{Falgari:2010sf,Schwienhorst:2010je,Brucherseifer:2013iv,Gao:2017goi}, since then.
Monte Carlo simulations that are accurate to NLO have been implemented 
consistently within several generators, namely \amc~\cite{Frixione:2005vw},
\powheg~\cite{Alioli:2009je}, and \sherpa~\cite{Bothmann:2017jfv}. They all
allow for the construction of fast interpolation grids, whereby partonic matrix 
elements are precomputed in such a way that the numerical convolution with 
generic input PDFs can be efficiently approximated by means of
interpolation techniques. Such grids are a fundamental ingredient in any PDF
fit, where the evaluation of the hadronic cross sections needs to be performed 
a large number of times. 

We generate fast interpolation grids, accurate to NLO in QCD, for all of the
$t$-channel single top-quark and top-antiquark data described in 
Sect.~\ref{sec:exp_data}. To this purpose, 
we use \amc~(v2.6.6~\cite{Alwall:2014hca})
interfaced to \applgrid~\cite{Carli:2010rw} 
with \amcfast~\cite{Bertone:2014zva}; the {\applgrid} 
output is processed with 
\apfelgrid~\cite{Bertone:2016lga} to construct fast interpolation grids
in a format suitable for an NNPDF fit.
The computation is performed in the 5FS with fixed factorisation and
renormalisation scales $\mu_f = \mu_r = m_t$, where $m_t$ is the
mass of the top quark, which we set equal to 172.5~GeV. The values of the 
bottom threshold, the $W$-boson mass, the $Z$-boson mass, and the strong
coupling are chosen to be $m_b=4.92$ GeV, $M_W=80.398$ GeV, $M_Z=91.1876$ GeV,
and $\alpha_s(M_Z)=0.118$, 
respectively~\cite{Ball:2017nwa}.
We require that the number of Monte Carlo events generated in the 
computation is sufficiently high to ensure that the residual
relative statistical fluctuations are smaller than one permille; that is, 
that they 
are negligible with respect to the experimental precision of the data and the 
theoretical accuracy of the corresponding predictions.
Our numerical results are benchmarked against the corresponding calculations 
obtained independently in Ref.~\cite{Berger:2017zof}. An excellent agreement, 
within the Monte Carlo statistical uncertainty, is found.

Very recently, NNLO QCD corrections to both total and differential cross 
sections have also been computed~\cite{Brucherseifer:2014ama,Berger:2016oht,
Berger:2017zof}. When the top decay is calculated, it is done in the 
narrow-width approximation, under which the QCD corrections to the 
top-(anti)quark production
and the decay are factorisable. The full QCD corrections are approximated by the
vertex corrections (a procedure known as the {\it structure function approach}
in the inclusive case~\cite{Han:1992hr}).
These calculations, however, are not available in a format suitable for their 
direct inclusion in a global fit of PDFs.
In order to make use of them, we therefore resort to the $C$-factor 
approximation~\cite{Ball:2011uy}: our NLO results computed with NNLO
PDFs are multiplied by a bin-by-bin factor
\begin{equation}
  C = \frac{\hat{\sigma}_{\rm NNLO} 
  \otimes \mathcal{L}_{\rm NNLO}}{\hat{\sigma}_{\rm NLO} 
  \otimes \mathcal{L}_{\rm NNLO}} ,
\label{eq:cfact}
\end{equation}
where $\hat\sigma_{\rm NNLO}$ ($\hat\sigma_{\rm NLO}$) is the partonic cross
section computed with NNLO (NLO) matrix elements, and $\mathcal{L}_{\rm NNLO}$ is
the corresponding parton luminosity evaluated with a reference set of NNLO PDFs.
The numerator and the denominator in Eq.~\eqref{eq:cfact} are computed with 
the code of Refs.~\cite{Berger:2016oht,Berger:2017zof} and the baseline 
NNPDF3.1 PDF set~\cite{Ball:2017nwa}.

\begin{figure}[t]
  \centering
  \includegraphics[angle=270,scale=0.22,clip=true,trim=0 9cm 0 0]{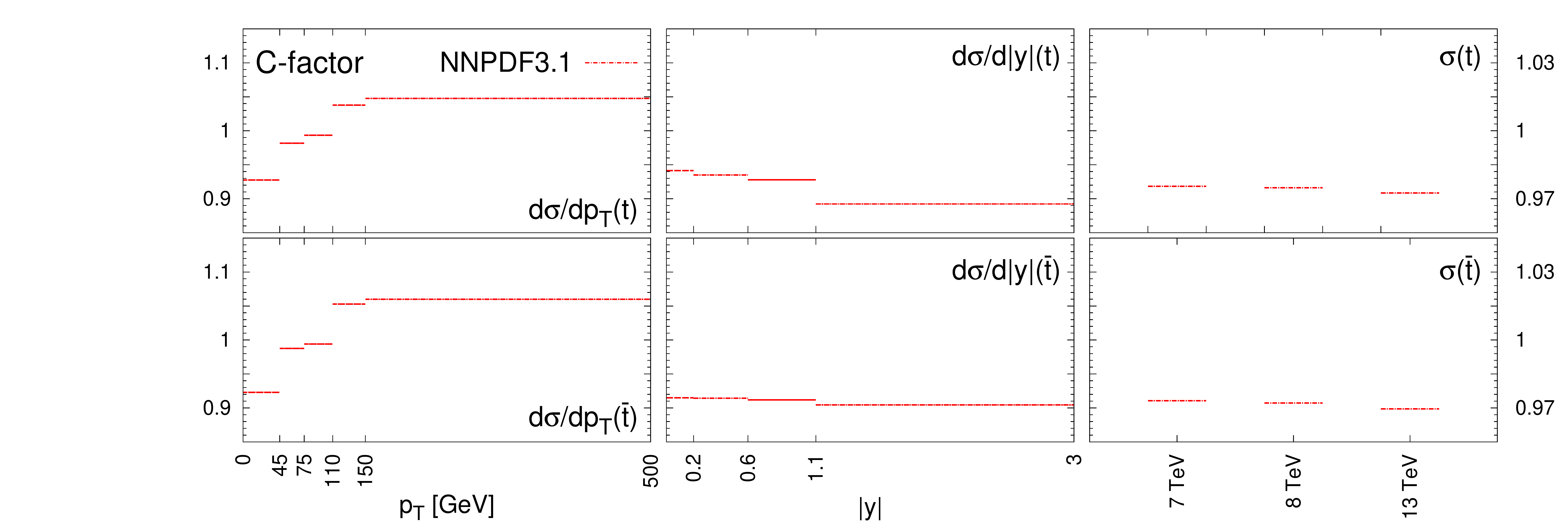}\\
  \caption{The $C$-factors, Eq.~\eqref{eq:cfact}, computed with the 
  {\tt NNPDF31\_nnlo\_as\_0118} PDF set, for the distributions differential in 
  the transverse momentum (left), in the absolute rapidity (centre), and for 
  the total cross sections (right) of top-quark (top) and top-antiquark 
  (bottom). Results are displayed at 7 TeV for differential distributions, 
  and at 7, 8, and 13 TeV for total cross sections. Note that the scale on the 
  vertical axis varies: it ranges from 0.9 to 1.1 for the transverse momentum 
  distributions, while it ranges from 0.97 to 1.03 for the rapidity 
  distributions and for the total cross sections.}
  \label{fig:cfact}
\end{figure}

In Fig.~\ref{fig:cfact} we display the $C$-factors computed according to 
Eq.~\eqref{eq:cfact} relative to the datasets discussed in 
Sect.~\ref{sec:exp_data} (see also 
Tables~\ref{tab:tot_measurements}--\ref{tab:distributions}):
specifically, for the differential distributions in 
the transverse momentum and absolute rapidity of the top quark and of the 
top antiquark at 7 TeV, and for the total cross sections for the top quark and
the top antiquark at 7, 8, and 13 TeV. We observe that NNLO corrections are 
comparable for top-quark and top-antiquark production. They display some
structure for the transverse momentum distributions, where they lead to an 
enhancement of the cross section (up to about 5\%) at large values of the 
transverse momentum. Conversely, NNLO corrections display little sensitivity to 
the kinematics for the absolute rapidity distributions and for the total cross 
section; in these cases they lead to a suppression of the cross section 
(of about 3\%) across all of the measured ranges in rapidity and centre-of-mass 
energy.

\begin{figure}[t]
  \centering
  \includegraphics[angle=270,scale=0.21,clip=true,trim=0 0 0 3.5cm]{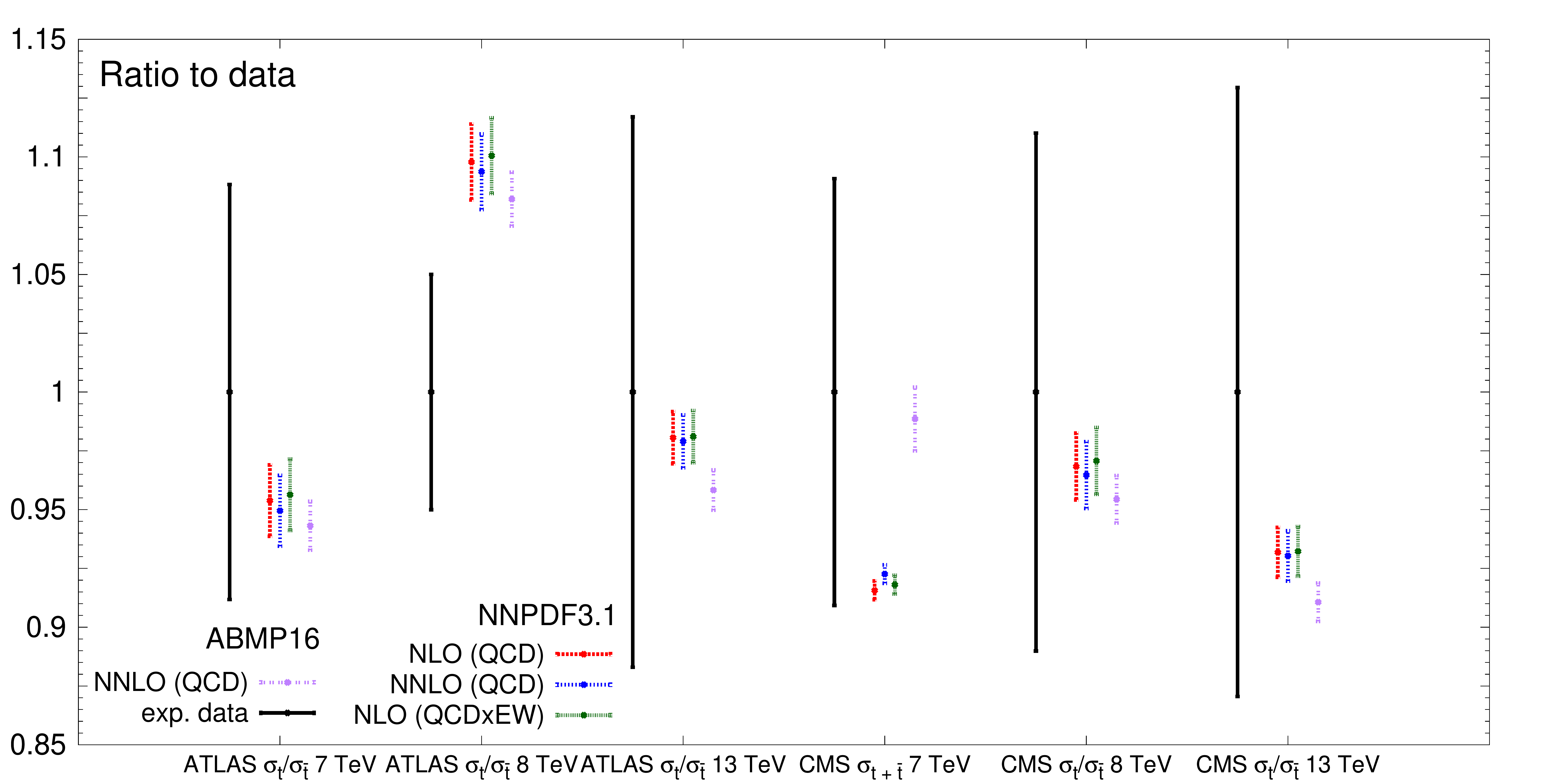}\\
  \caption{The total $t$-channel single top-quark and top-antiquark cross 
  sections and their ratios from ATLAS and CMS analysed in this work (see
  Table~\ref{tab:tot_measurements}). The measurements are compared to 
  theoretical predictions computed at NLO and NNLO in pure QCD, and at NLO in 
  mixed QCD and EW theories. The corresponding sets of PDFs from the NNPDF3.1
  determination are used: {\tt NNPDF31\_nlo\_as\_0118}, 
  {\tt NNPDF31\_nnlo\_as\_0118}, and {\tt NNPDF31\_nlo\_as\_0118\_luxqed},
  respectively. For NNLO QCD theory, an alternative PDF set that includes 
  some of the data in Table~\ref{tab:tot_measurements}, 
  {\tt ABMP16als118\_5\_nnlo}, is also used for comparison 
  (see text for details).
  Both the data and the theoretical predictions are 
  normalised to the central value of the data. Error bars on the data correspond
  to the square root of the sum in quadrature of statistical and systematic 
  uncertainties; error bars on the theoretical predictions correspond to the 
  68\% CL PDF uncertainty.}
\label{fig:tot_measurements}
\end{figure}

\begin{figure}[t]
  \centering
  \includegraphics[angle=270,scale=0.252]{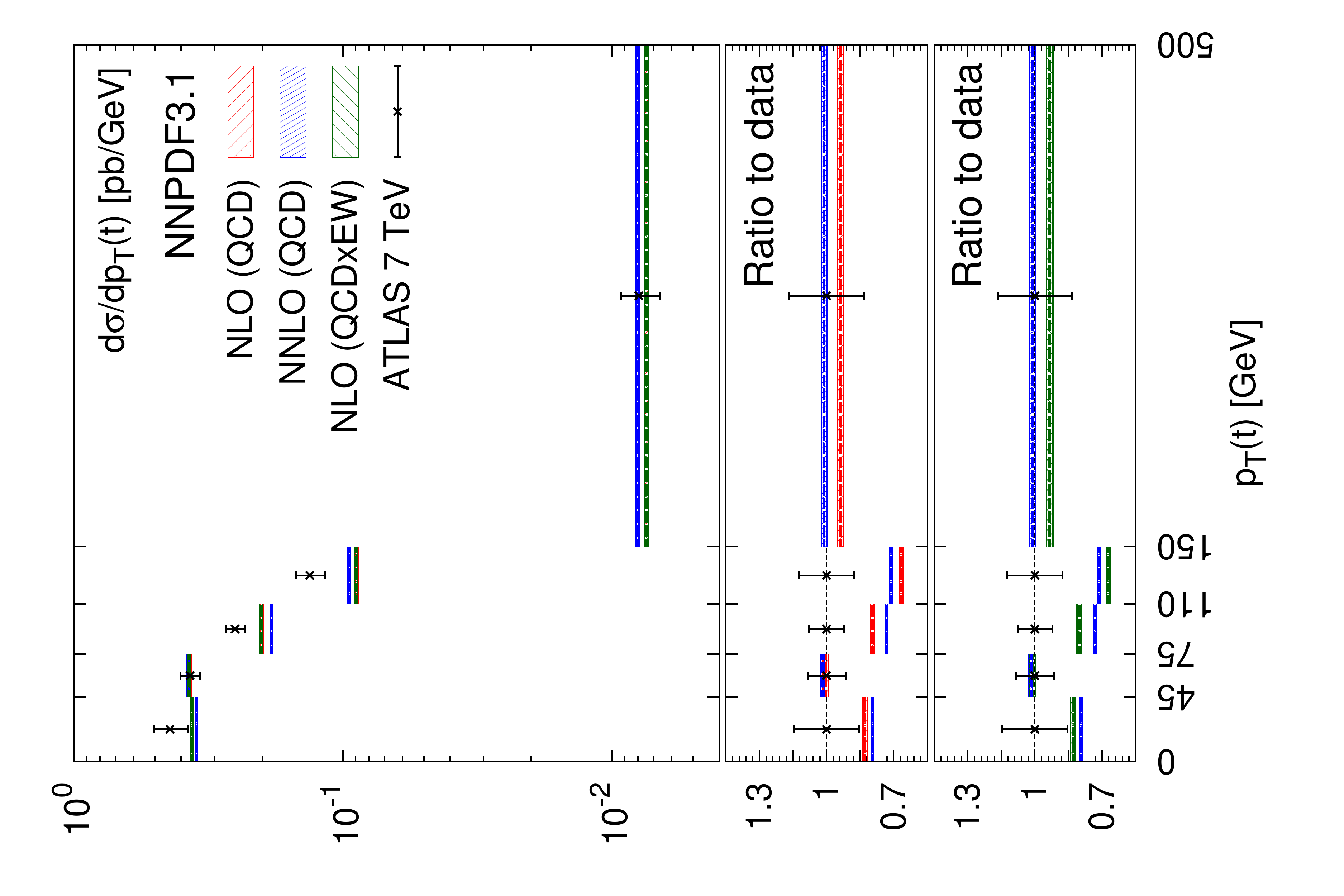}\ \ \ \ \
  \includegraphics[angle=270,scale=0.252]{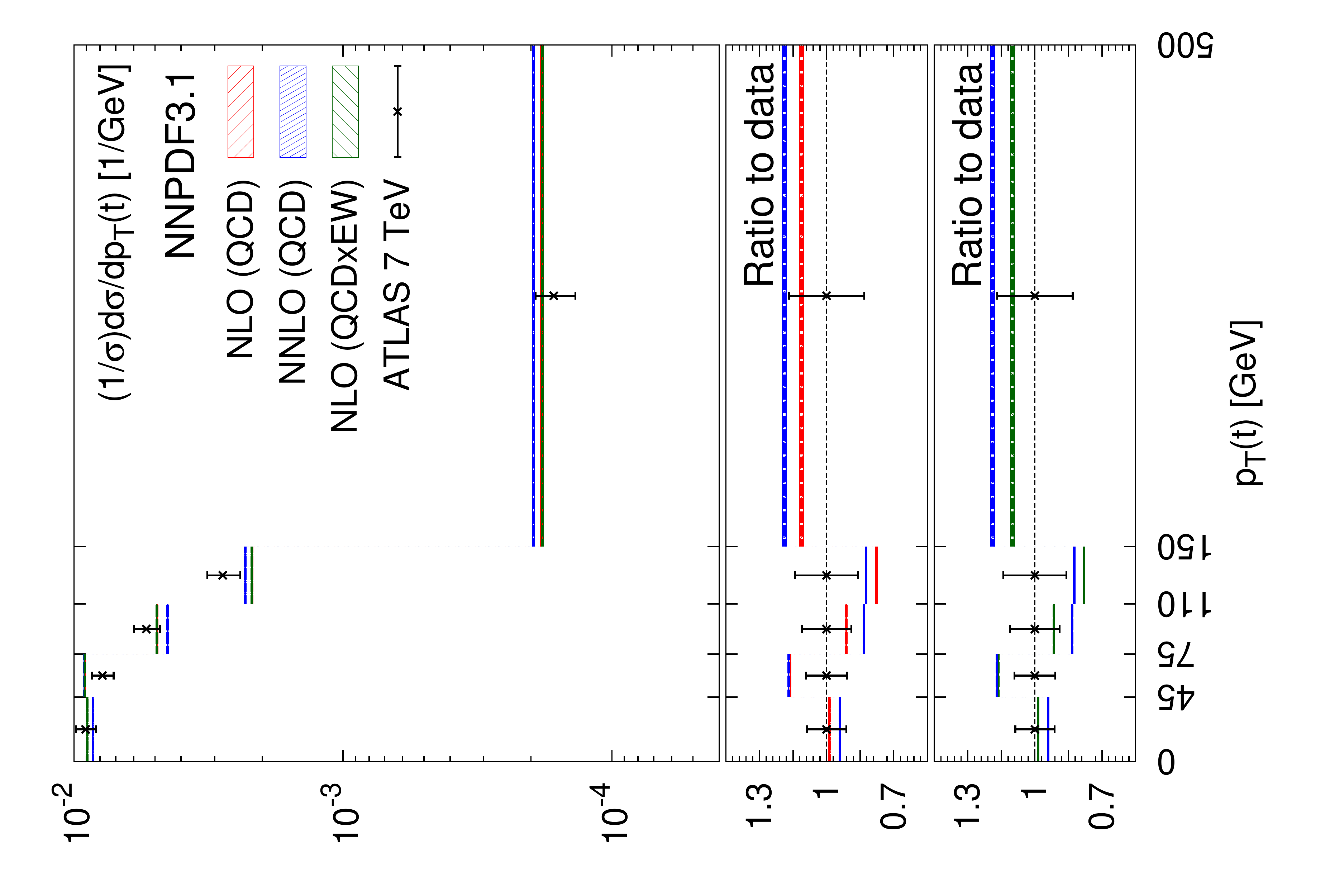}\\
  \includegraphics[angle=270,scale=0.252]{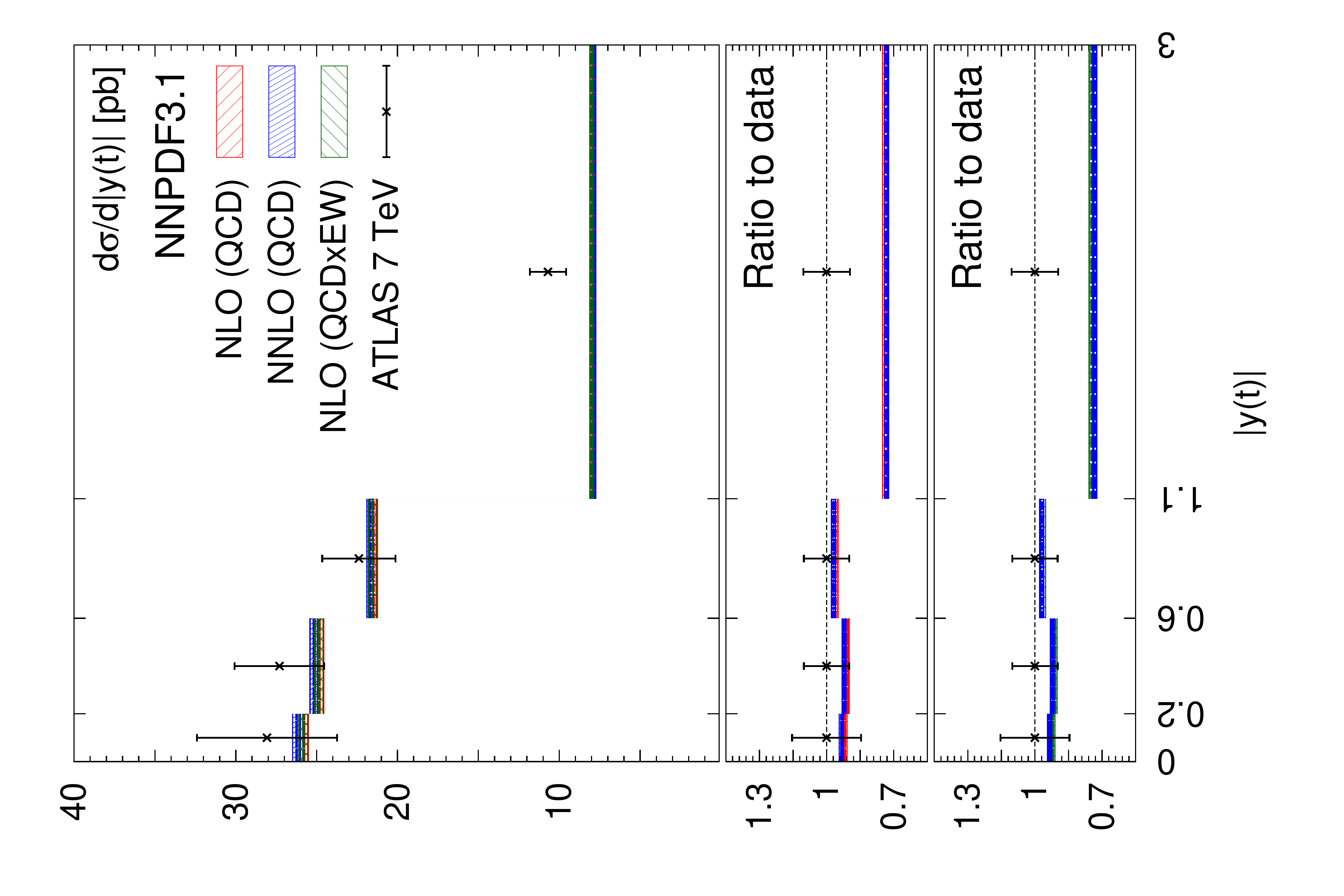} \ \ \ \ \
  \includegraphics[angle=270,scale=0.252]{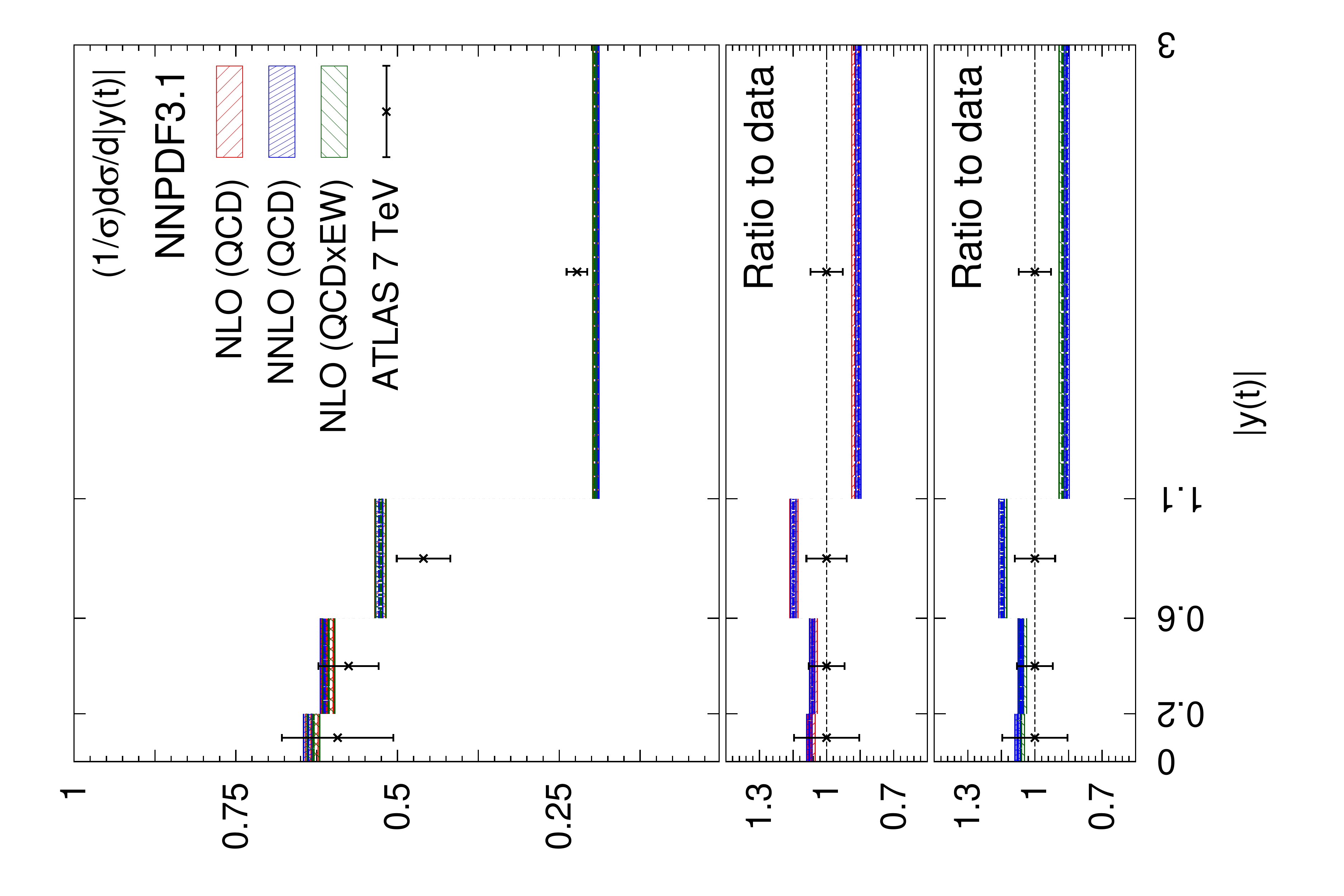}\\
  \caption{The $t$-channel single top-quark absolute (left) and normalised 
  (right) cross sections from ATLAS analysed in this work (see 
  Table~\ref{tab:distributions}). The measurements are compared to 
  theoretical predictions computed as in Fig.~\ref{fig:tot_measurements}.
  The two lower panels display the data and the theoretical predictions 
  normalised to the central value of the data.}
\label{fig:data_theory_top}
\end{figure}

\begin{figure}[!t]
  \centering
  \includegraphics[angle=270,scale=0.252]{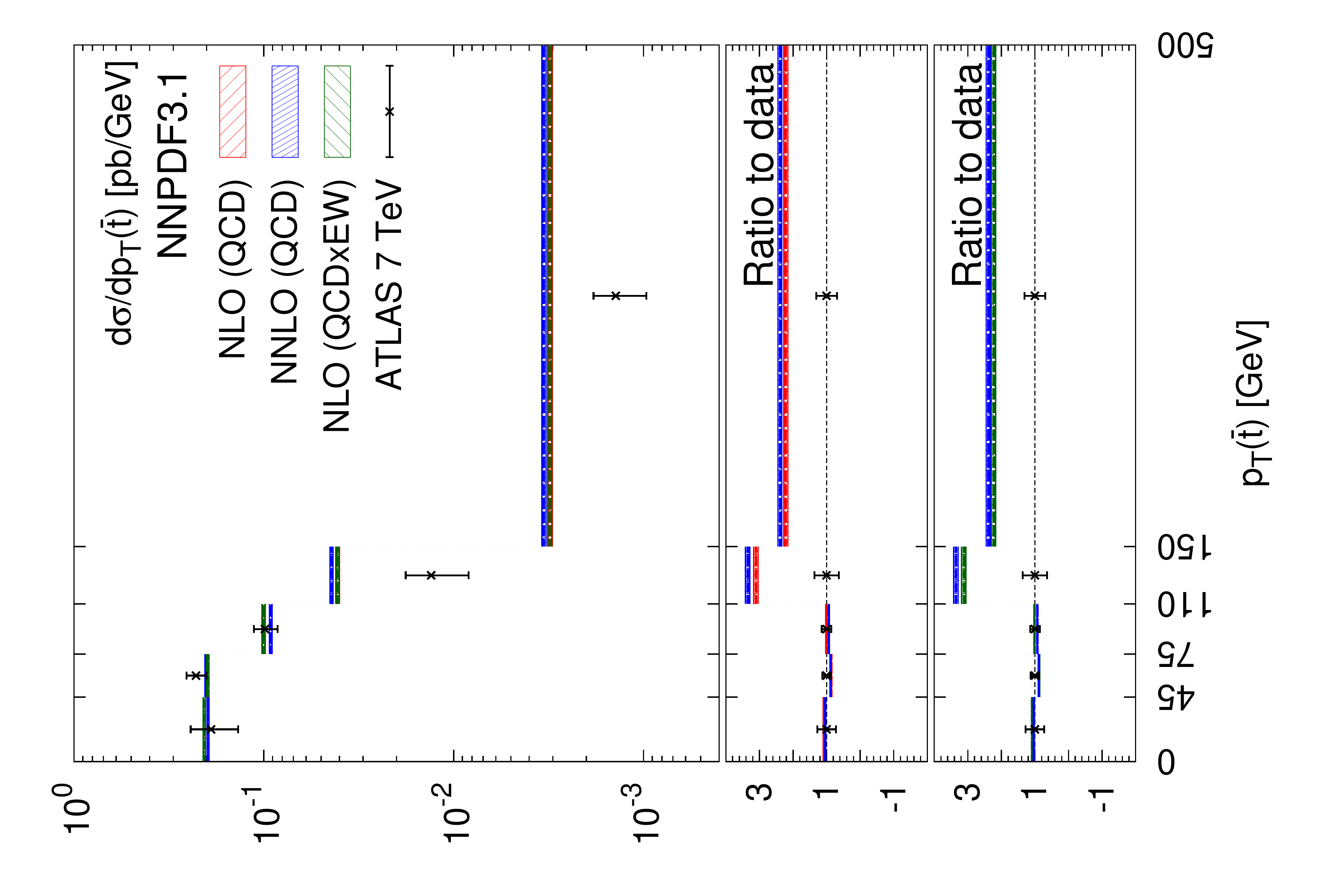}\ \ \ \ \
  \includegraphics[angle=270,scale=0.252]{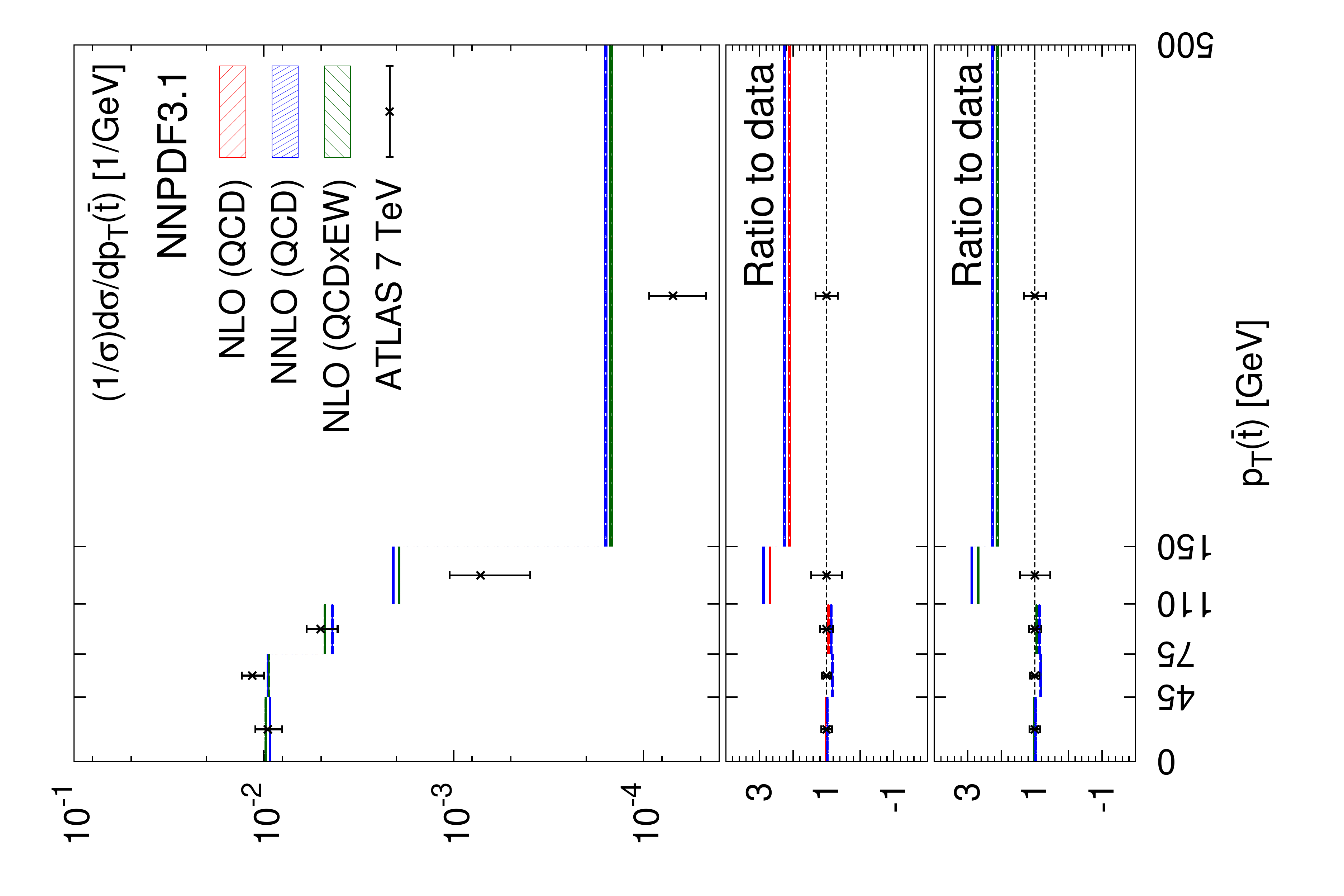}\\
  \includegraphics[angle=270,scale=0.252]{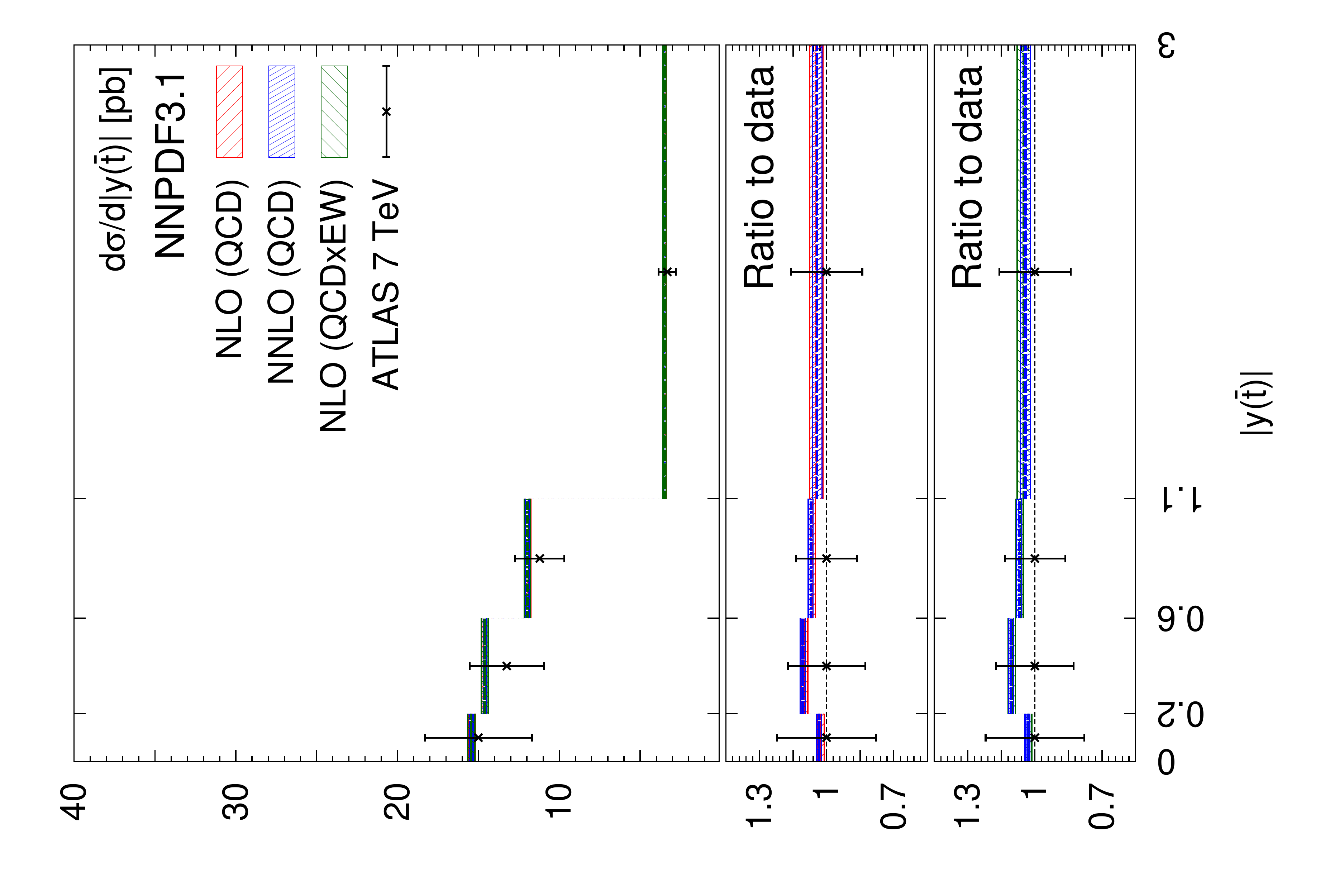} \ \ \ \ \
  \includegraphics[angle=270,scale=0.252]{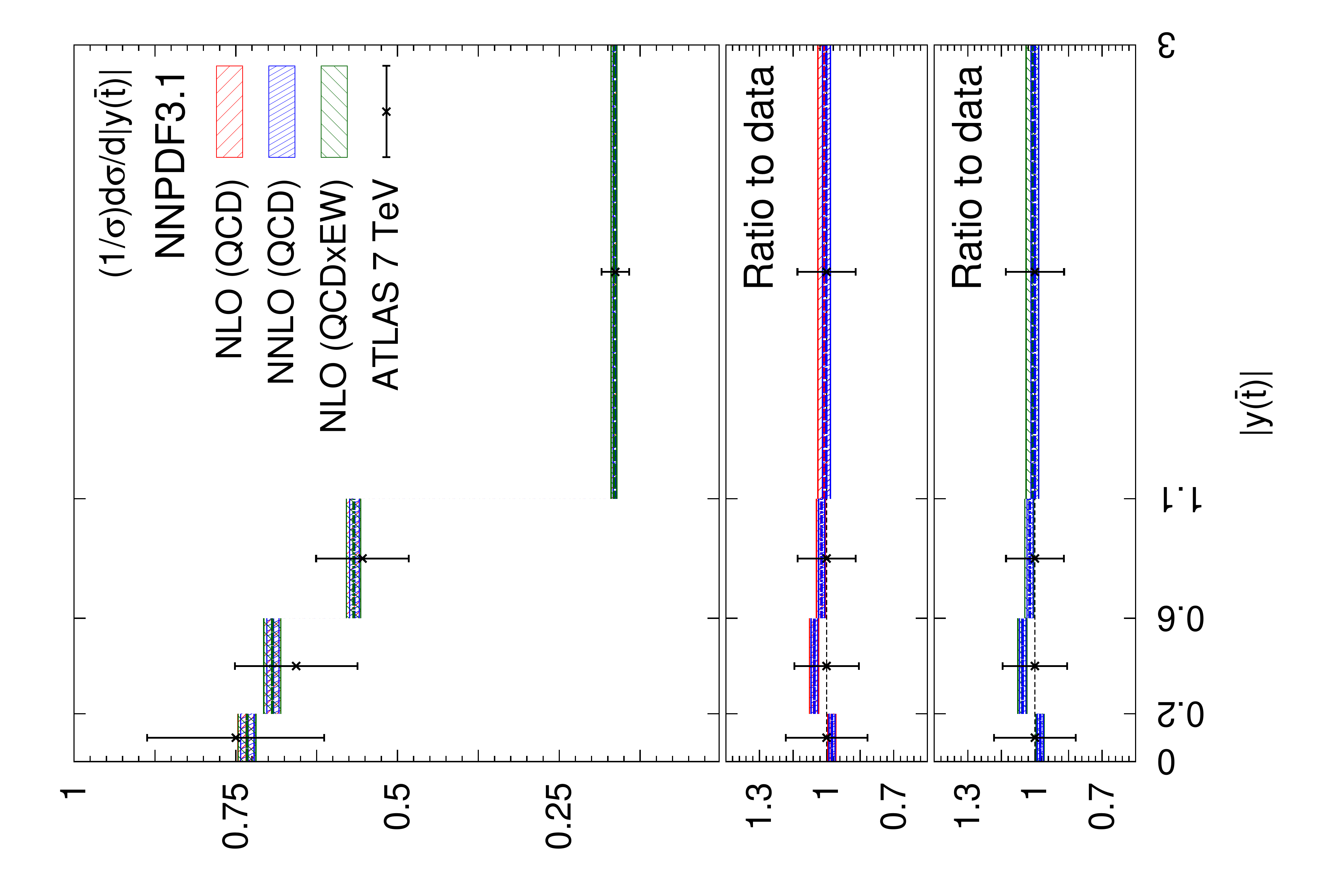}\\
  \caption{Same as Fig.~\ref{fig:data_theory_top}, but for top-antiquark 
  production.}
\label{fig:data_theory_atop}
\end{figure}

We now compare theoretical predictions, computed as explained above at NLO
and NNLO in pure QCD, to the experimental data listed in
Tables~\ref{tab:tot_measurements}--\ref{tab:distributions}:
in Fig.~\ref{fig:tot_measurements} we display the results for the total
cross sections, in each case normalised to the central value of the data point; 
in Figs.~\ref{fig:data_theory_top}--\ref{fig:data_theory_atop} we display the
results for the differential cross sections, both absolute (left) and 
normalised (right), for top-quark and top-antiquark production, respectively.
For data-theory comparison purposes, we do not remove the last bin of the 
normalised distributions, opposite to what we do in a fit (see
Sect.~\ref{sec:diffdistrs}), the reason being consistency with 
Ref.~\cite{Aad:2014fwa}.
For the differential distributions, the data and the theoretical predictions are
normalised to the central value of the data in the central inset of each plot.
The error bars on the data correspond to the square root of the diagonal 
elements of the experimental covariance matrix (see Eq.~\eqref{eq:cov_exp}). 
The error bars (or bands) on theoretical predictions correspond to 
the 68\% confidence level (CL) PDF uncertainty. The {\tt NNPDF31\_nlo\_as\_0118}
and {\tt NNPDF31\_nnlo\_as\_0118} PDF sets are used for NLO and NNLO
predictions, respectively.

We note from Figs.~\ref{fig:tot_measurements}--\ref{fig:data_theory_atop} that
NNLO QCD corrections are in general small, consistently with expectations
based on $C$-factors (see Fig.~\ref{fig:cfact}). 
For the ratios of single top-quark
to top-antiquark total cross sections their effect cancels out almost 
completely; for their sum (in the case of the CMS data point at 7 TeV), they 
increase the theoretical expectations by about 2\%, an amount which is smaller 
than both the data and the PDF uncertainties. For the distributions differential
in the transverse momentum of the top quark (top antiquark), they can be as 
large as 10\%, especially at large values of $p_T(t)$ ($p_T(\bar{t})$), while 
for the distributions differential in the rapidity of the top quark 
(top antiquark), they never exceed 1--2\%. In both cases the NNLO QCD 
corrections display a similar size for the absolute and the normalised 
distributions. They result in a shift of the theoretical predictions always 
smaller than the data uncertainty, which remains rather larger 
than the PDF uncertainty. For this reason, we anticipate a moderate impact of 
single top-quark and top-antiquark data on PDFs when they are included in a fit 
(see Sect.~\ref{sec:fits}).

That being said, a qualitative inspection of 
Figs.~\ref{fig:tot_measurements}--\ref{fig:data_theory_atop} reveals that
theoretical predictions based on the NNPDF3.1 PDF sets are overall in 
fair agreement with the data. Significant discrepancies are observed only in 
two cases, consistently with what was qualitatively reported in 
Ref.~\cite{Aad:2014fwa} (see in particular Figs.~15--16 therein).
First, the ATLAS single top-quark to top-antiquark production 
cross section ratio at 8 TeV differs from the theoretical expectation by 
about 10\%, corresponding to a 2$\sigma$ interval in units of the experimental
uncertainty. Such a discrepancy is not observed for the analogous 
measurement from CMS: after all, it is apparent from 
Table~\ref{tab:tot_measurements} that the ATLAS and CMS measurements at 8 TeV
have barely touching error bands, {\it i.e.}~a $\sqrt{2}\sigma$
discrepancy. Second, the single top-antiquark cross section differential in the
transverse momentum of the top-antiquark differs from the theoretical 
expectation by up to 300\% for the two bins at the largest values of 
$p_T(\bar{t})$. This discrepancy is independent from the normalisation of the 
distribution, as it is present for both the absolute and the normalised 
cross sections. Furthermore, it is not alleviated by the inclusion of NNLO QCD 
corrections in the computation of the theoretical expectations.

\begin{figure}[!t]
  \centering
  \includegraphics[angle=270,scale=0.252]{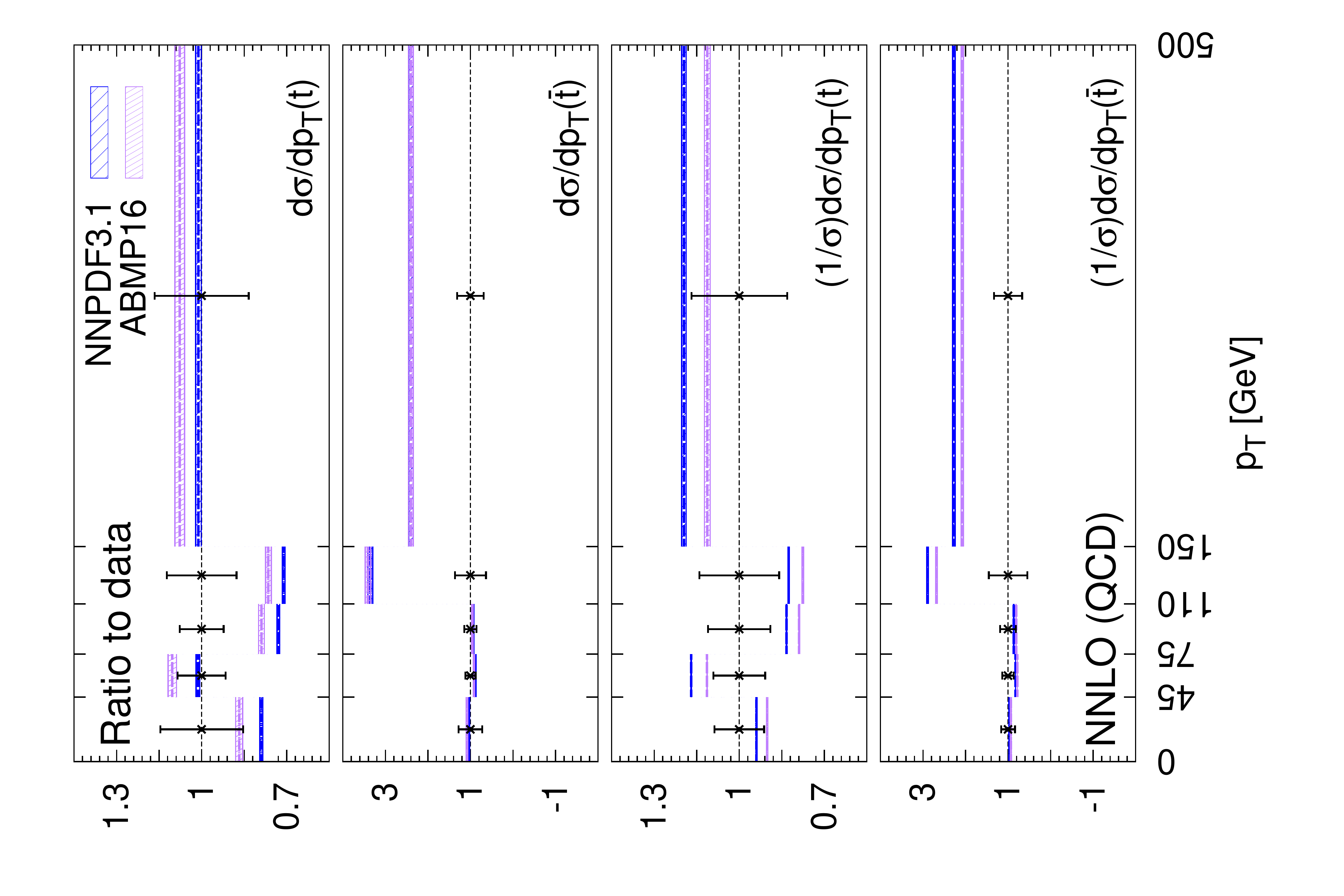}\ \ \ \ \
  \includegraphics[angle=270,scale=0.252]{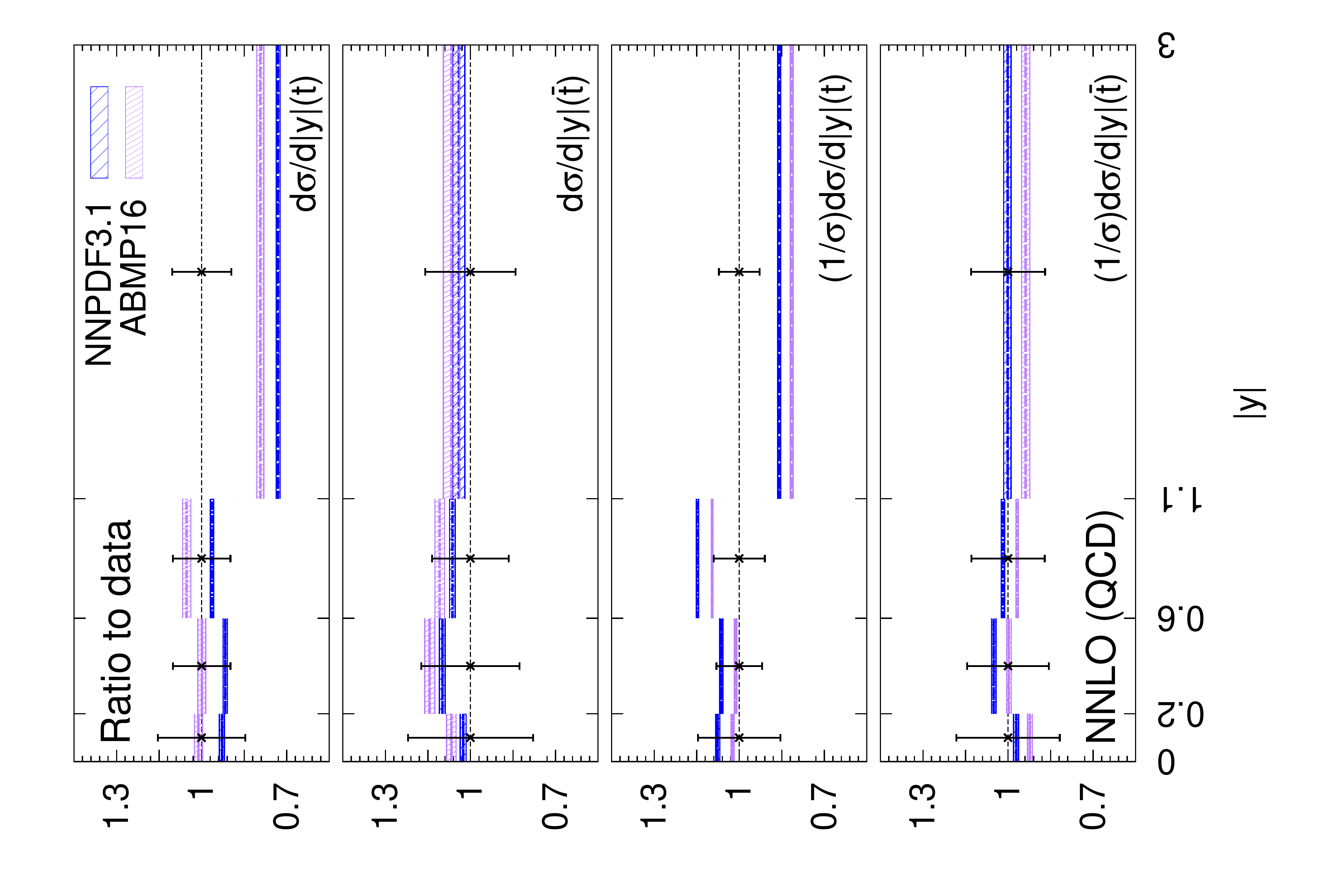}\\
  \caption{The $t$-channel absolute and normalised single top-(anti)quark 
  distributions differential in the transverse momentum (left) and rapidity
  (right) of the top-(anti)quark (left). The measurements are compared to 
  NNLO QCD theoretical predictions obtained either with the 
  {\tt NNPDF31\_nnlo\_as\_0118} or the {\tt ABMP16als118\_5\_nnlo} PDF sets.
  The data and the theoretical predictions are normalised to the central value
  of the data.}
\label{fig:data_theory_abmp}
\end{figure}

It might be interesting to investigate whether these discrepancies
persist if PDF sets other than NNPDF3.1 are used. We therefore look at the
data-theory comparison obtained alternatively with the 
MMHT~\cite{Harland-Lang:2014zoa}, CT18~\cite{Hou:2019qau}, or the PDF4LHC15 
combination~\cite{Butterworth:2015oua}. None of these PDF sets include any
single top data. Such comparisons appear to be almost identical to NNPDF3.1,
therefore we do not report them explicitly in
Figs.~\ref{fig:tot_measurements}--\ref{fig:data_theory_atop}. 
For completeness, we also look at data-theory comparisons obtained with 
the ABMP16 PDF set~\cite{Alekhin:2017kpj}: this is the only PDF determination 
that includes some single top data, namely the total cross section for single 
top-quark production from the combined D0 and CDF measurements at 
$\sqrt{s}=1.96$ TeV~\cite{Aaltonen:2015cra} 
and from ATLAS~\cite{Aad:2014fwa,Aaboud:2016ymp,Tepel:2014kna} and
CMS~\cite{Chatrchyan:2012ep,Khachatryan:2014iya,Sirunyan:2016cdg} 
measurements at $\sqrt{s}=7$, 8 and 13 TeV.
Specifically, we restrict ourselves to NNLO, and use the PDF set determined 
with five active flavours and a fixed value of $\alpha_s=0.118$, 
{\tt ABMP16als118\_5\_nnlo}. We report such comparisons in 
Fig.~\ref{fig:tot_measurements} for the total cross section, 
and in Fig.~\ref{fig:data_theory_abmp}
for the differential cross sections. In all cases, the data and the theoretical 
predictions are normalised to the central value of the data.

We see that the ABMP16 and the NNDPF3.1 predictions differ by an amount which 
depends on the specific distribution considered. For the total cross sections,
the largest differences are observed for the single top-quark to single
top-antiquark ratio at $\sqrt{s}=13$ TeV (about $\sqrt{2}\sigma$ in units
of the PDF uncertainty), and for the total top-quark and top-antiquark cross 
section from CMS at $\sqrt{s}=7$ TeV (about $3\sigma$). Such differences are
always smaller than the data uncertainty. For the differential 
cross sections, similar differences are observed for all distributions, 
with ABMP16 predictions generally being closer to the data than NNPDF3.1 
predictions (by about 2--3$\sigma$). Nevertheless, they still fail to provide
a good description of the ATLAS single top-quark to top-antiquark ratio 
measurement at $\sqrt{s}=8$ TeV, and of the last two bins of the 
single top-antiquark absolute and normalised distributions differential
in the transverse momentum of the top antiquark. We therefore conclude that
the same discrepancies persist across any available PDF set.

The impact on these discrepancies of other theoretical details, such as the 
inclusion of EW corrections, the dependence on the flavour scheme, 
and the variation of the factorisation and the renormalisation scales will be
studied in Sects.~\ref{sec:EW}--\ref{sec:4vs5FS}. A quantitative assessment of
the agreement between data and theory will be provided in Sect.~\ref{sec:chi2}.

\subsection{EW corrections}
\label{sec:EW}

Computations of NLO EW and NLO QCD$\times$EW corrections to $t$-channel single 
top-quark and top-antiquark production were performed in 
Refs.~\cite{Beccaria:2006ir,Mirabella:2008gj,Bardin:2010mz} in the context  
of supersymmetric extensions of the SM (a soft approximation to deal with 
real-emission contributions was employed in~\cite{Beccaria:2006ir}).
Corresponding Monte Carlo simulations have been implemented within the 
\amc~\cite{Frederix:2018nkq} (v3.0.1) generator, which we use to compute 
theoretical predictions accurate to NLO QCD$\times$EW for the data discussed 
in Sect.~\ref{sec:exp_data}. To this purpose, we adopt the same physical 
parameters and settings as in Sect.~\ref{sec:QCD}, and we take the PDFs 
(including the photon PDF) from the {\tt NNPDF31\_nlo\_as\_0118\_luxqed} 
set~\cite{Bertone:2017bme}.

Our results are compared to the NLO and NNLO QCD predictions in 
Fig.~\ref{fig:tot_measurements} for the ratio of the total top-quark to 
top-antiquark cross sections (or their sum, for the CMS data point at 7 TeV), 
and in Figs.~\ref{fig:data_theory_top}--\ref{fig:data_theory_atop} for the
top-quark and top-antiquark differential distributions, respectively. 
Results in Fig.~\ref{fig:tot_measurements} and in the lowest inset of 
Figs.~\ref{fig:data_theory_top}--\ref{fig:data_theory_atop} are normalised
to the central value of the data points. Error bars (or bands) on 
theoretical predictions account for the 68\% CL PDF uncertainty.

As is apparent from 
Figs.~\ref{fig:tot_measurements}--\ref{fig:data_theory_atop},
EW corrections are almost negligible for the data under consideration here.
For the ratios of single top-quark to top-antiquark total cross sections their 
effect cancels out completely; for their sum (in the case of the CMS data point 
at 7 TeV), they increase the NLO theoretical expectation by about 2\%, an 
amount which is comparable to the effect of the NNLO QCD correction, but still
significantly smaller than the data and the PDF uncertainties.
For the differential distributions, EW corrections account for a sub-percent 
effect, irrespective of whether the distributions are absolute or normalised, 
whether they are for top-quark or top-antiquark production, or whether
they are differential in the transverse momentum
or in the rapidity of the final particle. Such a result is
consistent with what was found in Ref.~\cite{Frederix:2019ubd}, where EW
corrections larger than 1\% were observed only for $p_T(t)$ and $p_T(\bar{t})$
bins above about 150 GeV, and at a centre-of-mass energy of 13 TeV. The size of 
such effects is much smaller than the statistical significance of the data,
and obviously cannot explain the data-theory discrepancies observed in
Sect.~\ref{sec:QCD}.

\subsection{Heavy flavour schemes}
\label{sec:4vs5FS}

The theoretical predictions presented so far were computed in the 5FS. This is 
a fixed-flavour number (FFN) scheme whereby all quarks except the top quark are
treated as massless partons. In this scheme, both the PDFs and the strong 
coupling evolve with five active flavours. The NNLO QCD corrections to  
$t$-channel single top production that we used to compute the $C$-factors in
Eq.~\eqref{eq:cfact} were obtained in this scheme.
The process could otherwise be analysed in the 
4FS~\cite{Campbell:2009ss}, in which
the bottom quark is produced at the matrix-element level and is not
part of the proton wave-function. The NNLO QCD corrections for
single top production in the 4FS are however not available, as
the process starts at ${\cal O}(\alpha_s)$, and higher-order corrections
including mass effects for the bottom quark are harder to compute. 
In this section, we investigate how predictions differ in
the two schemes at NLO, and study the effect of varying the bottom-quark
matching point in the 5FS.

The 5FS is used as a default in all NNPDF determinations to describe 
a wide range of hadronic data; a matched general-mass variable flavour number
(GM-VFN) scheme is used instead to describe deep-inelastic scattering 
(DIS) data~\cite{Ball:2014uwa}. The GM-VFN scheme adopted for DIS processes,
namely the FONLL scheme, was originally devised for the transverse momentum 
spectrum of bottom quarks produced in hadronic 
collisions~\cite{Cacciari:1998it}, and it was then generalised to 
DIS~\cite{Forte:2010ta} and to the 
$b\bar{b}\to H$~\cite{Forte:2015hba,Forte:2016sja} and 
$b\bar{b}\to Z$~\cite{Forte:2018ovl} processes. In principle, since the FONLL 
scheme is universally applicable, it could easily be generalised to 
single top production to combine the 4FS and 5FS calculations 
performed at any perturbative order.
Its availability would help overcome a drawback of the 5FS, 
namely that it neglects bottom-quark mass effects. In particular,
mass corrections of $\order(m_b^2/Q^2)$, where $m_b$ is the bottom-quark mass, 
are set to zero above the scale at which the bottom PDF is activated. 
Mass corrections neglected in the 5FS are immaterial for most 
processes, except for those that depend on the bottom PDFs. Single top
production, which is bottom-initiated at LO in the 5FS, is an example of such
a process, and its theoretical description could therefore be 
affected by the choice of flavour scheme.

\begin{figure}[!t]
  \centering
  \includegraphics[width=0.9\textwidth]{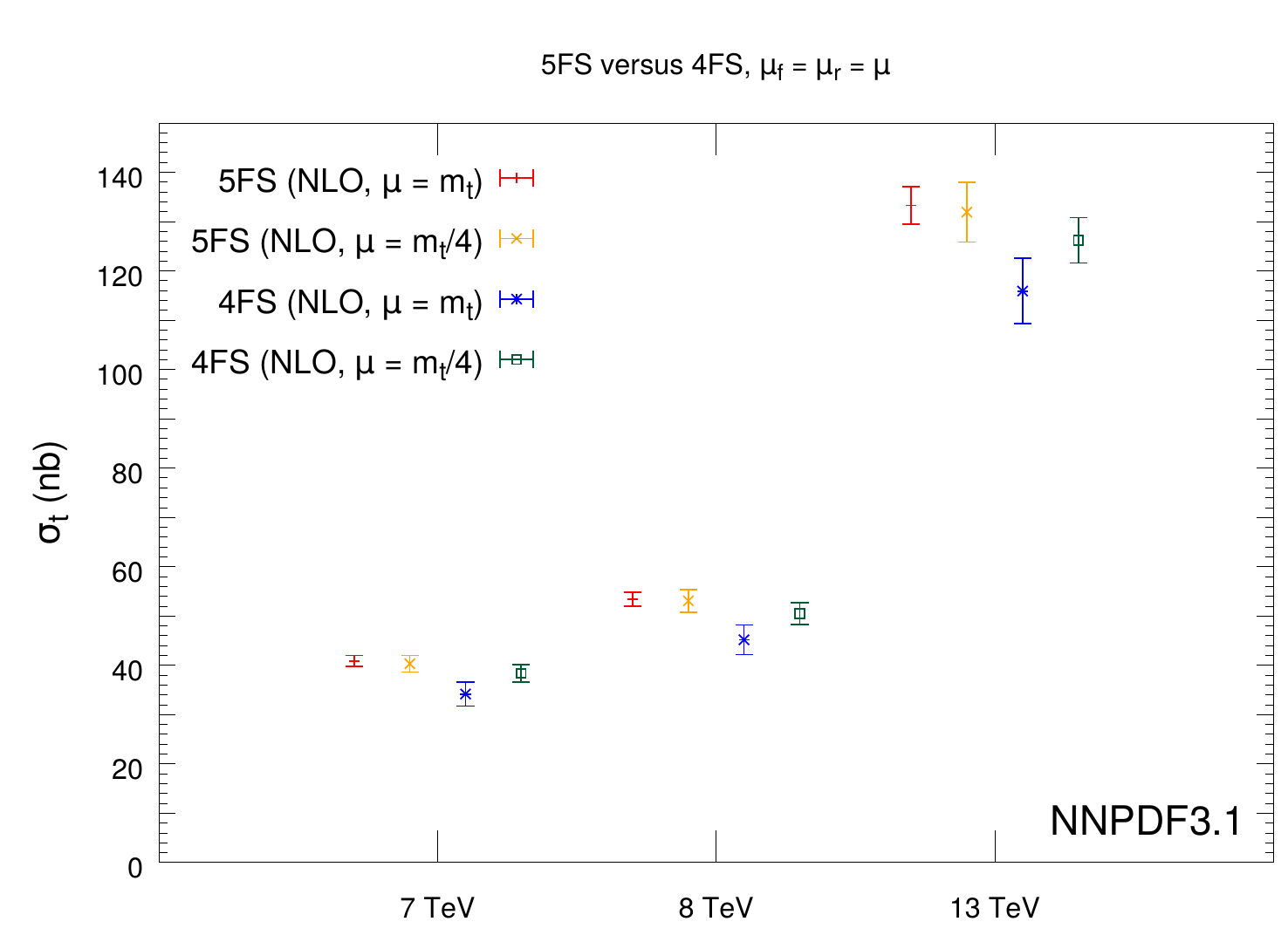}
  \caption{Comparison between total cross section predictions at NLO in the 5FS 
  and in the 4FS for $\mu_f=\mu_r=m_t$ and for $\mu_f=\mu_r=m_t/4$.
  The uncertainty of the theoretical predictions represent the envelope 
  of 7-point scale variations.}
\label{fig:4Fvs5F}
\end{figure}

To assess the size of such effects, in Table~\ref{tab:4Fvs5F} we compare the 
predictions for inclusive single top production obtained at NLO in the 4FS and 
5FS. The cross section is obtained by running \amc~with 4FS and 5FS 
settings, respectively. 
The NNPDF3.1 PDF input sets, {\tt NNPDF31\_nlo\_as\_0118} and 
{\tt NNPDF31\_nlo\_as\_0118\_nf\_4},
have been determined in a consistent scheme. We compare predictions both for
$\mu_r=\mu_f\equiv\mu=m_t$ and for a lower scale $\mu_r=\mu_f\equiv\mu=m_t/4$ 
motivated by the studies in~\cite{Campbell:2009ss,Maltoni:2012pa}. 
While the PDF uncertainty is similar in both schemes, the missing higher-order
uncertainty (MHOU) is larger in the 4FS than in the 5FS. Concerning 
predictions, the NLO total cross section in the 4FS is about 15\% lower than 
the NLO total cross section in the 5FS\footnote{The NLO corrections in the 4FS
feature diagrams that belong to the NNLO corrections in the
5FS. Therefore, one could compare the NNLO 5FS predictions to the NLO
4FS ones, in which case the difference would be slightly smaller, around 10\%.}.
However, the difference between the predictions obtained in the two
schemes, as well as between the size of the MHOUs, is considerably reduced
when lower renormalisation and factorisation scales are used as
central scale in the 4FS. The choice of such a scale is motivated by
the following evidence: the scale that appears in the
explicit collinear logarithms in the matrix element in the $m_b\to 0$ limit
is proportional to the hard scale of the process, $m_t$ in this case, but is
suppressed by a universal phase-space factor that is centered at about 0.25 at 
the LHC. Therefore the reduced difference between predictions, which amounts 
to about 5\%, is somewhat expected. 
The results are also displayed in Fig.~\ref{fig:4Fvs5F}. The trend of
the results is similar at 7, 8 and 13 TeV. 

\begin{table}[!t]
  \begin{center}
\scriptsize
\begin{tabular}{clll}
\toprule
 $\sqrt{s}$ [TeV] & scale & $\sigma^{\rm (5FS)}_{t}$ [pb] &       $\sigma^{\rm (4FS)}_{t}$ [pb] \\
\midrule
7  &  $\mu=m_t$    & 40.89 $^{+3.2\%} _{-2.3\%}\,\pm\,0.8\%$   & 34.17 $^{+7.0\%}_{-7.2\%}\,\pm\,0.8\%$  \\
   &  $\mu=m_t/4$ & 40.32 $^{+6.4\%} _{-2.0\%}\,\pm\,0.8\%$   & 38.37 $^{+4.1\%}_{-5.1\%}\,\pm\,0.8\%$  \\
\midrule
8  &  $\mu=m_t$   & 53.44 $^{+3.1\%} _{-2.3\%}\,\pm\,0.8\%$   & 45.17 $^{+6.5\%}_{-6.8\%}\,\pm\,0.8\%$ \\
   &  $\mu=m_t/4$ & 53.08 $^{+5.9\%} _{-2.7\%}\,\pm\,0.8\%$   & 50.50 $^{+4.0\%}_{-4.8\%}\,\pm\,0.8\%$ \\
\midrule
13 &  $\mu=m_t$   & 133.3 $^{+3.1\%} _{-2.6\%}\,\pm\,0.8\%$   & 115.9 $^{+6.2\%}_{-5.3\%}\,\pm\,0.8\%$ \\
   &  $\mu=m_t/4$ & 131.9 $^{+4.2\%} _{-5.0\%}\,\pm\,0.8\%$   & 126.2 $^{+3.3\%}_{-3.9\%}\,\pm\,0.8\%$ \\
\bottomrule
\end{tabular}
\end{center}

  \caption{Predictions for single top production at NLO in the 5FS and the 4FS. 
  The 5FS (4FS) predictions have been computed using the corresponding 
  NLO 5FS (4FS) input NNPDF3.1 PDF set. The first uncertainty quoted is
  the upper and lower missing higher-order uncertainty of the
  predictions, which is obtained by taking the envelope of the 7-point scale
  variation in which we vary the scales in the range $1/2\le\mu_f, \mu_r<2$.
  The second uncertainty represents the PDF uncertainty.}
  \label{tab:4Fvs5F}
\end{table}

\begin{figure}[!t]
  \centering
  \includegraphics[angle=270,scale=0.252]{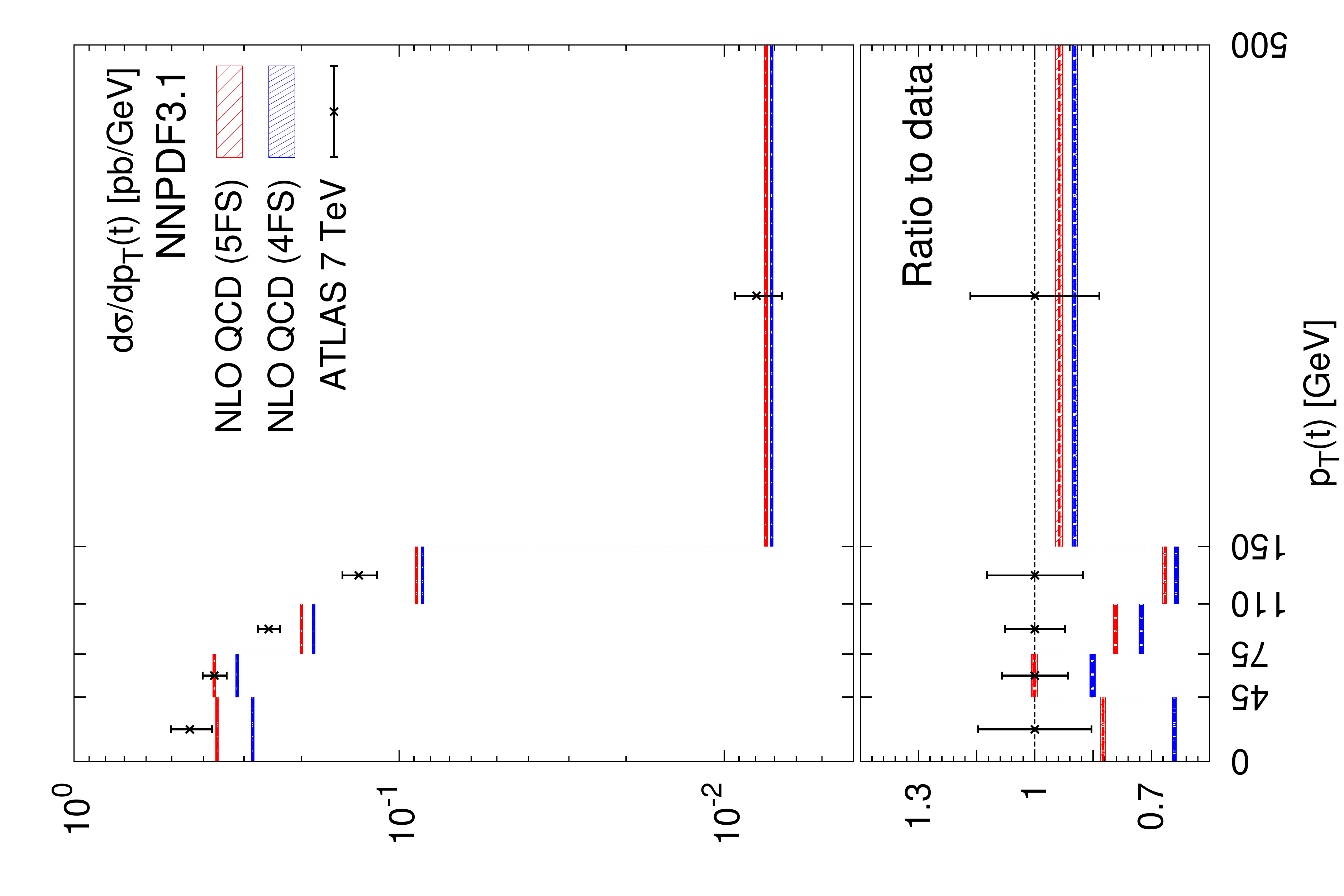}\ \
  \includegraphics[angle=270,scale=0.252]{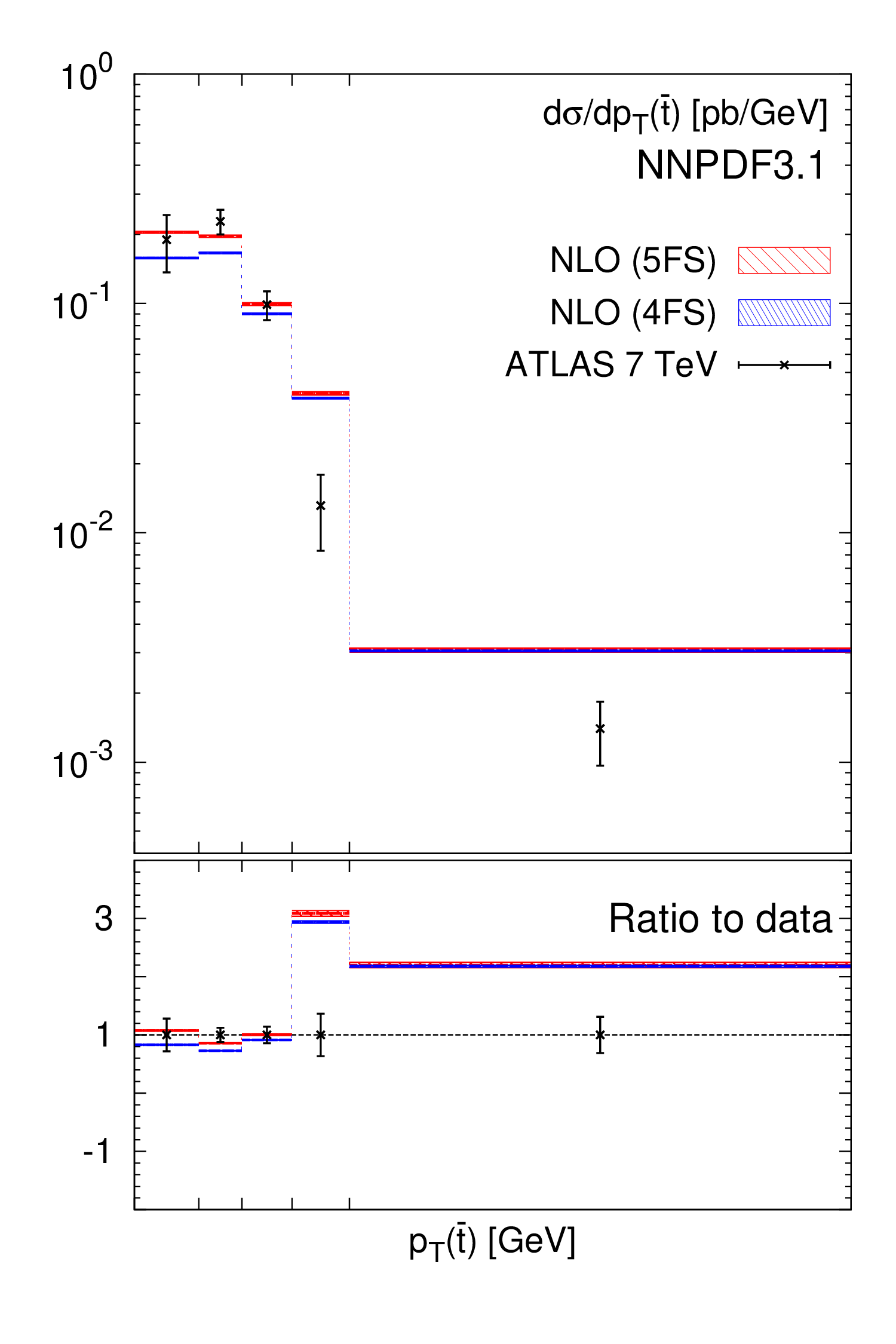}\\
  \includegraphics[angle=270,scale=0.252]{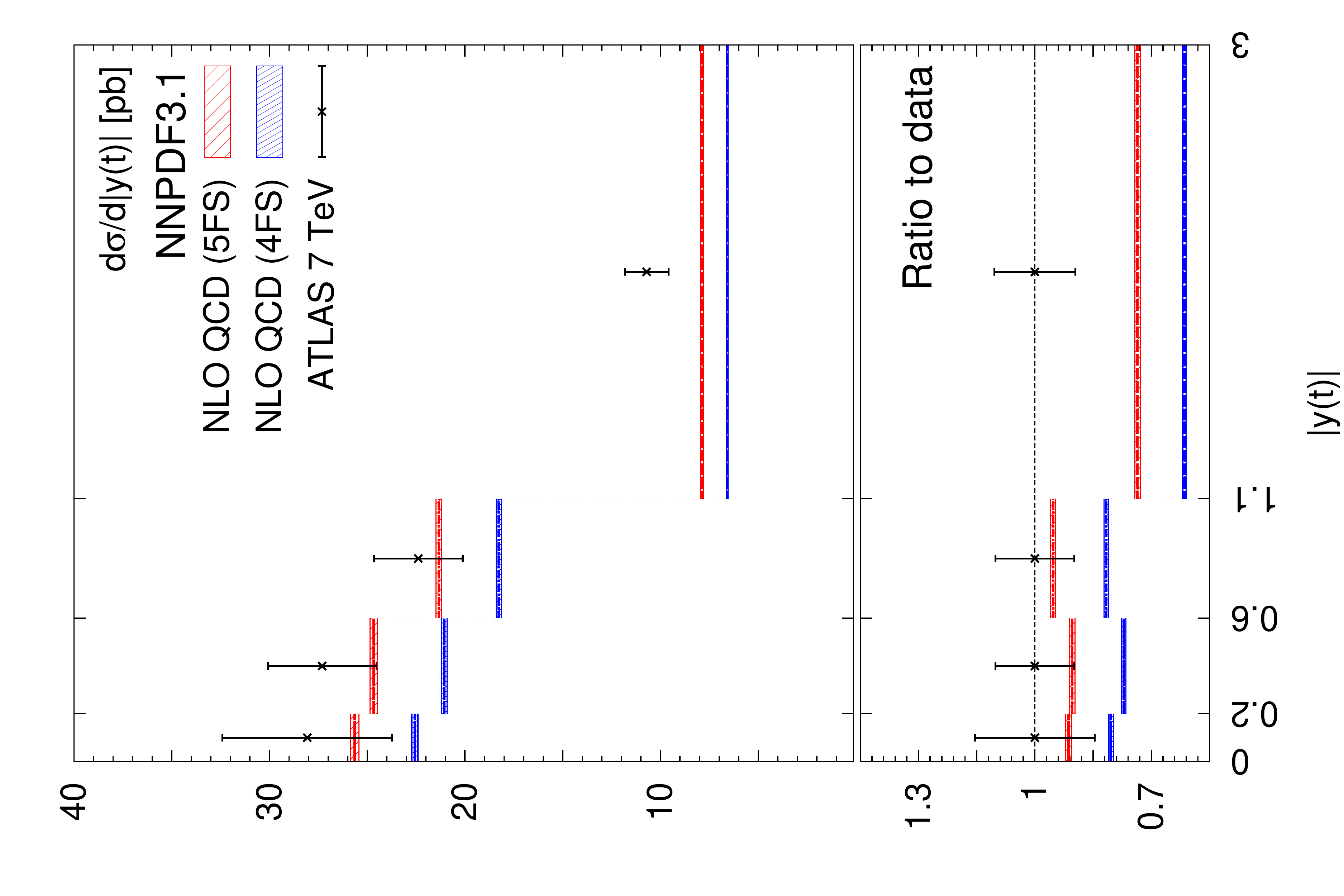} \ \
  \includegraphics[angle=270,scale=0.252]{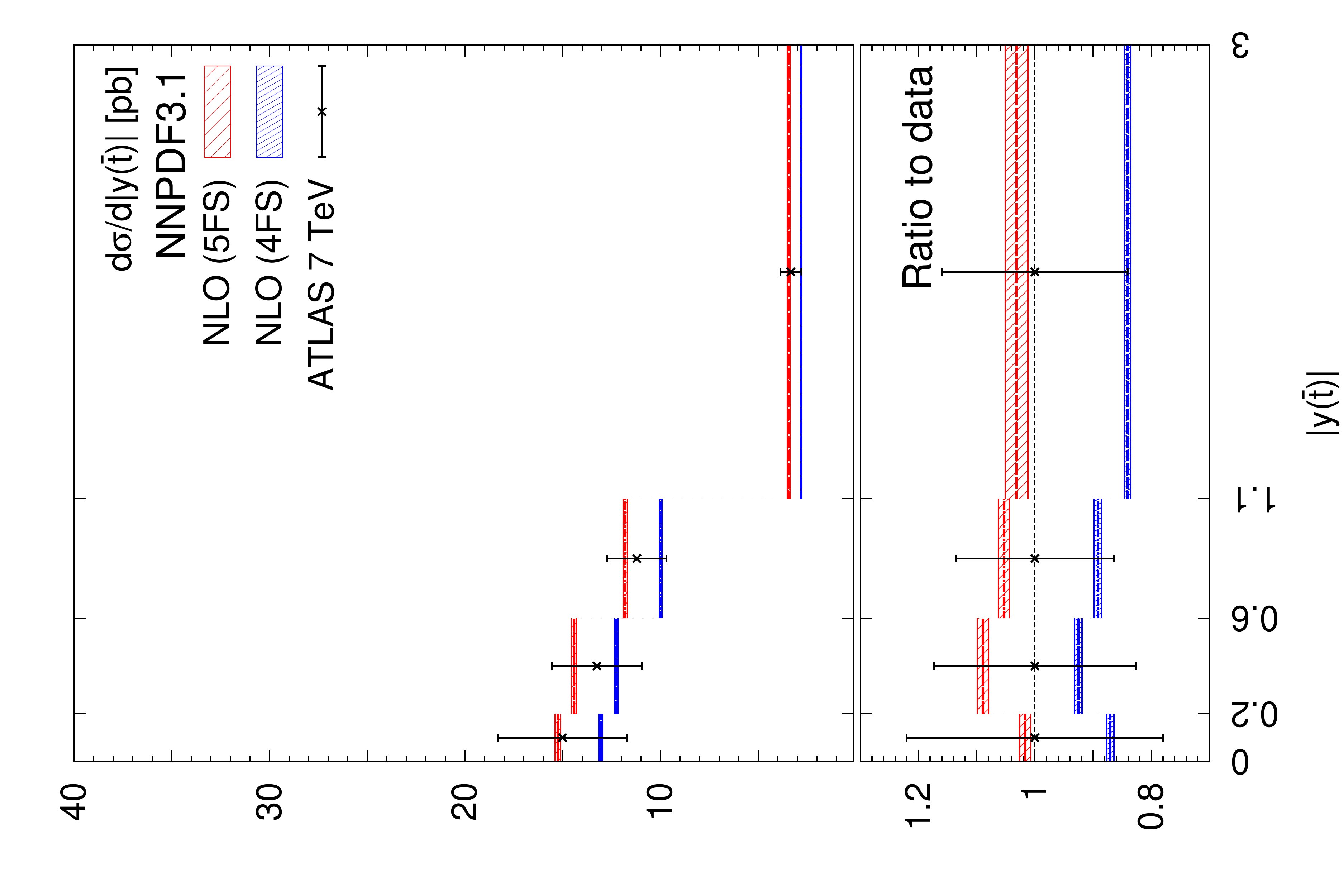}\\
  \caption{Same as Fig.~\ref{fig:data_theory_top}, but comparing NLO
  predictions in the 4FS and 5FS.}
\label{fig:data_theory_top_4F}
\end{figure}

In order to check the effect of the choice of scheme on the differential 
distributions, we display the data-theory comparison for the ATLAS absolute
distributions predicted at NLO in either the 4FS or the 5FS for
$\mu_f=\mu_r=m_t$, for which the difference between cross sections is
larger. In the case of lower factorisation and renormalisation central scales,
the differences are smaller, but similar features are exhibited.
The results for $\mu_f=\mu_r=m_t$ are displayed in 
Fig.~\ref{fig:data_theory_top_4F}. As far as the overall normalisation is
concerned, we notice that the 5FS predictions are closer to the data. As far 
as the shapes are concerned, they are quite similar for the 
rapidity distributions, while the two differ when the top quark has a small 
transverse momentum. This is not unexpected. As was noticed in the
comparison of differential cross sections for charged Higgs boson production in 
Ref.~\cite{Degrande:2015vpa}, the main difference between the two schemes is 
the normalisation, as far as observables that are inclusive in the $b$-quark 
degrees of freedom are concerned.
The top transverse momentum is more sensitive to the details of the 
$b$-quark transverse momentum
at small $p_T$, while it is unaffected by it at large $p_T$. 
Clearly, once the transverse momentum of the $b$ quark is precisely measured 
and included in a fit of PDFs, the 4FS and 5FS descriptions would display 
larger shape differences and the 4FS would possibly be more adequate.
However, an important conclusion that we can draw by looking at the plots in 
Fig.~\ref{fig:data_theory_top_4F} is that the discrepancy between data and 
theory in the transverse momentum spectrum of the top and antitop at 7 TeV
is not due to the choice of scheme. This conclusion will be made 
quantitative in Sect.~\ref{sec:chi2}. 

Finally, we assess the impact of the bottom matching point, $\mu_b$, which is 
the scale at which the $b$ PDF is activated in the 5FS.
Conventionally this is set equal to the bottom mass, that is $\mu_b = m_b$, 
however there is no physical reason why this should be the case. 
In Ref.~\cite{Bertone:2017djs} it was shown that the mass corrections
neglected in the 5FS can be reduced by utilising the freedom one 
has in setting $\mu_b$. In particular, by increasing the
bottom matching point in the range $m_b < \mu_b \leq 10 \,
m_b$. A strong dependence of a given cross section on the choice of the 
matching point might stem from the fact that matching conditions are only known 
to the lowest nontrivial order. As far as charm-initiated processes are 
concerned, it was shown 
that it is advantageous to treat the charm PDF on the same footing as 
light-quark PDFs; that is, to parametrise it and extract it from data, rather 
than to take it as radiatively generated through perturbative matching 
conditions~\cite{Ball:2015dpa,Ball:2015tna,Ball:2016neh}. 
In Ref.~\cite{Forte:2019hjc} a similar study was performed in the context 
of the bottom $b$ PDF and the FONLL scheme.

\begin{figure}[!t]
  \centering
  \includegraphics[width=0.9\textwidth]{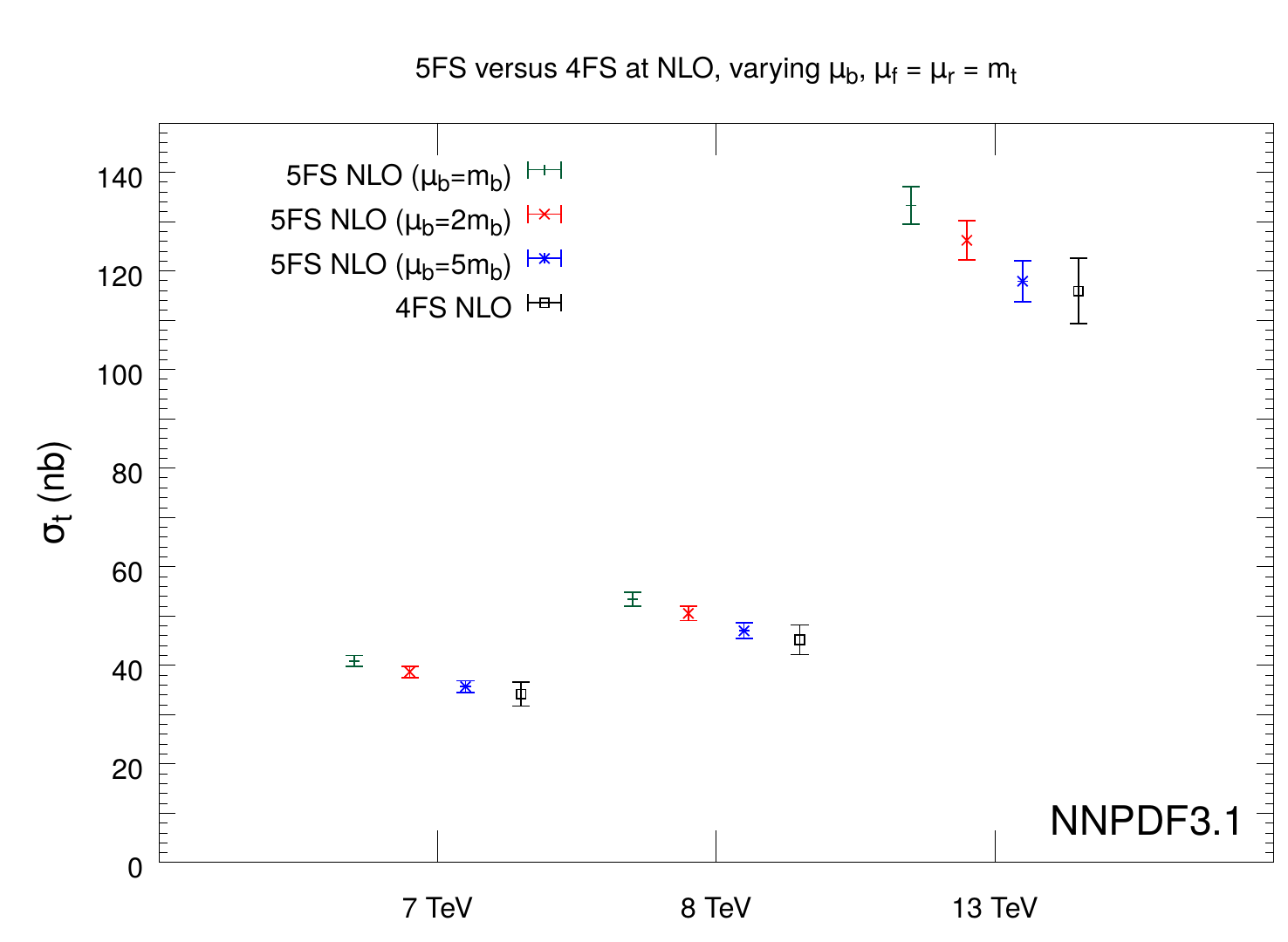}
  \caption{Comparison between 4FS and 5FS NLO predictions for $\mu_f=\mu_r=m_t$ 
  and varying $\mu_b$. We display predictions at NLO for $\mu_b=m_b,\,2m_b$ and 
  $5m_b$, and compare them to the 4FS NLO predictions.}
\label{fig:muBth}
\end{figure}

Here we consider three NLO PDF sets obtained by varying the bottom matching 
points as follows: $\mu_b=m_b$; $\mu_b=2 \, m_b$; and $\mu_b=5 \, m_b$. The PDF 
set with $\mu_b=m_b$ coincides with the standard NNPDF3.1 PDF set with 
$\alpha_s(M_Z)=0.118$. The PDFs are evolved from the initial scale $Q_0$ 
up to the scale $\mu_f=m_t$, at which we compute the single top total cross
 section, by crossing the bottom matching point $\mu_b$.
Results are displayed in Fig.~\ref{fig:muBth}. We notice that the dependence 
of the single top cross section on the bottom-quark matching point is quite 
strong, with predictions decreasing by 5--7\% (10--12\%) as
the bottom matching point is raised to twice (five times) the mass of the 
bottom quark. Furthermore, as $\mu_b$ increases, the 5FS predictions get 
closer to the 4FS ones. We have checked that, using the NNLO PDF sets obtained 
by varying the bottom matching point, the NLO predictions still decrease but 
the effect is much smaller, of order 2\% (3-4\%) as the bottom matching point 
is raised to twice (five times) the mass of the bottom quark. To fully assess 
this effect at NNLO, we would need to have the NNLO calculations obtained with 
higher bottom matching points.
These results suggest that the choice of a bottom matching point higher than
$m_b$ has a non negligible effect on the theoretical predictions, whose size is 
comparable to that of missing power corrections of $m_b$ in a 5FS. Such effects 
are well below the experimental uncertainty of the data discussed here. 
However, when more precise data become available, it will be crucial
to either perform dedicated studies to determine the optimal bottom matching 
point $\mu_b$ that minimises missing power corrections of $m_b$ in a 5FS,
or to explicitly compute matched predictions {\`a} la FONLL for single top 
observables, in order to directly include $b$-quark mass corrections via the 
matching of the 5FS to the 4FS cross sections.

\subsection{Quantitative assessment of the agreement between theory and data}
\label{sec:chi2}

A quantitative assessment of the agreement between theoretical expectations and
the data presented above requires an appropriate statistical measure, which 
should take into account experimental correlations. We therefore define 
the $\chi^2$ per data point (called $\chi^2$ henceforth)
\begin{equation}
\chi^2\{\mathcal{T}[f],\mathcal{D}\}
=
\frac{1}{N_{\rm dat}}
\sum_{i,j}^{N_{\rm dat}}(T_i[f]-D_i) \, C_{ij}^{-1} \,(T_j[f]-D_j)\,,
\label{eq:chi2}
\end{equation}
which depends on the dataset, $\mathcal{D}$, and on the theoretical predictions
computed with the PDFs $f$, $\mathcal{T}[f]$. The indices $i$ and $j$ run over 
the experimental data points, $D_i$ are their measured central values, $T_i$ 
are the corresponding theoretical predictions (computed with a given theory 
and a given set of PDFs), and $C_{ij}$ is the covariance matrix, constructed 
from the available information on experimental statistical and systematic 
uncertainties. In Eq.~\eqref{eq:chi2}, we specifically use the experimental 
covariance matrix
\begin{equation}
C_{ij}=C_{ij}^{\rm exp}
\equiv
\delta_{ij}(s_i^{\rm stat})^2
+
\left[
\sum_{\alpha=1}^{N_{\rm sysA}} \sigma_{i,\alpha}^{\rm sysA}\sigma_{j,\alpha}^{\rm sysA}
+
\sum_{\beta=1}^{N_{\rm sysM}} \sigma_{i,\beta}^{\rm sysM}\sigma_{j,\beta}^{\rm sysM}
\right]
D_i D_j
\,,
\label{eq:cov_exp}
\end{equation}
where $s_i^{\rm stat}$ is the uncorrelated absolute uncertainty of the data point 
(obtained by adding in quadrature statistical and uncorrelated systematic
uncertainties), and $\sigma_i^{\rm sysA}$ ($\sigma_i^{\rm sysM}$) are the 
$N_{\rm sysA}$ ($N_{\rm sysM}$) correlated additive (multiplicative) 
relative systematic uncertainties. 

The values of the $\chi^2$  computed using Eq.~\ref{eq:cov_exp} 
are collected in Table~\ref{tab:chi2} for all of the datasets presented in
Tables~\ref{tab:tot_measurements}--\ref{tab:distributions}. The last bin 
of the normalised distributions is excluded from the computation.
They are presented
for predictions obtained within three different theoretical scenarios: 
NLO and NNLO in the 5FS and NLO in the 4FS. The input PDF sets are chosen 
consistently: {\tt NNPDF31\_nlo\_as\_0118}, {\tt NNPDF31\_nnlo\_as\_0118} and 
{\tt NNPDF31\_nlo\_as\_0118\_nf\_4}, respectively. In the case of NNLO QCD
within the 5FS, we also report the values of the $\chi^2$  
obtained with the
{\tt ABMP16als118\_5\_nnlo} PDF set, to complete the data-theory comparisons 
displayed in Figs.~\ref{fig:tot_measurements}--\ref{fig:data_theory_abmp}.
We have explicitly checked that $\chi^2$ variations induced by the use of other 
PDF sets, such as MMHT~\cite{Harland-Lang:2014zoa}, CT18~\cite{Hou:2019qau}, or 
the PDF4LHC15 combination~\cite{Butterworth:2015oua}, and theoretical
expectations accurate to NLO QCD$\times$EW (computed with the 
{\tt NNPDF31\_nlo\_as\_0118\_luxqed} PDF set) are tiny, and do not
alter the interpretation of these results.
The values of the $\chi^2$  obtained upon these variations are 
therefore not shown in Table~\ref{tab:chi2}. In order to assess the impact of 
correlations in the 5FS, we report in parentheses the $\chi^2$  
values computed by neglecting bin-by-bin correlations. 

\begin{table}[!t]
\centering
\scriptsize
\begin{tabular}{lcccllll}
\toprule       
Experiment & $\sqrt{s}$ [TeV] & $\mathcal{O}$ & $N_{\rm dat}$  
           & $\chi^{2\ {\rm (5FS)}}_{\rm NLO}$ 
           & $\chi^{2\ {\rm (5FS)}}_{\rm NNLO}$ 
           & $\chi^{2\ {\rm (4FS)}}_{\rm NLO}$
           & $\chi^{2\ {\rm (5FS)}}_{\rm NNLO,ABMP16}$\\
\midrule
ATLAS &  7 & $\sigma_t/\sigma_{\bar{t}}$    
           & 1 & 0.28 (---) & 0.30 (---) & 0.41 & 0.42\\
      &  8 & $\sigma_t/\sigma_{\bar{t}}$    
           & 1 & 3.71 (---) & 3.67 (---) & 3.06 & 2.69\\
      & 13 & $\sigma_t/\sigma_{\bar{t}}$    
           & 1 & 0.03 (---) & 0.03 (---) & 0.02 & 0.13\\
\midrule
CMS   &  7 & $\sigma_{t+\bar{t}}$           
           & 1 & 0.82 (---) & 0.66 (---) & 6.28 & 0.02\\
      &  8 & $\sigma_t/\sigma_{\bar{t}}$    
           & 1 & 0.09 (---) & 0.09 (---) & 0.14 & 0.17\\
      & 13 & $\sigma_t/\sigma_{\bar{t}}$    
           & 1 & 0.29 (---) & 0.29 (---) & 0.25 & 0.48\\
\midrule
ATLAS &  7 & $d\sigma/dp_T(t)$                  
           & 5 & 2.16 (1.24) & 2.88 (1.45) & 2.22 & 3.15\\
      &    & $d\sigma/dp_T(\bar{t})$            
           & 5 & 9.49 (3.31) & 11.9 (4.15) & 12.4 & 11.5\\
      &    & $d\sigma/d |y(t)|$                 
           & 4 & 1.16 (0.97) & 1.29 (0.93) & 1.99 & 1.36\\
      &    & $d\sigma/d|y(\bar{t})|$            
           & 4 & 0.05 (0.07) & 0.06 (0.08) & 0.13 & 0.11\\
      &    & $(1/\sigma)d\sigma/dp_T(t)$        
           & 4 & 1.89 (0.99) & 2.79 (1.29) & 1.76 & 2.32\\
      &    & $(1/\sigma)d\sigma/dp_T(\bar{t})$  
           & 4 & 5.14 (2.76) & 6.26 (3.34) & 7.14 & 6.81\\
      &    & $(1/\sigma)d\sigma/d|y(t)|$        
           & 3 & 0.92 (1.11) & 0.96 (1.18) & 1.19 & 2.13\\
      &    & $(1/\sigma)d\sigma/d|y(\bar{t})|$  
           & 3 & 0.07 (0.04) & 0.07 (0.04) & 0.05 & 1.36\\
\bottomrule
\end{tabular}

\caption{The values of the $\chi^2$ , Eq.~\eqref{eq:chi2}, 
  computed for all 
  of the datasets described in Sect.~\ref{sec:exp_data} and collected in 
  Tables~\ref{tab:tot_measurements}--\ref{tab:distributions}. For ease of
  reference, we indicate the centre-of-mass energy of each measurement, 
  $\sqrt{s}$, the corresponding observable, $\mathcal{O}$, and the number of 
  data points, $N_{\rm dat}$. Theoretical predictions are computed at either NLO
  or NNLO accuracy in the 5FS or at NLO accuracy in the 4FS, always 
  in pure QCD; the input PDFs are taken from the 
  {\tt NNPDF31\_nlo\_as\_0118}, {\tt NNPDF31\_nnlo\_as\_0118} and
  {\tt NNPDF31\_nlo\_as\_0118\_nf\_4} sets, 
  respectively. Values in parentheses do not include bin-by-bin 
  experimental correlations.}
\label{tab:chi2}
\end{table}

The results collected in Table~\ref{tab:chi2} quantitatively confirm the 
pattern of data-theory inconsistencies already qualitatively displayed in 
Figs.~\ref{fig:tot_measurements}--\ref{fig:data_theory_abmp}. While most of the
data are fairly well described by the theoretical predictions, irrespective of their
accuracy, there are three notable exceptions. The first is the ratio of single
top-quark to top-antiquark total cross sections from ATLAS at 8 TeV 
($\chi^2\sim 3-4$); the second is the distribution differential in the 
top-antiquark transverse momentum $p_T(\bar{t})$ ($\chi^2\sim 10-12$); and the 
third is the corresponding normalised distribution ($\chi^2\sim 6$). 
In the first case, we note that the analogous measurement from CMS, which 
differs from the ATLAS one by approximately $\sqrt{2}\sigma$,
is instead very well described. We therefore conclude that the two points
might be somewhat in tension, for reasons that do not seem to depend on the 
accuracy of the theoretical predictions. In the second and third cases, 
the high $\chi^2$ values are a consequence of the large data-theory 
discrepancy (up to 300\%) observed in the fourth and fifth transverse momentum
bins. Despite a good $\chi^2$  being reported for this 
distribution in Ref.~\cite{Aad:2014fwa} at NLO (see Table VIII therein), also 
in these cases, we are not able to envision theoretical effects that could 
explain such a difference (nor the $\chi^2$  values reported in 
Ref.~\cite{Aad:2014fwa}), as we further discuss below.

Overall, NNLO QCD corrections have a moderate impact. In the case of ratios of 
single top-quark to top-antiquark total cross sections their effect on
the $\chi^2$  cancels out almost completely; for their sum 
(in the case of the CMS dataset at 7 TeV), they instead tend to improve the 
description of the data point a little (the $\chi^2$  decreases 
from 0.82 to 0.66). In the case of 
differential distributions, NNLO QCD corrections generally lead to a worsening 
of the $\chi^2$, which is significant in particular for the distributions
differential in the transverse momentum of the top quark or top antiquark.

The 4FS leads to a worse $\chi^2$ value than the 5FS for most of 
the datasets, consistently with the qualitative behaviour observed in
Fig.~\ref{fig:data_theory_top_4F}. However, while the increase in the $\chi^2$
can be substantial for the sum of top-quark and top-antiquark cross sections, 
it is moderate for the differential distributions and only barely noticeable 
for their ratio: the 4FS modifies the individual cross sections in such a way 
that their variations cancel out in the ratio.

More strikingly, experimental correlations have a large effect:
neglecting them leads to an acceptable $\chi^2$ for all of the differential
distributions, except, again, for those differential in the transverse momentum
of the top-antiquark $p_T(\bar{t})$ (for which the $\chi^2$ improvement is 
nevertheless the largest). 

This state of affairs leads us to wonder whether the poor description of some 
single top-quark and top-antiquark datasets follows from internal 
inconsistencies or from inconsistencies with the rest of the dataset. 
As far as the ATLAS differential distributions
at $\sqrt{s}=7$ TeV are concerned, we hypothesise that they are internally 
inconsistent. Our hypothesis is driven by the fact that the data-theory 
discrepancies persist no matter what theoretical framework or PDF set we use. 
Furthermore, we also investigate how other single
top-quark and top-antiquark datasets, not considered in this section, are
described. We do so with the ATLAS differential measurements at a centre-of-mass
energy of 8 TeV~\cite{Aaboud:2017pdi} in Appendix~\ref{app:ATLAS8TeVdiff}.
Even though no information on correlations is available for this dataset,
as explained in Sect.~\ref{sec:exp_data}, we do not observe any significant
data-theory discrepancy. 

Of course, only the inclusion of the 7 TeV data in
a PDF fit will tell us whether they are inconsistent with the rest of the 
dataset, and if the observed discrepancies can be resolved 
by altering the shape of the PDFs (see Sect.~\ref{sec:fits}).
To include the data presented in 
Tables~\ref{tab:tot_measurements}--\ref{tab:distributions} in a global fit, 
we shall decide which theoretical framework is the most convenient for their
analysis. Because one of our goals is to assess the
interplay of single top-quark and top-antiquark data with the rest of the 
dataset, NNLO corrections are mandatory to achieve the best overall description 
we can. However, the use of NNLO QCD forces us to the 5FS, as higher-order 
corrections to single top-quark and top-antiquark data are not known in the 4FS.
Theoretical predictions will therefore be computed accordingly in 
Sect.~\ref{sec:fits}.

We shall however note that the experimental covariance matrix defined in 
Eq.~\eqref{eq:cov_exp}, used in this section, is not suitable for PDF fits.
The reason being that it would lead to a fit that systematically undershoots 
the data: this is the so-called D'Agostini bias~\cite{DAgostini:1993arp}, 
which is related to an inappropriate treatment of multiplicative uncertainties. 
We therefore define an alternative covariance matrix that removes such a bias
\begin{equation}
C_{ij}=C_{ij}^{t_0}
\equiv
\delta_{ij}\left(s_i^{\rm stat} \right)^2
+
\left(
\sum_{\alpha}^{N_{\rm sysA}}\sigma_{i,\alpha}^{\rm sysA} \sigma_{j,\alpha}^{\rm sysA}
\right)D_iD_j
+
\left(  
\sum_{\beta}^{N_{\rm sysB}}\sigma_{i,\beta}^{\rm sysB} \sigma_{j,\beta}^{\rm sysB}
\right)T_i^{(0)}T_j^{(0)}\,,
\label{eq:cov_t0}
\end{equation}
where multiplicative uncertainties are multiplied by the theoretical 
predictions $\{ T_i^{(0)}\}$. This is the $t_0$-prescription~\cite{Ball:2009qv}, 
which we use in all of the PDF fits presented in Sect.~\ref{sec:fits}.

\section{Impact of single top data on PDFs} 
\label{sec:fits}

In this section we assess the impact on PDFs of the single top data discussed 
previously. Firstly, we study the correlations between the 
data and PDFs. Then we outline the settings of our PDF fits, and we detail the 
baseline dataset on top of which we add single top data. Next, the results 
of the PDF fits are presented. We then study the correlations between 
single top data and other measurements used in the fits, before finishing
with a discussion of the impact of single top data on LHC phenomenology.

\subsection{Observable-PDF correlations}
\label{obs-pdf-corrs}

Before assessing the impact of single top data in a PDF fit, it can be 
instructive to compute the correlation coefficient 
$\rho$~\cite{Guffanti:2010yu} between the PDFs and the single top data. 
We calculate $\rho$ between a given observable $\mathcal{O}$ and a PDF $f$ as
\begin{equation}
  \rho(\mathcal{O}, f) 
  = 
  \frac{\langle \mathcal{O} f \rangle - \langle \mathcal{O} \rangle \langle f \rangle}{\Delta_{\mathcal O} \Delta_f} \, ,
  \label{eq:pdf_obs_corr}
\end{equation}
where the angled brackets denote the mean and $\Delta$ denotes the uncertainty 
on $\mathcal{O}$ or $f$, both computed over an ensemble of PDF replicas. For 
fixed energy scale $Q$ and Bjorken-$x$, $\rho$ is a number between $-1$ and $1$,
with its magnitude indicating the degree to which $\mathcal{O}$ is sensitive to 
$f$. If the single top data have any impact on $f$ in a PDF fit, then the 
regions of high correlation are the regions where we would expect to see 
changes in $f$.
Note that these correlations are inherently dependent on the data that are
already included in the fit of the PDFs $f$. That is, the correlations depend 
on the way in which single top production is initiated at the parton level and 
the kinematics of single top production, but also on the experimental 
constraints included in the
determination of $f$. This means that the maxima of $|\rho|$ are in general not
solely dependent on what the most favoured initial state is, and in which
kinematic region this dependence is most prominent. As a result, the maxima are
likely to be displaced from the values that one might naively expect.

In what follows we restrict our analysis to the up-quark ($u$), the down-quark 
($d$) and the gluon ($g$) PDFs, since these are three of the main partons that 
initiate single top production. Note that we ignore the up antiquark 
($\overline{u}$) and the down antiquark ($\overline{d}$) here because the 
sensitivity of the single top data to them can largely be inferred from the 
behaviour of the $u$ and the $d$. We compute $\rho$ using the 
{\tt NNPDF31\_nnlo\_as\_0118} PDF set at an energy scale equal to the top mass, 
{\it i.e.} $Q = m_t = 172.5$~GeV. 

We begin by looking at the inclusive total cross section measurements, 
$\sigma_t + \sigma_{\overline{t}}$, and at the ratio $\sigma_t/\sigma_{\overline{t}}$.
In Refs.~\cite{Aad:2014fwa,Alekhin:2015cza,Alekhin:2017kpj} it has been 
suggested that measurements of $\sigma_t/\sigma_{\overline{t}}$ could be used to 
constrain the ratio $u/d$ in the region $0.02 \lesssim x \lesssim 0.5$. 
This has been stated as a reason in favour of using measurements of 
$\sigma_t/\sigma_{\overline{t}}$ in PDF determinations, in addition to the fact 
that common systematic uncertainties are likely to at least partially cancel 
when the ratio is taken, which makes these measurements comparatively 
precise~\cite{Aad:2014fwa,Aaboud:2017pdi,Aaboud:2016ymp}.

\begin{figure}[!t]
\centering
  \includegraphics[width=0.49\textwidth]{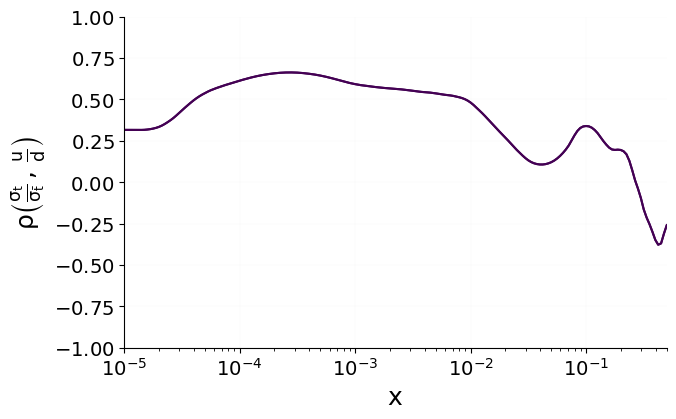}
  \includegraphics[width=0.49\textwidth]{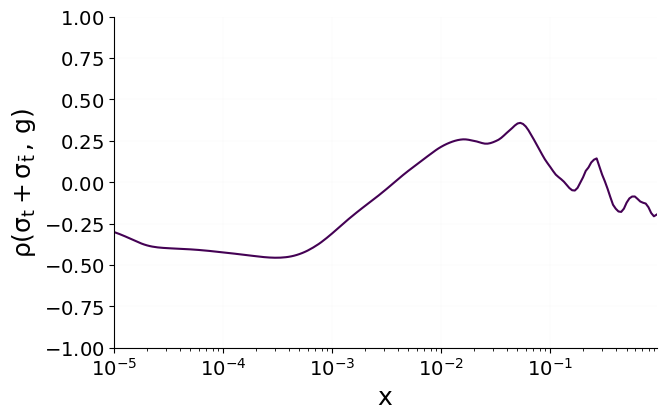}
  \caption{Correlation coefficient $\rho$ between the up-quark/down-quark PDF 
  ratio and the ATLAS measurement of $\sigma_t/\sigma_{\overline{t}}$ at 13 TeV 
  (left panel), and the gluon and the CMS measurement of 
  $\sigma_t + \sigma_{\overline{t}}$ at 7 TeV (right panel). The correlation
  coefficient $\rho$ is computed using the {\tt NNPDF31\_nnlo\_as\_0118} 
  PDF set at $Q = m_t = 172.5$~GeV. Note that the $x$-axis in the left panel 
  runs between $x=10^{-5}$ and $x=0.5$. This truncation of large values of $x$ 
  is due to the fact that as $x \rightarrow 1$, $u/d \rightarrow 0/0$, which 
  leads to numerical instabilities in $\rho$.}
  \label{fig:ratio_tot_corr}
\end{figure}

Fig.~\ref{fig:ratio_tot_corr} displays two representative plots of the 
correlation coefficient as a function of $x$. The left panel shows the 
correlation between $u/d$ and the ATLAS measurement of 
$\sigma_t/\sigma_{\overline{t}}$ at 13 TeV, while the 
right panel shows the correlation between $g$ and the CMS measurement of 
$\sigma_t + \sigma_{\overline{t}}$ at 7 TeV.
We see that for the ATLAS ratio measurement the maximum correlation is observed 
for $x \lesssim 10^{-3}$, with $\rho$ peaking at around 0.7 for $x = 10^{-4}$. 
The same general behaviour is observed for each of the five 
$\sigma_{t}/\sigma_{\overline{t}}$ measurements that are detailed in 
Sect.~\ref{sec:exp_data}, and is therefore not explicitly displayed in 
Fig.~\ref{fig:ratio_tot_corr}. In fact, the same characteristic minima and 
maxima are seen for each ratio measurement, with two maxima observed between 
$x=0.1$ and $x=0.2$ and a region of maximum anti-correlation observed at 
$x \simeq 0.4$. For the corresponding measurements at lower values of 
$\sqrt{s}$, the sensitivity weakens slightly as $\sqrt{s}$ decreases.
These observations back up the claim made in Ref.~\cite{Aad:2014fwa} that 
measurements of $\sigma_{t}/\sigma_{\overline{t}}$ are sensitive to $u/d$, although
the highest sensitivity is seen at lower values of $x$ than was originally 
suggested. We find that, in contrast, these measurements are mostly 
insensitive to the gluon PDF.

In comparison, the right panel of Fig.~\ref{fig:ratio_tot_corr} shows that the 
inclusive total cross section $\sigma_t + \sigma_{\overline{t}}$ is somewhat 
sensitive to the gluon, in particular around $x \simeq 5 \times 10^{-4}$
($\rho \simeq -0.5$) and $x \simeq 0.05$ ($\rho \simeq 0.5$). However, while 
$\sigma_t + \sigma_{\overline{t}}$ is more sensitive to the gluon, it is less 
sensitive to the $u$ and $d$ --- and it has very 
low sensitivity to $u/d$ --- as we have explicitly checked by looking at the 
corresponding values of $\rho$. We therefore conclude that the gluon 
sensitivity largely 
cancels when the top-quark and top-antiquark cross sections are
divided, as in $\sigma_{t}/\sigma_{\overline{t}}$, but not when they
are summed, as in $\sigma_t + \sigma_{\overline{t}}$. The reverse is true for $u/d$. This is 
because $\sigma_t + \sigma_{\overline{t}}$ contains no information on the relative 
strength of the $u$ and $d$ contributions to the process, while 
$\sigma_t/\sigma_{\overline{t}}$ does.

It might at first appear surprising that $\sigma_t + \sigma_{\overline{t}}$ is  
sensitive to the gluon, since at LO in the 5FS single top production is not 
initiated by the gluon. However, firstly, this is not true at NLO in the 5FS, 
and secondly, single top production at LO in the 5FS is initiated by the 
bottom quark, which is generated 
perturbatively from the gluon PDF. Therefore, single top production 
is in general sensitive to the gluon, even at LO.

\begin{figure}[!t]
\centering
  \includegraphics[width=0.49\textwidth]{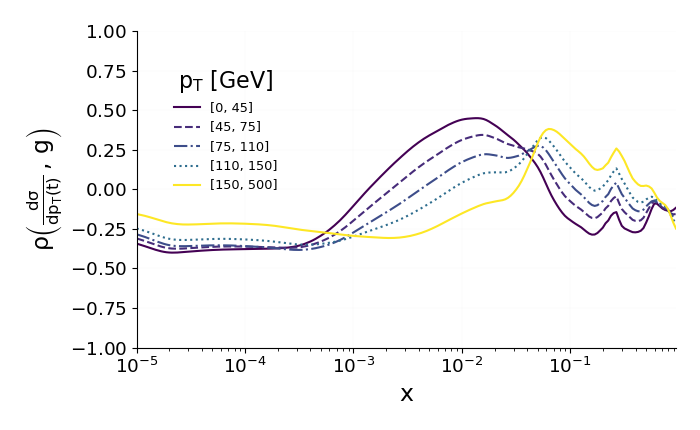}
  \includegraphics[width=0.49\textwidth]{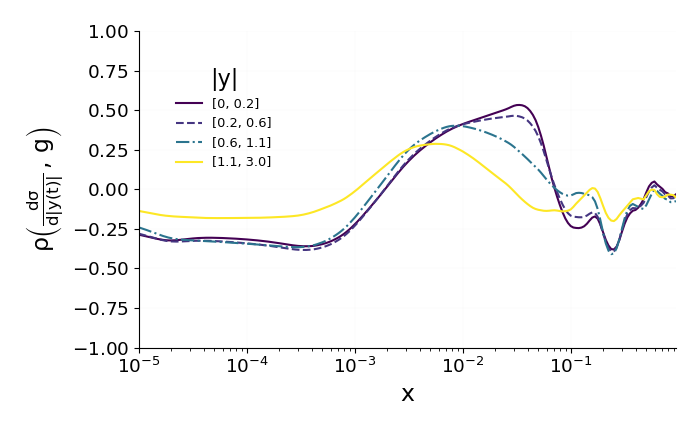}
  \caption{Same as Fig.~\ref{fig:ratio_tot_corr} but for the absolute ATLAS 7 
  TeV differential measurement of the top-quark transverse momentum (left panel)
  and the absolute ATLAS 7 TeV differential measurement of the top-quark 
  rapidity (right panel). Both panels show the correlation coefficient for the 
  gluon. The items in the legend indicate the bins that each of the lines 
  corresponds to.}
  \label{fig:diff_corr}
\end{figure}

We now look at the correlation coefficient between the PDFs and bins of the 
single top differential distributions. Fig.~\ref{fig:diff_corr} shows
the correlation coefficient between the gluon and two single top distributions. The left panel is for the 
ATLAS 7 TeV measurement of the transverse momentum of the top quark, while the 
right panel is for the ATLAS 7 TeV single top measurement of the rapidity of 
the top quark. Both distributions are absolute. For the $p_T(t)$ distribution, 
we observe that each of its bins is sensitive to the gluon, but that the shape 
of $\rho$ is somewhat bin-dependent. The largest correlation ($\rho\simeq 0.5$) 
is seen around $x\simeq 10^{-2}$, in particular for the lowest $p_T$ bin.
For the $|y(t)|$ distribution, we see that it displays a correlation with the
gluon PDF similar to the $p_T(t)$ distribution. The lowest rapidity bins are 
most sensitive to the gluon, as they display a correlation peak 
($\rho\simeq 0.5$) at $x\simeq 0.03$.

We systematically looked at the correlations between the $p_T(t)$ and $|y(t)|$
distributions and the $u$ and $d$ PDFs, as well as at the correlations between 
the top-antiquark distributions and all partons. Such correlations generally
appear to be weaker than those displayed in Fig.~\ref{fig:diff_corr}, and
are therefore not explicitly shown. The only notable exceptions are, as 
expected, correlations of about $\rho\simeq 0.5$ between the $u$ ($d$) and the
$p_T(t)$ distribution around $x\simeq 0.5$ ($x\simeq 0.6$). In this region, 
the higher the transverse momentum of the bin, the higher the sensitivity.
The correlation pattern between the gluon and any of the top-antiquark 
distributions is similar to the one displayed in Fig.~\ref{fig:diff_corr}.

\subsection{Fit settings}
\label{fit-settings}

Having assessed the sensitivity of PDFs to the single top data presented in 
Sect.~\ref{sec:exp_data}, we can now study whether this sensitivity leads to 
any change in the PDFs. We do so by including the data in a PDF fit on top of a 
baseline dataset. Specifically, we consider the NNPDF3.1 dataset of 
Ref.~\cite{Ball:2018iqk}. This includes fixed-target neutral-current (NC) DIS 
structure function data from NMC~\cite{Arneodo:1996kd,Arneodo:1996qe}, 
SLAC~\cite{Whitlow:1991uw} and BCDMS~\cite{Benvenuti:1989rh,Benvenuti:1989fm}; 
charged-current (CC) DIS structure function data from 
CHORUS~\cite{Onengut:2005kv} and NuTeV~\cite{Goncharov:2001qe,Mason:2006qa}; 
HERA measurements from their combined 
datasets~\cite{Abramowicz:2015mha}, including charm-production cross 
sections~\cite{Abramowicz:1900rp}
and $b$-tagged structure functions~\cite{Aaron:2009af,Abramowicz:2014zub}; 
fixed-target 
Drell-Yan data from E866~\cite{Webb:2003ps,Webb:2003bj,Towell:2001nh}
and E605~\cite{Moreno:1990sf}; Tevatron data from 
CDF~\cite{Aaltonen:2010zza} and D0~\cite{Abazov:2007jy,
Abazov:2013rja,D0:2014kma}; and LHC data from 
ATLAS~\cite{Aad:2013iua,Aad:2014qja,Aad:2011dm,Aaboud:2016btc,Aad:2015auj,
Aad:2011fc,Aad:2014kva,Aaboud:2016pbd,Aad:2015mbv}, CMS~\cite{Chatrchyan:2012xt,
Chatrchyan:2013mza,Chatrchyan:2013tia,Khachatryan:2016pev,Khachatryan:2015oaa,
Chatrchyan:2012bja,Khachatryan:2016mqs,Khachatryan:2015uqb,Khachatryan:2015oqa} 
and LHCb~\cite{Aaij:2012vn,Aaij:2012mda,Aaij:2015gna,Aaij:2015zlq}. 
In total this baseline dataset contains $N_{\rm dat}$ = 3979 data points. 

To avoid double counting, the single top data considered here cannot be all 
included at the same time (see Sect.~\ref{sec:exp_data}). We therefore perform
a series of fits, where various combinations of single top 
measurements are added on top of the baseline dataset outlined above. These 
combinations, which define Fit 2 - Fit 5, are summarised in 
Table~\ref{tab:fits_run}. A fit without any single top data (Fit 1) 
and a fit containing the optimal choice of single top measurements 
according to our investigations below (Optimal fit) are performed as reference.
All fits are iterated to reach convergence
of the $t_0$-method (see Sect.~\ref{sec:chi2}).

The realisation of the fits otherwise closely follows the NNPDF3.1 
methodology~\cite{Ball:2017nwa}.
In this analysis, we use $\alpha_s(M_Z)=0.118$, in line with 
the current recommended value from the PDG~\cite{Tanabashi:2018oca}. 
The charm-quark, bottom-quark and top-quark pole masses are set as in the 
NNPDF3.1 default analysis to $m_c=1.51~$GeV, $m_b=4.92~$GeV
and $m_t=172.5~$GeV, respectively, in accordance with the values recommended by 
the Higgs Cross Section Working Group~\cite{deFlorian:2016spz}. 
The PDFs are parametrised and fitted at $Q_0=1.65$~GeV and then evolved at NNLO
with~\textsc{APFEL}~\cite{Bertone:2013vaa}. The {\sc ReportEngine} 
software~\cite{zahari_kassabov_2019_2571601} is used to analyse each fit.

\begin{table}[!t]
  \begin{center}
\scriptsize
\begin{tabular}{lcccccc}
\toprule
Dataset & Fit 1 & Fit 2 & Fit 3 & Fit 4 & Fit 5 & Optimal fit\\
\midrule
Global baseline 
& {\bf y} & {\bf y} & {\bf y} & {\bf y} & {\bf y} & {\bf y}\\
\midrule
ATLAS $\sigma_t/\sigma_{\overline{t}}$ 7 TeV 
& n & n & n & {\bf y} & {\bf y} & {\bf y}\\
ATLAS $\sigma_t/\sigma_{\overline{t}}$ 8 TeV 
& n & {\bf y} & {\bf y} & {\bf y} & {\bf y} & n\\
ATLAS $\sigma_t/\sigma_{\overline{t}}$ 13 TeV 
& n & {\bf y} & {\bf y} & {\bf y} & {\bf y} & {\bf y}\\
CMS $\sigma_t/\sigma_{\overline{t}}$ 8 TeV 
& n & {\bf y} & {\bf y} & {\bf y} & {\bf y} & {\bf y} \\
CMS $\sigma_t/\sigma_{\overline{t}}$ 13 TeV 
& n & {\bf y} & {\bf y} & {\bf y} & {\bf y} & {\bf y}\\
CMS $\sigma_{t+\overline{t}}$ 7 TeV 
& n & {\bf y} & {\bf y} & {\bf y} & {\bf y} & {\bf y}\\
ATLAS $d\sigma/dp_T(t)$ 7 TeV 
& n & {\bf y} & n & n & n & n\\
ATLAS $d\sigma/dp_T(\overline{t})$ 7 TeV 
& n & {\bf y} & n & n & n & n\\
ATLAS $d\sigma/d |y(t)|$ 7 TeV 
& n & n & {\bf y} & n & n & n\\
ATLAS $d\sigma/d|y(\overline{t})|$ 7 TeV 
& n & n & {\bf y} & n & n & n\\
ATLAS $(1/\sigma)d\sigma/dp_T(t)$ 7 TeV            
& n & n & n & {\bf y} & n & n\\
ATLAS $(1/\sigma)d\sigma/dp_T(\overline{t})$ 7 TeV 
& n & n & n & {\bf y} & n & n\\
ATLAS $(1/\sigma)d\sigma/d|y(t)|$ 7 TeV            
& n & n & n & n & {\bf y} & {\bf y}\\
ATLAS $(1/\sigma)d\sigma/d|y(\overline{t})|$ 7 TeV 
& n & n & n & n & {\bf y} & {\bf y} \\
\bottomrule
\end{tabular}
\end{center}

  \caption{The combinations of data that are included (y) or not (n) 
  in each fit.}
  \label{tab:fits_run}
\end{table}

\subsection{Fit results}
\label{global-fit-results}

We now present the results of the fits summarised in Table~\ref{tab:fits_run}.
Table~\ref{tab:global_fits} shows the $\chi^2$  values for the 
datasets included in the various fits, computed according to 
Eqs.~\eqref{eq:chi2}-\eqref{eq:cov_exp}. Numbers in boldface (square brackets)
denote the single top measurements (not) included in each fit. 
The Baseline fit has a $\chi^2=1.19$  which remains stable upon 
the inclusion of single top data in the fit, irrespective of the single top 
distribution considered. 

\begin{table}[!t]
  \begin{center}
\scriptsize
\begin{tabular}{lcccccccc}
\toprule
Dataset & Ref. & $N_{\rm dat}$ & Fit 1 & Fit 2 & Fit 3 & Fit 4 & Fit 5 & Optimal fit \\
\midrule
NMC 												& \cite{Arneodo:1996kd, Arneodo:1996qe} & 325 &  1.31 &  1.31 &  1.32 &  1.31 &  1.30 &  1.31 \\
SLAC                                       			& \cite{Whitlow:1991uw} &  67 & 0.76 &  0.75 &  0.76 & 0.78 &  0.75 &  0.75 \\
BCDMS                                       		& \cite{Benvenuti:1989rh,Benvenuti:1989fm} & 581 & 1.18 & 1.18 & 1.19 & 1.19 & 1.19 & 1.19 \\
CHORUS                                      		& \cite{Onengut:2005kv} & 832 & 1.13 & 1.12 & 1.13 & 1.13 & 1.13 & 1.12 \\
NuTeV                                       		& \cite{Goncharov:2001qe, Mason:2006qa} & 76 & 0.83 & 0.81 & 0.86 & 0.81 & 0.85 & 0.87 \\
HERA $\sigma_{\rm NC,CC}^p$                         & \cite{Abramowicz:2015mha} & 1145 & 1.16 & 1.16 & 1.15 & 1.15 & 1.16 & 1.16 \\
HERA $\sigma_{\rm NC}^c$                            & \cite{Abramowicz:1900rp} & 37 & 1.47 & 1.50 & 1.49 & 1.42 & 1.47 & 1.46 \\
HERA $F_2^b$                                        & \cite{Aaron:2009af,Abramowicz:2014zub} & 29 & 1.12 & 1.12 & 1.12 & 1.12 & 1.12 & 1.12 \\
E886                                                & \cite{Webb:2003ps, Webb:2003bj, Towell:2001nh} & 104 & 1.27 & 1.25 & 1.29 & 1.28 & 1.30 & 1.31 \\
E605                                        		& \cite{Moreno:1990sf} & 85 & 1.23 & 1.22 & 1.24 & 1.26 & 1.19 & 1.22 \\
CDF                                         		& \cite{Aaltonen:2010zza} & 29 & 1.47 & 1.44 & 1.52 & 1.42 & 1.44 & 1.48 \\
D0                                          		& \cite{Abazov:2007jy, D0:2014kma, Abazov:2013rja} & 45 & 1.17 & 1.15 & 1.17 & 1.16 & 1.17 & 1.16 \\
ATLAS high-mass DY 7 TeV                          	& \cite{Aad:2013iua} & 5 & 1.52 & 1.53 & 1.50 & 1.51 & 1.52 & 1.51 \\
ATLAS low-mass DY 7 TeV                          	& \cite{Aad:2014qja} & 6 & 0.89 & 0.90 & 0.88 & 0.89 & 0.90 & 0.89 \\
ATLAS $W,Z$ 7 TeV 2010                              & \cite{Aad:2011dm} & 30 & 0.97 & 0.96 & 0.98 & 0.97 & 0.96 & 0.96 \\
ATLAS $W,Z$ 7 TeV 2011                              & \cite{Aaboud:2016btc} & 34 & 2.21 & 2.22 & 2.17 & 2.24 & 2.17 & 2.06 \\
ATLAS $Z$ $p_T$ 8 TeV                               & \cite{Aad:2015auj} & 92 & 0.92 & 0.92 & 0.90 & 0.91 & 0.92 & 0.91 \\
ATLAS jets 2011 7 TeV                               & \cite{Aad:2011fc} & 31 & 1.11 & 1.13 & 1.11 & 1.11 & 1.12 & 1.12 \\
ATLAS $t\overline{t}$                               & \cite{Aad:2014kva, Aaboud:2016pbd, Aad:2015mbv} & 13 & 1.29 & 1.23 & 1.31 & 1.28 & 1.24 & 1.24 \\
CMS $W$ asymmetry                                   & \cite{Chatrchyan:2012xt, Chatrchyan:2013mza} & 22 & 1.28 & 1.26 & 1.26 & 1.27 & 1.27 & 1.25 \\
CMS Drell-Yan 2D 2011                               & \cite{Chatrchyan:2013tia} & 110 & 1.27 & 1.26 & 1.26 & 1.26 & 1.26 & 1.26 \\
CMS $W$ rapidity 8 TeV                              & \cite{Khachatryan:2016pev} & 22 & 1.03 & 1.01 & 1.00 & 0.98 & 1.03 & 0.99 \\
CMS $Z$ $p_T$ 8 TeV $(p_T^{ll},y_{ll})$             & \cite{Khachatryan:2015oaa} & 28 & 1.31 & 1.30 & 1.34 & 1.32 & 1.31 & 1.31 \\
CMS jets 7 TeV 2011                                 & \cite{Chatrchyan:2012bja} & 133 & 1.03 & 0.94 & 0.95 & 0.96 & 0.96 & 0.93 \\
CMS $t\overline{t}$                                 & \cite{Khachatryan:2016mqs, Khachatryan:2015uqb, Khachatryan:2015oqa} & 13 & 0.86 & 0.87 & 0.87 & 0.87 & 0.87 & 0.85 \\
LHCb $Z$ 940 pb                                   	& \cite{Aaij:2012vn} & 9 & 1.43 & 1.45 & 1.47 & 1.41 & 1.46 & 1.45 \\
LHCb $Z \to ee$ 2 fb                              	& \cite{Aaij:2012mda} & 17 & 1.14 & 1.12 & 1.14 & 1.12 & 1.12 & 1.13 \\
LHCb $W,Z \to \mu$ 7 TeV                          	& \cite{Aaij:2015gna} & 29 & 1.93 & 1.83 & 1.88 & 1.84 & 1.84 & 1.86 \\
LHCb $W,Z \to \mu$ 8 TeV                          	& \cite{Aaij:2015zlq} & 30 & 1.64 & 1.56 & 1.54 & 1.58 & 1.55 & 1.52 \\
\midrule
ATLAS $\sigma_t/\sigma_{\overline{t}}$ 7 TeV      	& \cite{Aad:2014fwa} & 1 & [0.32] & [0.39] & [0.37] & \textbf{0.33} & \textbf{0.34} & \textbf{0.30} \\
ATLAS $\sigma_t/\sigma_{\overline{t}}$ 8 TeV      	& \cite{Aaboud:2017pdi} & 1 & [3.54] & \textbf{3.13} & \textbf{3.15} & \textbf{3.34} & \textbf{3.25} & [3.72] \\
ATLAS $\sigma_t/\sigma_{\overline{t}}$ 13 TeV     	& \cite{Aaboud:2016ymp} & 1 & [0.04] & \textbf{0.04} & \textbf{0.05} & \textbf{0.04} & \textbf{0.04} & \textbf{0.04} \\
CMS $\sigma_t/\sigma_{\overline{t}}$ 8 TeV        	& \cite{Khachatryan:2014iya} & 1 & [0.10] & \textbf{0.13} & \textbf{0.13} & \textbf{0.11} & \textbf{0.12} & \textbf{0.09} \\
CMS $\sigma_t/\sigma_{\overline{t}}$ 13 TeV       	& \cite{Sirunyan:2016cdg} & 1 & [0.31] & \textbf{0.31} & \textbf{0.33} & \textbf{0.31} & \textbf{0.30} & \textbf{0.30} \\
CMS $\sigma_{t+\overline{t}}$ 7 TeV               	& \cite{Chatrchyan:2012ep} & 1 & [0.63] & \textbf{0.59} & \textbf{0.63} & \textbf{0.62} & \textbf{0.59} & \textbf{0.62} \\
ATLAS $d\sigma/dp_T(t)$ 7 TeV                     	& \cite{Aad:2014fwa} & 5 & [2.87] & \textbf{2.87} & [2.87] & [2.87] & [2.88] & [2.87] \\
ATLAS $d\sigma/dp_T(\overline{t})$ 7 TeV          	& \cite{Aad:2014fwa} & 5 & [12.0] & \textbf{12.0} & [12.1] & [12.0] & [12.0] & [11.9] \\
ATLAS $d\sigma/d|y(t)|$ 7 TeV                     	& \cite{Aad:2014fwa} & 4 & [1.31] & [1.32] & \textbf{1.31} & [1.30] & [1.31] & [1.31] \\
ATLAS $d\sigma/d|y(\overline{t})|$ 7 TeV          	& \cite{Aad:2014fwa} & 4 & [0.07] & [0.07] & \textbf{0.06} & [0.07] & [0.07] & [0.07] \\
ATLAS $(1/\sigma)d\sigma/dp_T(t)$ 7 TeV     		& \cite{Aad:2014fwa} & 4 & [2.79] & [2.79] & [2.79] & \textbf{2.79} & [2.79] & [2.79] \\
ATLAS $(1/\sigma)d\sigma/dp_T(\overline{t})$ 7 TeV  & \cite{Aad:2014fwa} & 4 & [6.28] & [6.30] & [6.32] & \textbf{6.29} & [6.29] & [6.28] \\
ATLAS $(1/\sigma)d\sigma/d|y(t)|$ 7 TeV    			& \cite{Aad:2014fwa} & 3 & [0.97] & [0.97] & [0.97] & [0.98] & \textbf{0.94} & \textbf{0.96} \\
ATLAS $(1/\sigma)d\sigma/d|y(\overline{t})|$ 7 TeV 	& \cite{Aad:2014fwa} & 3 & [0.07] & [0.06] & [0.06] & [0.06] & \textbf{0.06} & \textbf{0.06} \\
\midrule
Total & & 4017 & 1.19 & 1.19 & 1.19 & 1.19 & 1.19 & 1.19 \\
\bottomrule
\end{tabular}
\end{center}

  \caption{The $\chi^2$  for the NNLO PDF fits summarised in 
  Table~\ref{tab:fits_run}, as well as for the PDF fit to the combination
  of data that we find to be optimal. Numbers in boldface denote the single top
  measurements included in each fit. Numbers in square brackets denote
  datasets not included in the fit, and are computed from each fit.}
  \label{tab:global_fits}
\end{table}

Turning our attention to the results for individual single top datasets, 
we observe, overall, a remarkable stability in the values of the
$\chi^2$  
before and after the fit, irrespective of the specific single top 
measurement included in the fit. In each case, the $\chi^2$ 
fluctuates around its baseline value in a statistically insignificant way.
Note that we consider a shift in the $\chi^2$ as statistically significant
only if it is larger than one standard deviation of the $\chi^2$ distribution; 
that is the new $\chi^2$  lies outside the limits defined by
$\frac{\chi^2 N_{\text{dat}} \pm \sqrt{2N_{\text{dat}}}}{N_{\text{dat}}}$,
where $\chi^2$  is the value for the Baseline fit, 
and $N_{\text{dat}}$ is the number of data points which have been used to 
normalise the $\chi^2$. Overall, single top data are fairly well
described, albeit with the following exceptions, already observed in 
Sect.~\ref{sec:data-theory-comparison}.

First, the poor pre-fit $\chi^2$  observed in 
Sect.~\ref{sec:data-theory-comparison} for the transverse momentum differential 
distributions does not improve at all when the data is included in a fit. 
By comparison with the rapidity distributions, which instead show an equally 
acceptable $\chi^2$  before and after the fit, we cannot envision 
any theoretical reason that could explain such a discrepancy. We therefore 
conclude that, unless this is understood, the differential distribution in the 
transverse momentum of the top antiquark should not be included in a PDF fit. 
Second, the comparatively large pre-fit $\chi^2$  for the 
ATLAS 8 TeV $\sigma_{t}/\sigma_{\overline{t}}$ measurement 
decreases each time single top data is included in the fit, although it reaches 
a not particularly good minimum of 3.13 in Fit 2. We therefore conclude that 
the ATLAS single top-quark to top-antiquark ratio at 8 TeV should not be 
included in a fit.

Having established which datasets are not suitable for inclusion in a fit, 
we then assessed which are the most constraining on the PDFs
among the remaining ones. A careful analysis of the results of all of the fits 
reported in Table~\ref{tab:fits_run} led us to the conclusion that the ATLAS 7 
TeV normalised rapidity distributions are such datasets. We therefore included
them in our Optimal fit, together with the ATLAS top-quark to top-antiquark
ratios at 7 and 13 TeV, the CMS total cross section at 7 TeV, and the CMS
top-quark to top-antiquark ratios at 8 and 13 TeV. Note that the Optimal
fit is equivalent to Fit 5, except for the exclusion of the ATLAS top-quark
to top-antiquark ratio at 8 TeV.

\begin{figure}[!t]
\centering
  \includegraphics[width=0.49\textwidth]{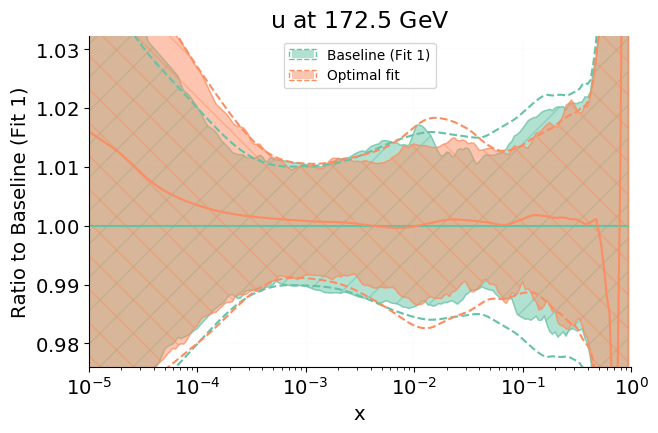}
  \includegraphics[width=0.49\textwidth]{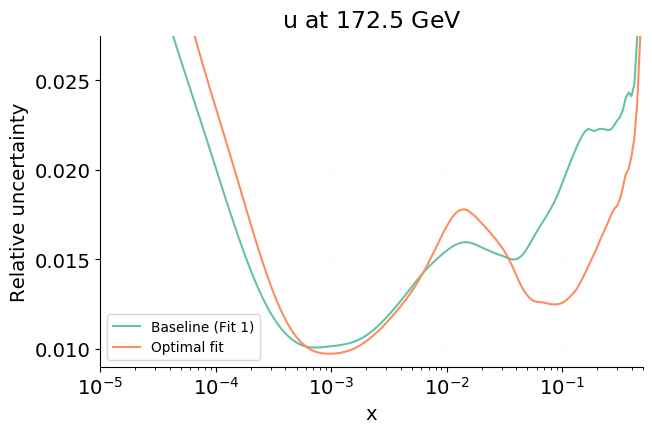}
  \includegraphics[width=0.49\textwidth]{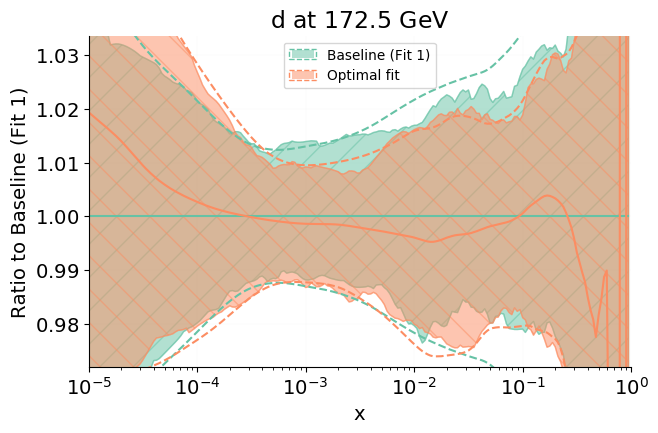}
  \includegraphics[width=0.49\textwidth]{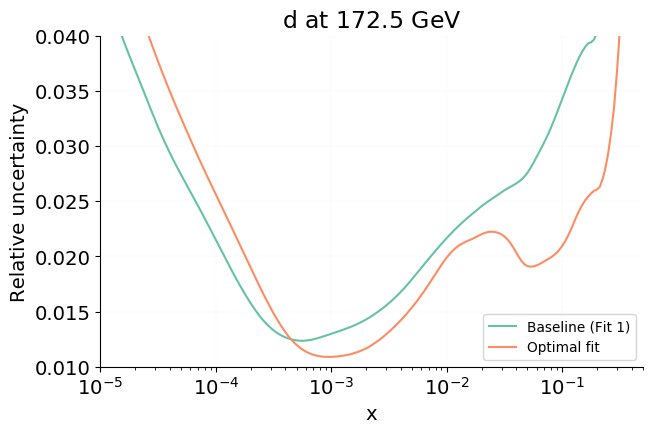}
  \includegraphics[width=0.49\textwidth]{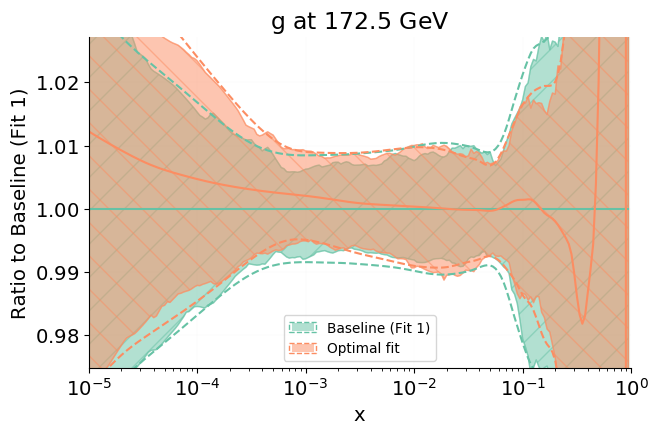}
  \includegraphics[width=0.49\textwidth]{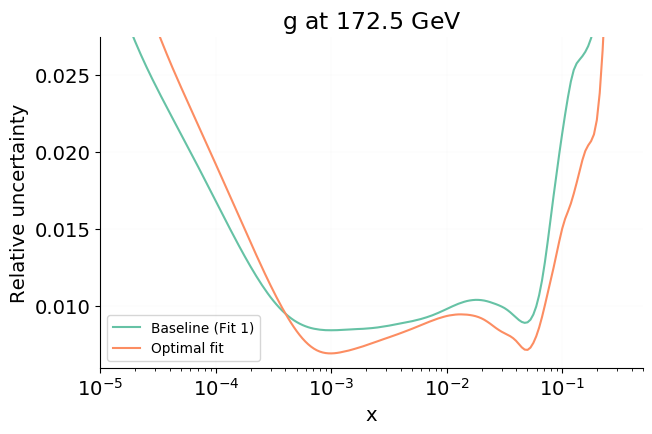}
  \caption{(Left) PDF plots of the up quark (top), down quark (middle) and 
  gluon (bottom) for two global fits: Fit 1 (baseline) and the Optimal fit,
  as labelled in Tab.~\ref{tab:global_fits}. Each plot is normalised to the  
  central value of Fit 1, and is shown at $Q=m_t=172.5$~GeV. The PDF 
  uncertainty is shown as both a 68\% CL hatched band and a 1$\sigma$ dashed 
  line. (Right) The 1-$\sigma$ PDF uncertainties, normalised to their 
  respective PDF central values.}
  \label{fig:pdf_plots}
\end{figure}

In Fig.~\ref{fig:pdf_plots} we compare the up, down and gluon PDFs for
the Optimal fit to those for the Baseline fit (Fit 1). Specifically, we show
the PDFs and their uncertainties (both 1$\sigma$ and $68\%$ CL bands) 
normalised to the central value of the Baseline fit (left panel); and the 
relative uncertainties of the PDFs (right panel). Results are shown at 
$Q=m_t=172.5$~GeV.
As can be seen from Fig.~\ref{fig:pdf_plots}, the central values remain
essentially stable, while PDF uncertainties are reduced. The largest reduction
(by about one third of the baseline uncertainty) is observed for the up and 
down quark PDFs at $x\gtrsim 10^{-3}$. The regions of $x$ 
in which we see constraining power from the single top data is consistent 
with what we expected to see from our analysis of the observable-PDF
correlations in Sect.~\ref{obs-pdf-corrs}. For a discussion of the impact of
single top data on the $u/d$ ratio, see Sect.~\ref{sec:pheno}.

Turning our attention to the other fits in our study, we find that the fit 
including absolute $p_T$ distributions is the least constraining (Fit 2), 
and that the fits including the normalised $p_T$ distributions (Fit 4) and 
the absolute rapidity distributions (Fit 3) are similarly constraining, 
with the former tending to be slightly more constraining. We attribute these
results to two sources: firstly, we would expect that the fits including 
rapidity distributions are more constraining than those including transverse 
momentum distributions, because the former are more consistent with theoretical 
predictions than the latter are (see Sect.~\ref{sec:data-theory-comparison});
secondly, we would expect that the fits including normalised distributions are 
more constraining than those including absolute distributions, since the former 
benefit from uncertainty cancellations between the numerator and the 
denominator.

Having assessed the impact of the various datasets on PDFs, we can now revisit 
the question of whether the poor $\chi^2$  for the top-antiquark 
differential distributions and for the ATLAS top-quark to top-antiquark 
ratio at 8 TeV could be driven by tensions with other measurement. To this 
purpose, we check whether a statistically significant shift in the 
$\chi^2$  of any of the baseline datasets has occurred when 
single top data has been fitted. We do this as defined at the start of 
Sect.~\ref{global-fit-results} and we find that no datasets exhibit such a 
shift. We do however see shifts of around $0.8\sigma$ towards a lower $\chi^2$ 
 for the CMS jets at 7 TeV~\cite{Chatrchyan:2012bja}
in two cases: firstly, when fitting the absolute transverse momentum 
distributions (Fit 2), and secondly, when fitting the normalised rapidity
distributions (Fit 5). Note that the shift is $0.7\sigma$ for the Optimal fit.
This suggests that `CMS jets 7 TeV 2011' is highly
correlated with some of the single top data. The next section 
will discuss such inter-dataset correlations in more detail.

We also repeat our fits by adding the single top data on top of a 
maximally consistent baseline dataset made up of HERA data 
only~\cite{Abramowicz:2015mha,Abramowicz:1900rp}. Despite the more restrictive
dataset, we run into the same inconsistencies reported above, namely we are
not able to achieve an acceptable $\chi^2$  for the 7 TeV 
top-antiquark 
transverse momentum differential distributions. This result demonstrates that 
the poor description of this dataset does not depend on tensions with other LHC 
measurements. We therefore conclude that the ATLAS differential distributions
in the transverse momentum of the top antiquark are internally inconsistent
(for further evidence obtained by analysing differential distributions at 
8 TeV, see Appendix~\ref{app:ATLAS8TeVdiff}).
Such an inconsistency should possibly require the experimental analysis to
be reassessed in the future.

All of these findings allow us to summarise which combination of single top data 
is optimal for inclusion in the next generation of PDF fits: namely, the data 
used in the Optimal fit, that is the CMS total cross section at 7 TeV, the CMS
top-quark to top-antiquark ratios at 8 and 13 TeV, the ATLAS top-quark to 
top-antiquark ratios at 7 and 13 TeV, and the ATLAS normalised rapidity 
distributions for the top quark and the top antiquark at 7 TeV. 
This combination is best able to constrain the PDFs relevant 
to single top production, avoids data with unsatisfactorily poor
theoretical descriptions, and minimises any possible residual theoretical ambiguities.

\subsection{Observable-observable correlations}
\label{obs-obs-corrs}

We now study the correlation between single top data and other datasets that 
are included in the Baseline fit. To calculate this correlation 
we replace $f$ with a second observable in Eq.~\ref{eq:pdf_obs_corr}. 
In what follows we compute $\rho$ between single top data and two other classes 
of measurement: firstly, those which are largely gluon-initiated, and secondly,
those which are EW-induced, and in particular are mediated by a $W$ boson. 
We choose the first class because single top production is sensitive to the 
gluon PDF, as was seen in Sect.~\ref{obs-pdf-corrs}, and we choose the second 
class because single top production in the $t$-channel is mediated by a 
$W$ boson. We consistently compute observables at NNLO in pure QCD at a 
reference scale $Q=m_t=172.5$~GeV, and use
the {\tt NNPDF31\_nnlo\_as\_0118} PDF set. We explicitly checked that 
correlation patterns are not affected if any of the fits presented in 
Sect.~\ref{sec:fits} is used instead.

\begin{figure}[!t]
\centering
  \includegraphics[width=0.49\textwidth]{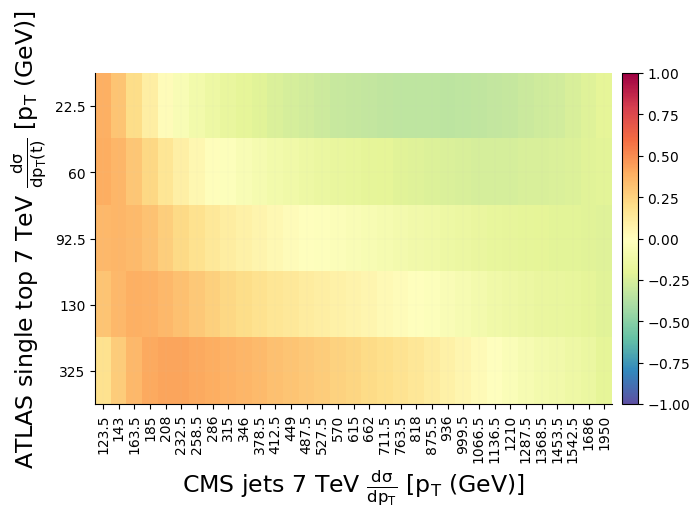}
  \includegraphics[width=0.49\textwidth]{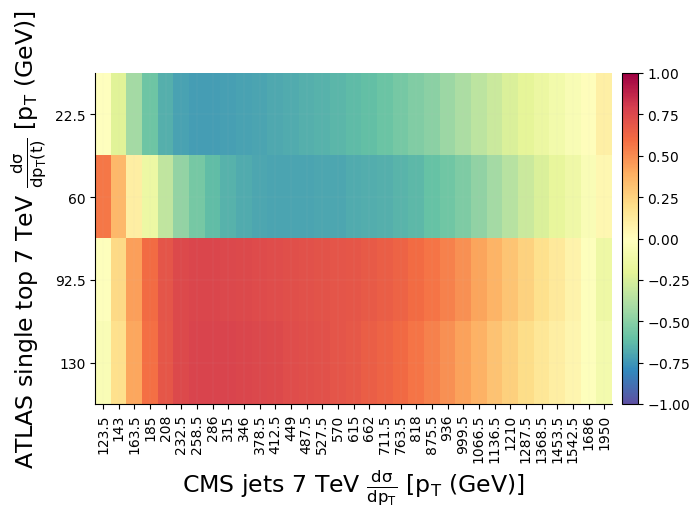}
  \caption{The correlations between bins of the CMS 7 TeV inclusive jet dataset,
  which is binned in the $p_T$ of the jet, and both the ATLAS 7 TeV single 
  top-quark $p_T$ absolute (left panel) and normalised (right panel) 
  distributions. The CMS 7 TeV inclusive jet dataset is made up of five 
  separate distributions, each of which represents a different rapidity bin. 
  One bin is shown here, which is for $|y|<0.5$. These plots have been produced 
  using theoretical predictions at NNLO in QCD with the 
  {\tt NNPDF31\_nnlo\_as\_0118} PDF set 
  and $Q = m_t = 172.5$~GeV. The colour bar indicates the correspondence 
  between $\rho$ and the colours in the plot.}
  \label{fig:cmsjets11-corr}
\end{figure}

We start by looking at the jets data. Fig.~\ref{fig:cmsjets11-corr} shows the 
correlations between the ATLAS 7 TeV top-quark $p_T$ 
distribution, with the single top data unnormalised 
(left panel) and normalised (right panel), and 
the CMS 7 TeV jet data~\cite{Chatrchyan:2012bja}, for which we saw a 
0.8$\sigma$ improvement
in the $\chi^2$ upon fitting single top data. 
For the absolute single top distribution we see reasonably strong correlation 
for low values of the jet $p_T$, which turns into anti-correlation as the jet 
$p_T$ increases. Upon normalising the single top data, the correlations become 
starker and a different correlation pattern emerges. The pattern becomes one 
where the lowest two rapidity bins of the single top data are generally 
strongly anti-correlated with the jet data, while the highest three rapidity 
bins are generally strongly correlated, with the correlation reaching 
$|\rho| \simeq 1$ in each case. Similar patterns of correlation are seen for the 
companion ATLAS 7 TeV jet data~\cite{Aad:2011fc}, although no shift in the 
$\chi^2$ was seen for this dataset upon fitting single top data. Note that our fits in 
Sect.~\ref{global-fit-results} include five separate rapidity bins for the 
CMS jets data, the lowest of which we show here. A general trend of lower 
rapidities leading to stronger correlations is seen.
%
%
Such a  correlation has its origin in the 
gluon.

\begin{figure}[!t]
\centering
  \includegraphics[width=0.49\textwidth]{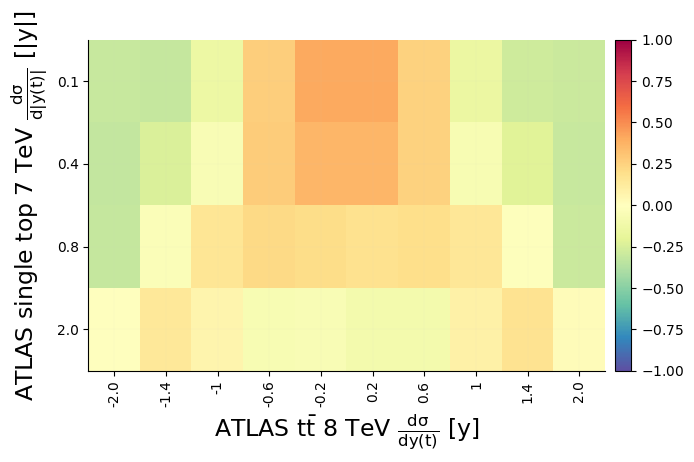}
  \includegraphics[width=0.49\textwidth]{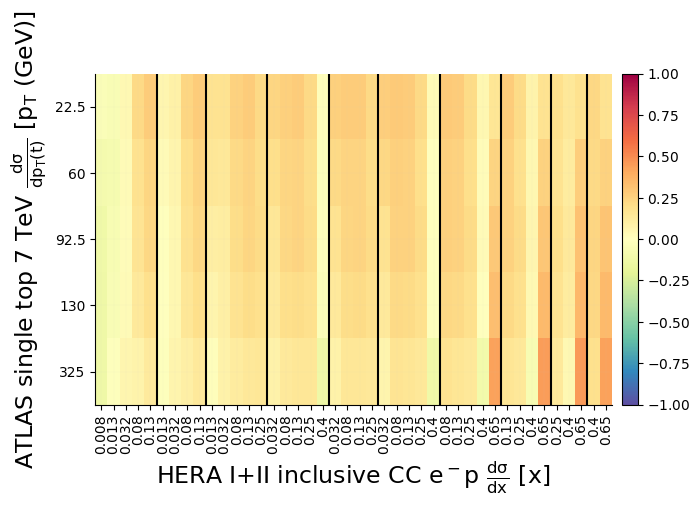}
  \caption{(Left) The correlations between bins of the ATLAS 8~TeV 
  $t\overline{t}$ dataset, which is binned in the rapidity of the top quark, 
  and the ATLAS 7~TeV single top-quark absolute rapidity distribution. 
  (Right) The correlations between the HERA combination $e^{-}p$ dataset and 
  the absolute ATLAS 7 TeV single top-quark transverse momentum distribution. 
  The HERA data is binned in $Q^2$ and then within these bins, it is binned in 
  Bjorken-$x$. The vertical black lines indicate the edges of the $Q^2$ bins. 
  The lowest of these bins is for $Q^2 = 300$ GeV$^2$ and the highest is for 
  $Q^2 = 30 \times 10^3$ GeV$^2$. Both plots have been 
  produced using theoretical predictions at NNLO in QCD with the 
  {\tt NNPDF31\_nnlo\_as\_0118} PDF set and $Q = m_t = 172.5$~GeV. The colour 
  bar indicates the correspondence between $\rho$ and the colours in the plot.}
  \label{fig:atlasttbar-corr}
\end{figure}

To study this further, we now look at another gluon-sensitive process: 
$t\overline{t}$ production. Fig.~\ref{fig:atlasttbar-corr} (left) shows the 
correlation between the ATLAS 7 TeV top-quark $|y|$ distribution and the 
ATLAS 8 TeV $t\overline{t}$ rapidity distribution~\cite{Aad:2015mbv}. 
Here we see some correlation in some bins, most notably with the low rapidity bins of each 
distribution exhibiting strong correlation ($\rho \simeq 0.5$) and the high 
rapidity bins of the $t\overline{t}$ data showing notable anti-correlation 
with the lowest three bins of the single top distribution 
($\rho \simeq 0.3-0.5$). However, the correlation with the $t\overline{t}$ data 
seems weaker than with the jets data. Clearly, while both single top production and $t\overline{t}$ are 
gluon-sensitive, $t\overline{t}$ tends to probe higher values of $x$ than both 
single top production and jets production. Hence, the inter-dataset correlation 
is stronger in the case of the latter datasets.

Next we look at correlations between single top data and other EW processes. 
Naively, we might expect that single top production data would exhibit 
non-negligible correlations with datasets such as weak boson production 
from LHCb~\cite{Aaij:2015gna} and ATLAS~\cite{Aaboud:2016btc} at 7 TeV, 
because these datasets contain many data 
points for $W$-boson mediated processes, and therefore they may depend on 
similar contributions from the PDFs. However, for these datasets we do not see 
any notable correlations with the single top data. This is due to the
fact that single top production probes $u$ and $d$ in different regions 
of $x$ to either of these two datasets.

To investigate this further we consider another EW dataset, but this time from 
the HERA collider instead of the LHC. Here we look at HERA CC data from $e^{-}p$
collisions~\cite{Abramowicz:2015mha}. 
Fig.~\ref{fig:atlasttbar-corr} (right) shows the correlations 
between these data and the ATLAS 7 TeV single top-quark $p_T$ distribution. 
Here we see regions of large correlation, up to $\rho \simeq 0.5-0.75$, 
contrary to what was seen for the LHC data we looked at previously. 
In particular, the correlation is largest for bins where $x=0.65$, with the 
correlation increasing with the single top-quark $p_T$. The correlation is 
relatively weak for the $x=0.4$ bins but is moderate for the $x \simeq 0.1$ 
and the $x=0.25$ bins. We therefore conclude that single top production data is 
indeed correlated with other EW observables, but that this effect is very 
dependent on the region of $x$ that the data probe. The HERA CC dataset covers
similar regions of $x$ to the single top data, whereas the same is possibly not 
true for the aforementioned LHC data.

We conclude by noting that the picture we see here is similar to that which we 
saw in Sect.~\ref{obs-pdf-corrs}. Namely, that single top production has a 
non-negligible dependence on the gluon PDF through its $b$-dependence, except 
when we consider measurements of $\sigma_{t}/\sigma_{\overline{t}}$ in which the 
gluon dependence largely cancels. This sensitivity, which we see via the 
correlation between single top data and both jets data and top-quark pair 
production data, supplements the up-quark and down-quark sensitivity that we 
expect. It also leads to notable correlations with HERA CC data, in which a 
similar $x$-region is probed.

This tells us that strong correlation is a necessary but not a sufficient 
condition for the data having an impact at the fit-level. For single top 
production we observe both non-negligible PDF-observable correlations and 
notable correlations with HERA CC data, $t\overline{t}$ data and LHC jets data. 
However, the CMS jets data is the only data to show an improved data-theory 
agreement upon the inclusion of single top data in a fit, and this is true only 
in three of the five fits in which the single top data were included.
Such a picture is expected to be modified as soon as more abundant and more 
precise single top data will become available.

\subsection{Phenomenological impact} 
\label{sec:pheno}

Here we discuss the ability of single top data to constrain quantities that are
relevant for LHC phenomenology. We first consider the effect of single top data
on the up-quark to down-quark PDF ratio, to which end we analyse the change in
PDF uncertainties in predictions of the $W^+/W^-$ ratio. We then turn our attention to the
effect that single top data have on constraining parton luminosities.

\begin{figure}[!t]
\centering
  \includegraphics[width=0.49\textwidth]{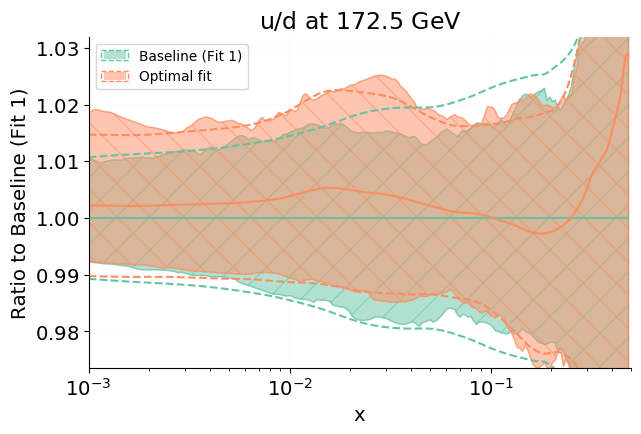}
  \includegraphics[width=0.49\textwidth]{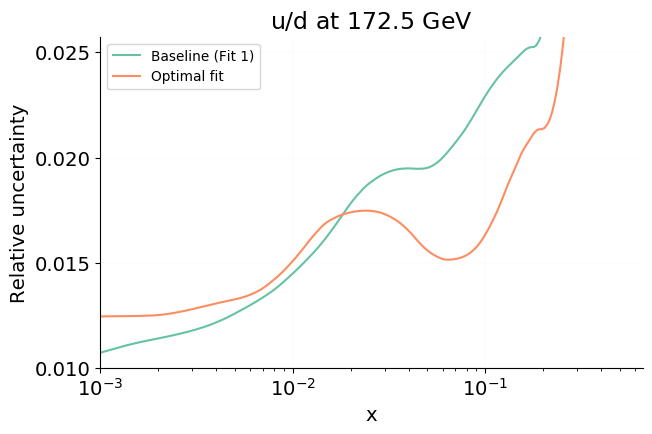}
  \caption{Same as Fig.~\ref{fig:pdf_plots} but for the up-quark to down-quark PDF ratio.}
  \label{fig:pdf_plots_u_d_ratio}
\end{figure}

Fig.~\ref{fig:pdf_plots_u_d_ratio} shows how the $u/d$ ratio changes upon the
inclusion of single top data in the PDF fit, where specifically we compare the
Baseline fit with the Optimal fit, as defined in Sect.~\ref{tab:global_fits}. We
show how the PDF central value and its uncertainty changes upon including single
top data (left-hand panel) and in addition how the relative uncertainty changes
(right-hand panel). We find that this combination of data leads to a reduction
in PDF uncertainties for the $u/d$ ratio, as we also saw for the PDF flavours
discussed in Sect.~\ref{global-fit-results}. In particular, a reduction in
uncertainties is seen in the range $10^{-2} \lesssim x \lesssim 0.5$, which is
largest (up to 25\%) at $x \simeq 0.003$. This corroborates the discussion in
Sect.~\ref{obs-pdf-corrs} in which we said that single top data, and in
particular measurements of $\sigma_t/\sigma_{\overline{t}}$, are indeed
sensitive to $u/d$, and that this is true for a slightly larger range than was
found in Ref.~\cite{Aad:2014fwa}.

\begin{figure}[!t]
\centering
  \includegraphics[width=0.98\textwidth]{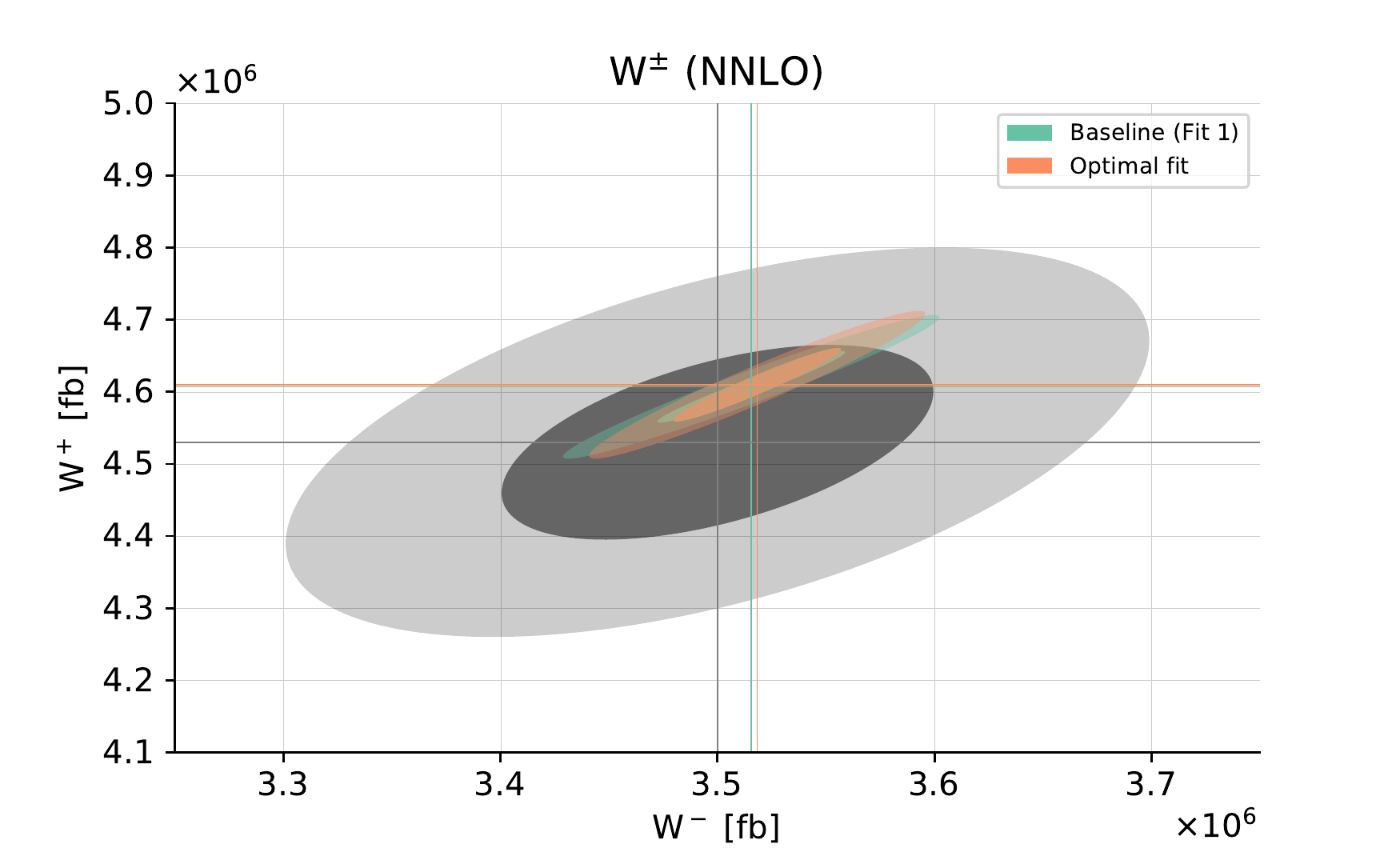}
  \caption{Theoretical predictions for the $W^+$ and $W^-$
cross section at NNLO obtained with the Baseline fit (green) and the Optimal fit (orange), 
compared to the 13 TeV measurement performed by ATLAS~\cite{Aad:2016naf}. 
The 68\% CL and 90\% CL ellipses for the theoretical predictions are represented in dark 
and light shades, respectively. Correspondingly, the same CLs are plotted
for the experimental measurement in dark and light grey.}
  \label{fig:wratio}
\end{figure}

The $u/d$ ratio is an important quantity for predictions of observables such as 
$W$ and $Z$ production, which have been measured very precisely at the 13 TeV 
LHC run by ATLAS~\cite{Aad:2016naf}. In particular the measurement of the 
$W^+/W^-$ inclusive cross section ratio is very sensitive to the $u/d$ ratio. 
To quantify the difference that including single top data in PDF fits makes on 
such predictions, we compute the $W^+/W^-$ ratio at NNLO at $\sqrt{s}=13$ TeV 
using the Baseline and the Optimal fit including the constraints provided by 
single top data and we compare them to the ATLAS experimental result.
Results are displayed in Fig.~\ref{fig:wratio}. We see that the central value 
is slightly shifted by the inclusion of the single top data, but it is still 
well within the experimental 1$\sigma$ ellipse. The PDF uncertainty on the other
hand remains almost unchanged. Still it is significantly smaller than the data
uncertainty: a consistency test could be made more stringent only by increasing
the precision of the experimental measurement.

\begin{figure}[!t]
\centering
  \includegraphics[scale=0.38]{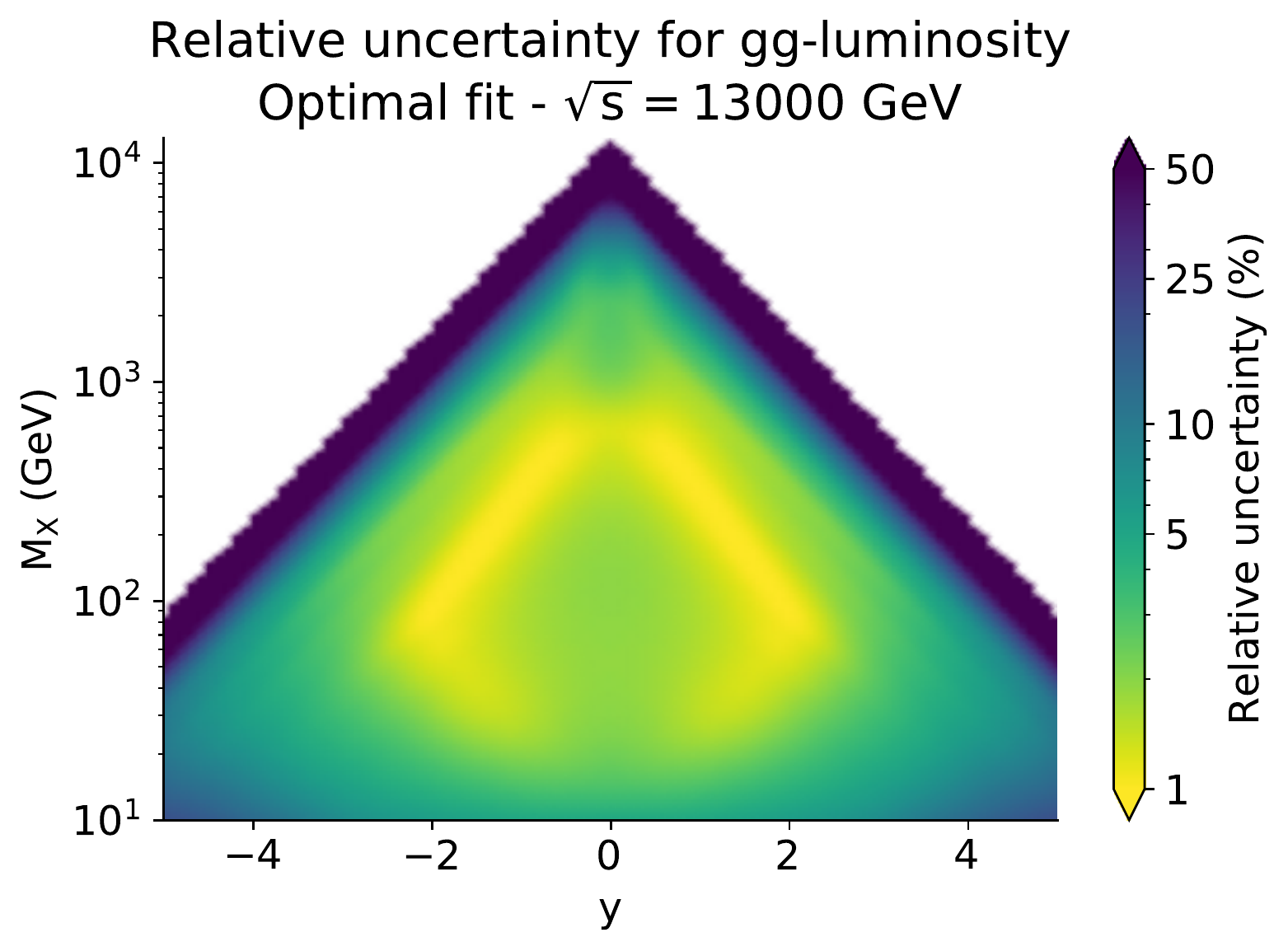}
  \includegraphics[scale=0.38]{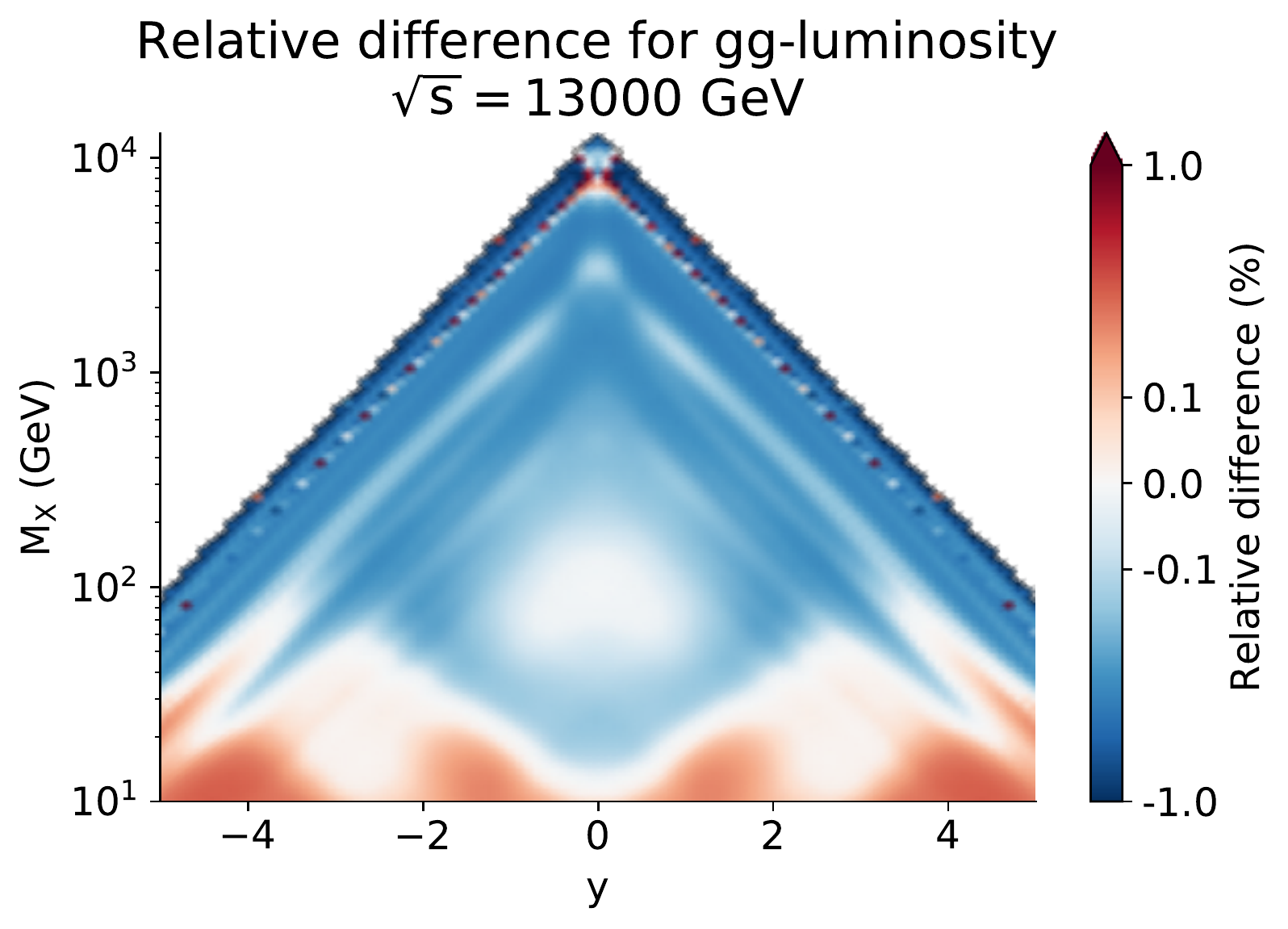}\\
  \includegraphics[scale=0.38]{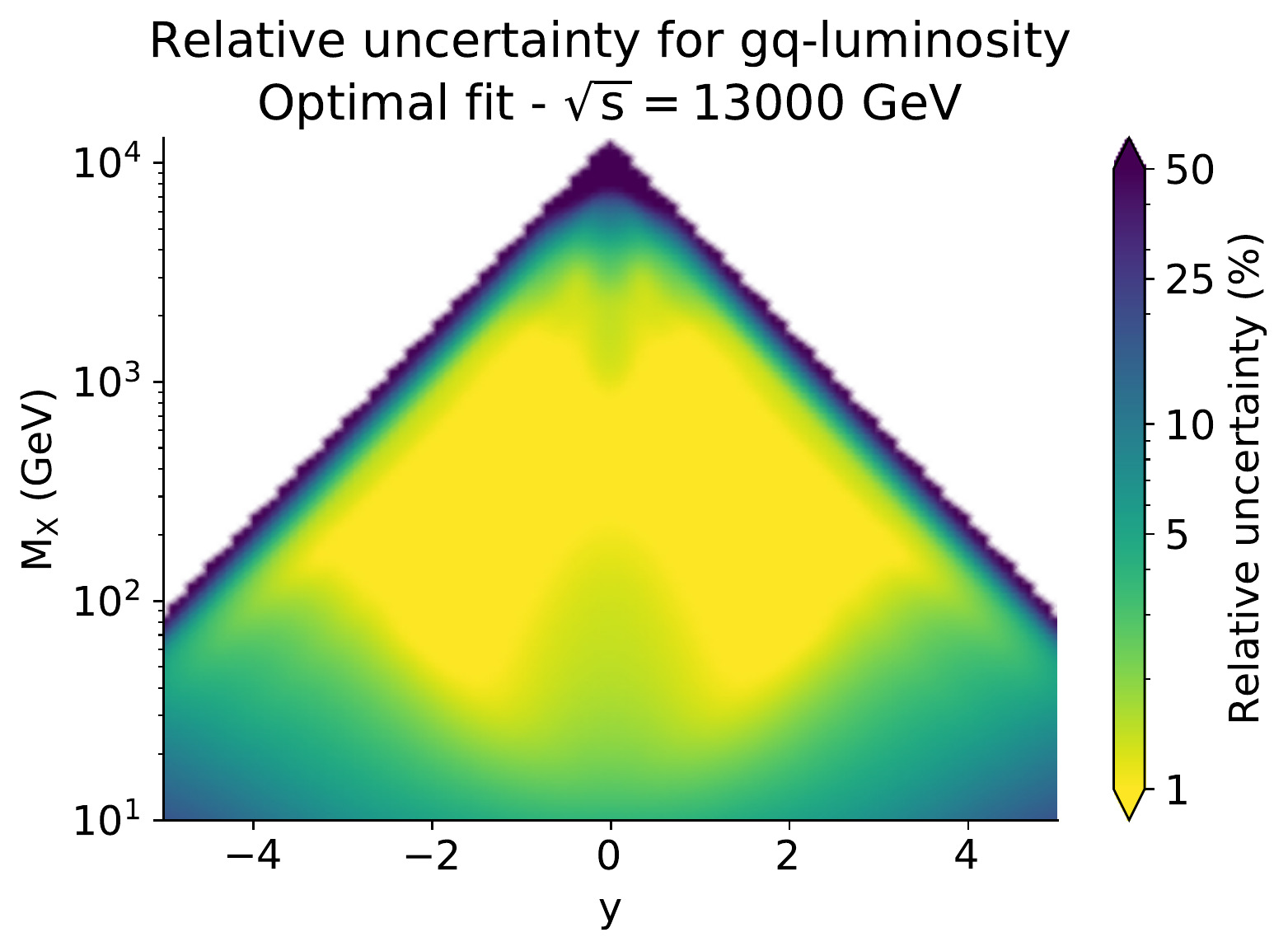}
  \includegraphics[scale=0.38]{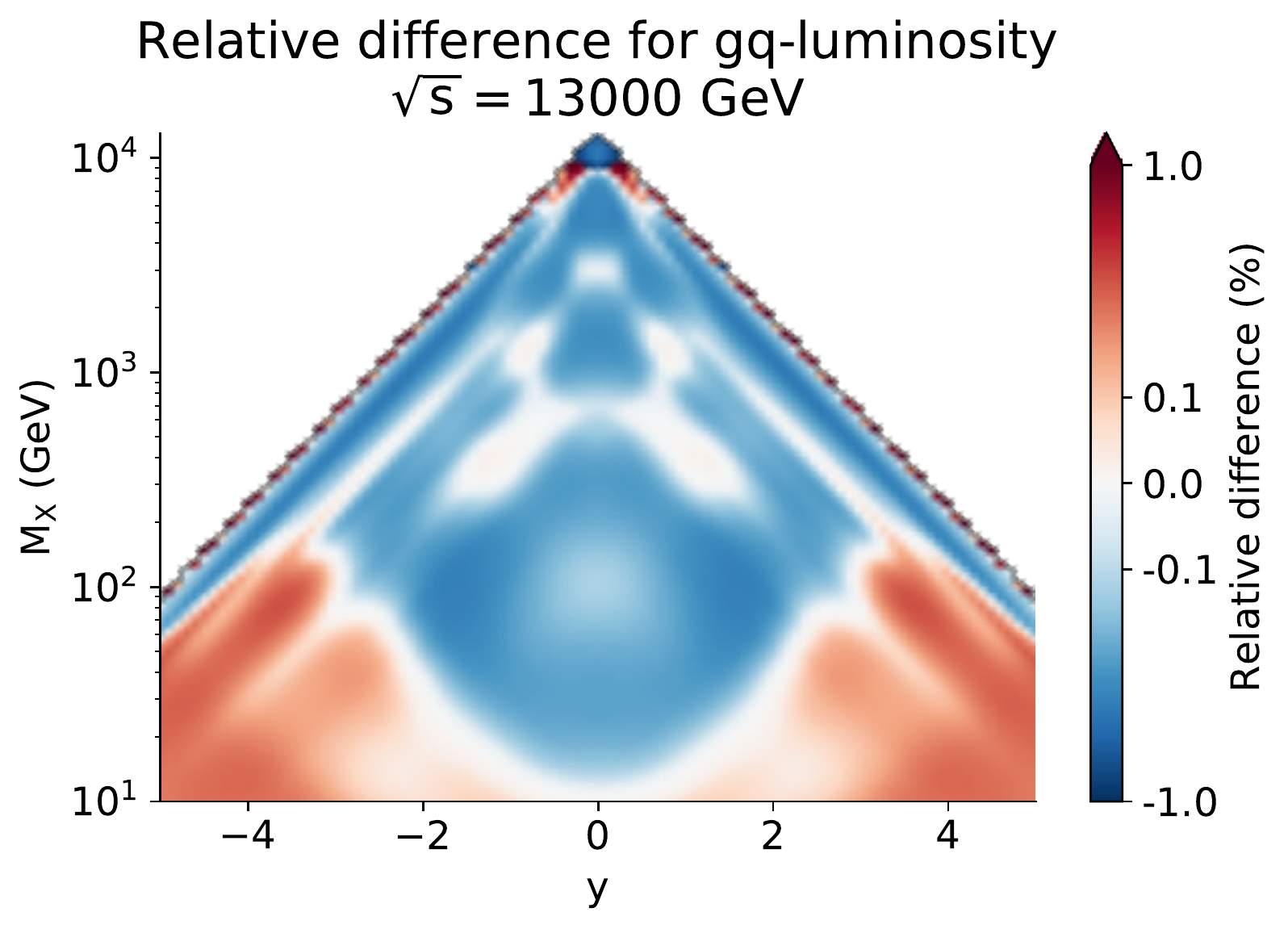}\\
  \includegraphics[scale=0.38]{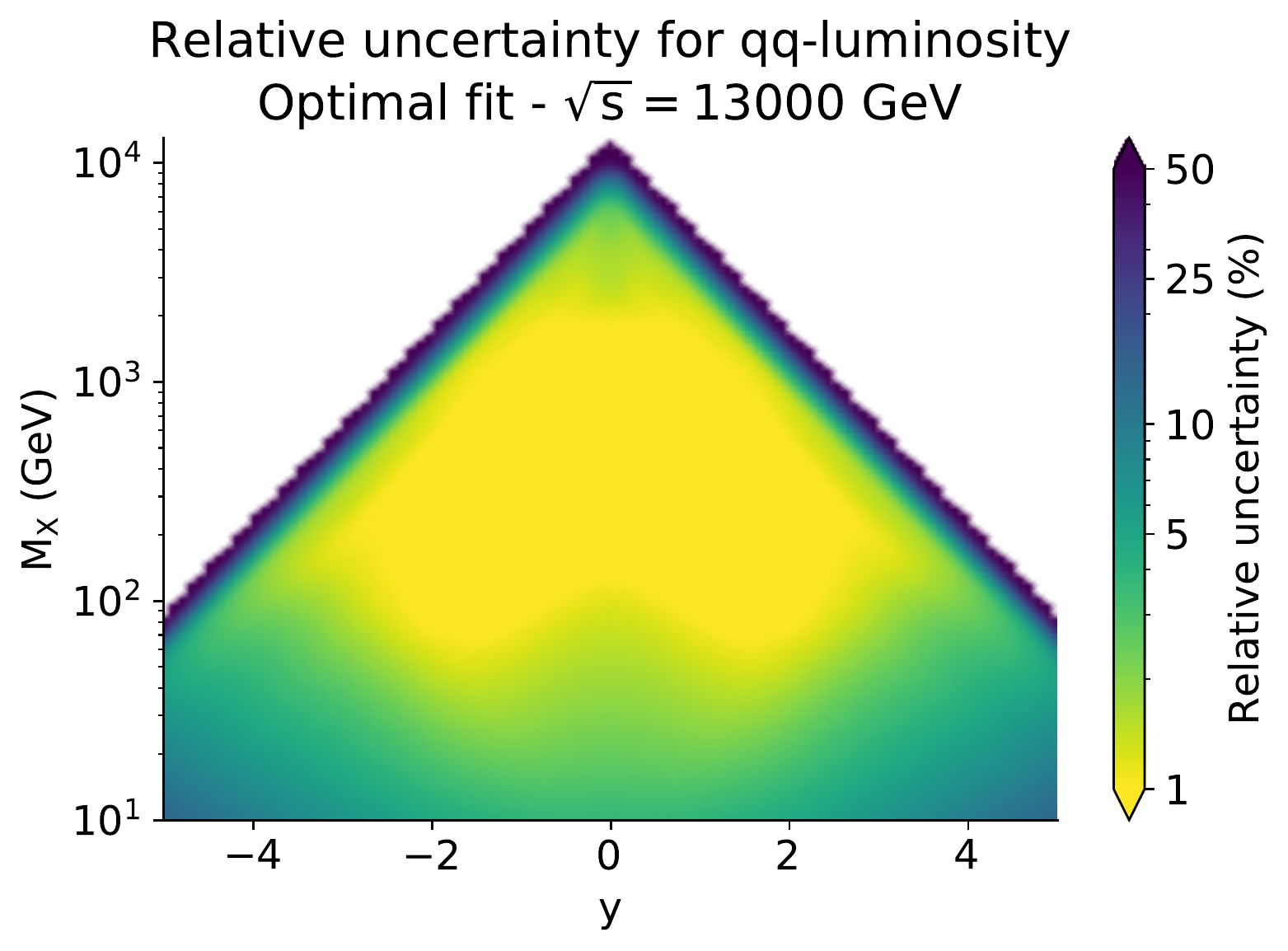}
  \includegraphics[scale=0.38]{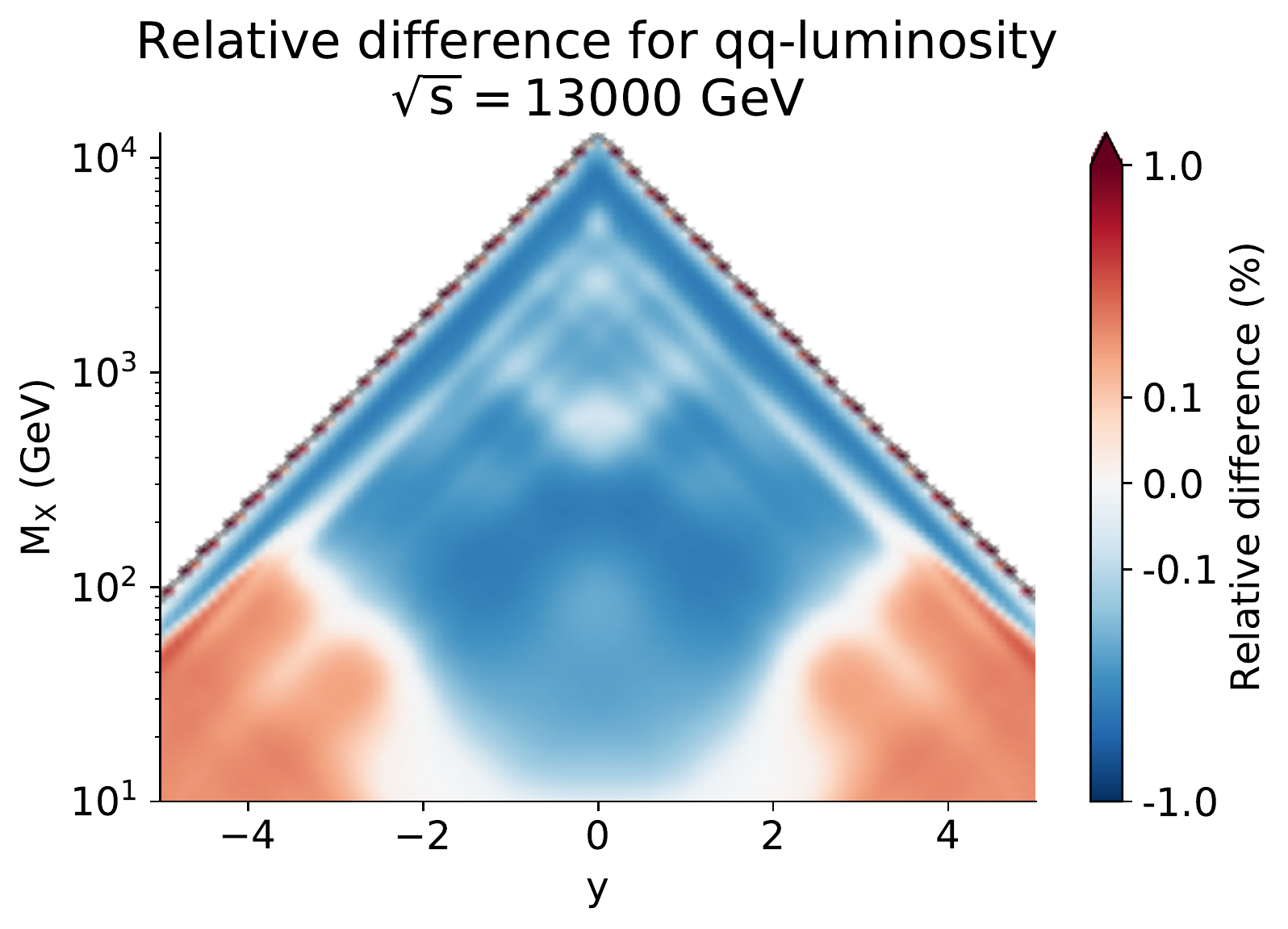}\\
  \includegraphics[scale=0.38]{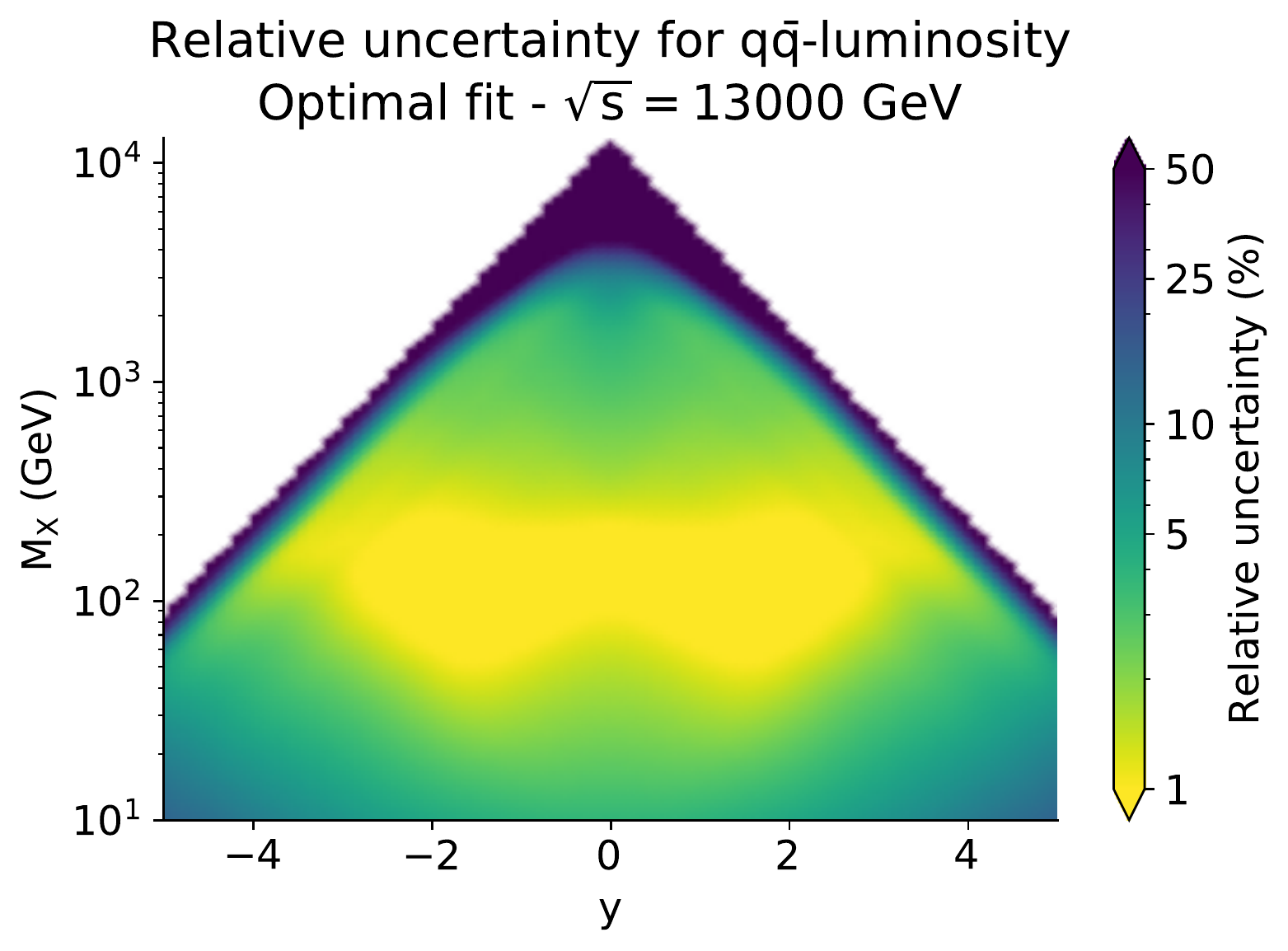}
  \includegraphics[scale=0.38]{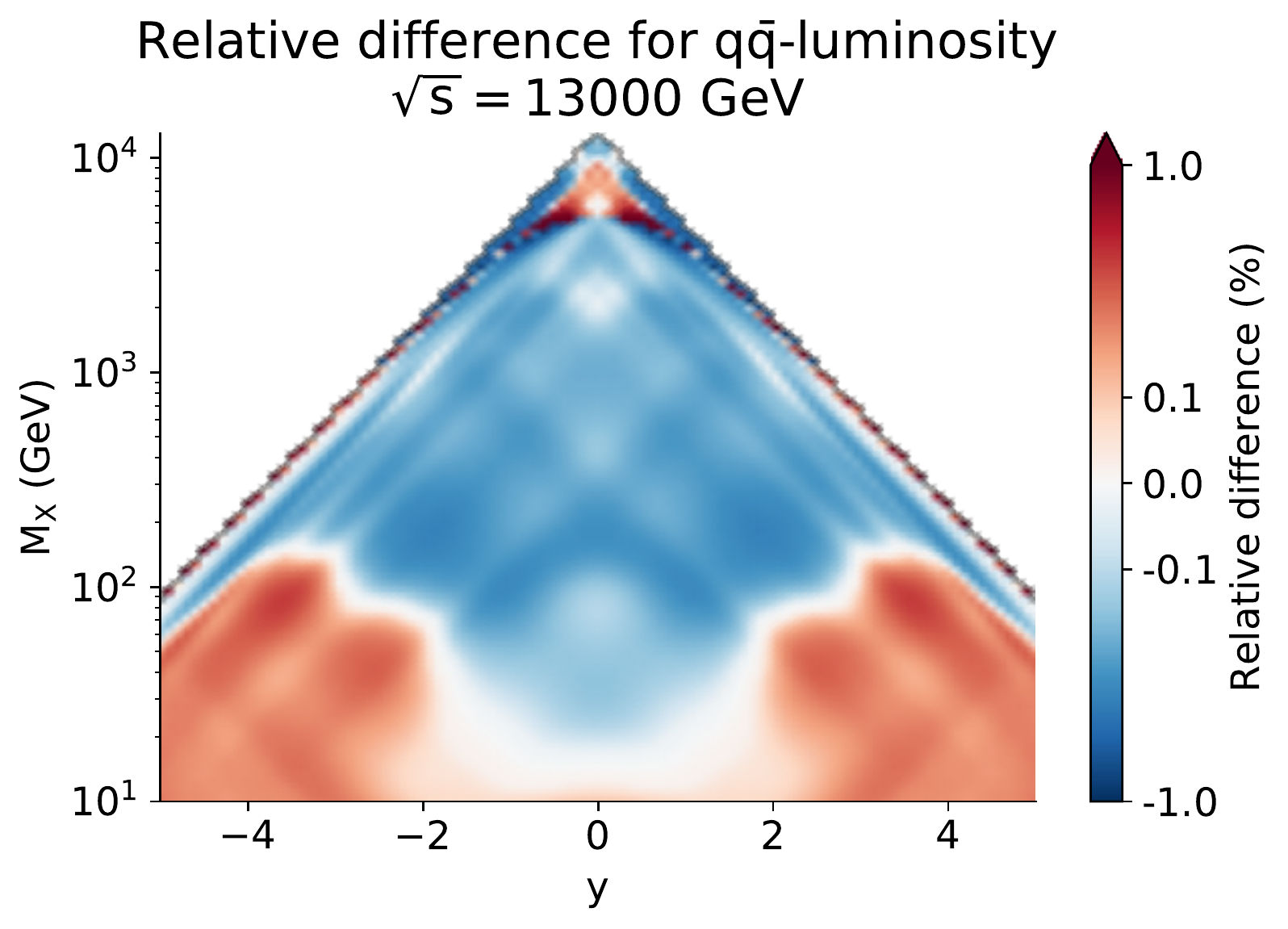}\\
  \caption{The parton luminosities as a function 
  of the invariant mass $M_X$ and the rapidity $y$ of the final state at 
  $\sqrt{s}=13$~TeV. From top to bottom, the results shown are for the 
  gluon-gluon, gluon-quark, quark-quark, and quark-antiquark luminosities. 
  The results on the left-hand side show the relative uncertainties on the 
  parton luminosities for the Optimal fit. For these plots the colour bars 
  indicate the relative uncertainty, with yellow representing uncertainties of 
  around 1\%. The results on the right-hand side show the relative difference 
  in uncertainties between the Baseline Fit (Fit 1) and the Optimal fit. The 
  relative difference is found by taking the uncertainty for the Baseline fit 
  away from the uncertainty for the Optimal fit, and normalising to the former. 
  Therefore, relative decreases in the uncertainty are shown in blue and
  relative increases are shown in red.}
  \label{fig:2d-lumi-uncertainties}
\end{figure}

\begin{figure}[!t]
\centering
  \includegraphics[scale=0.38]{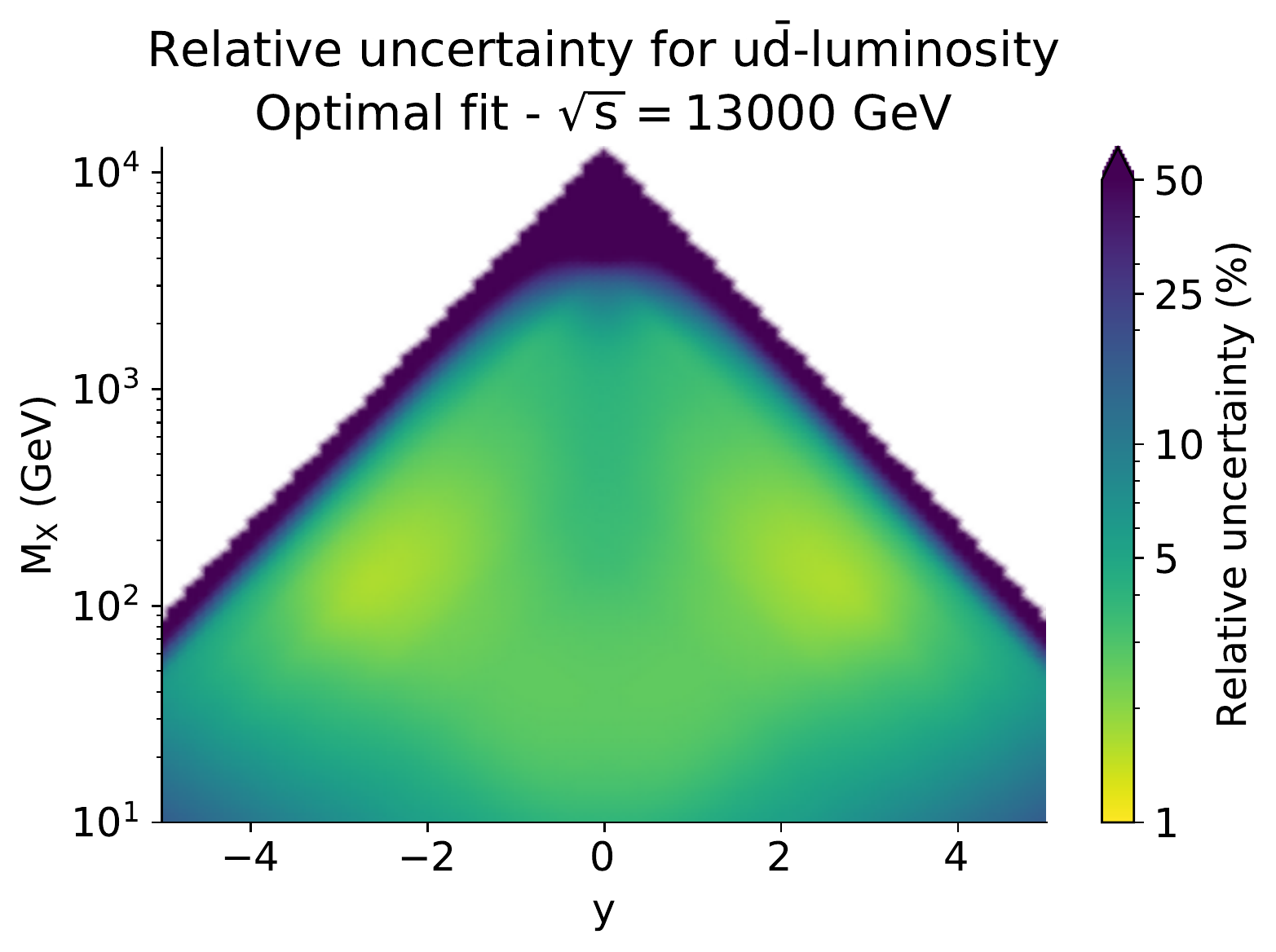}
  \includegraphics[scale=0.38]{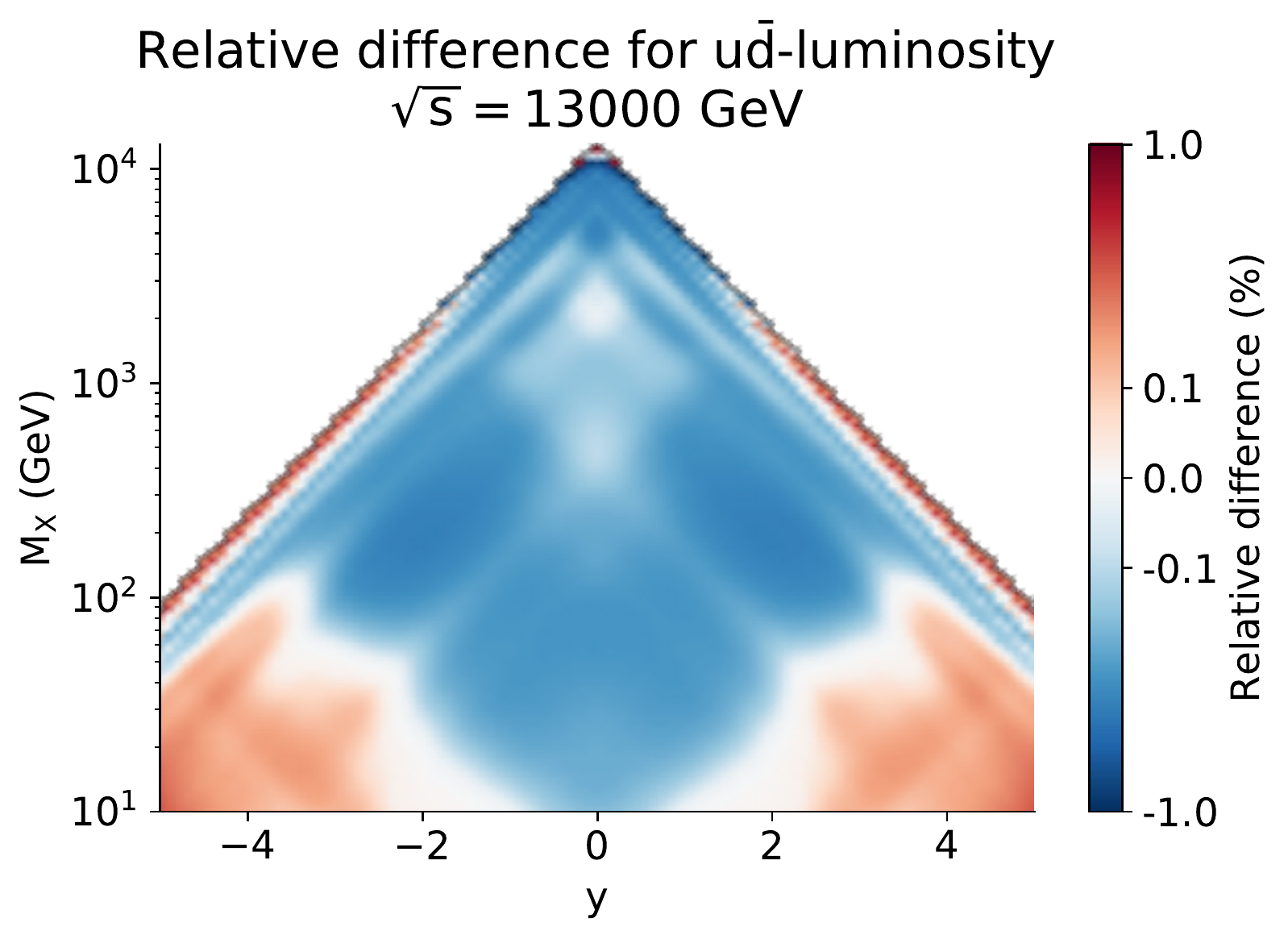}
\caption{Same as Fig.~\ref{fig:2d-lumi-uncertainties} but for the $u\bar{d}$ 
 luminosity.}
\label{fig:2d-lumi-uncertainties-ud}
\end{figure}

We now broaden the scope of our discussion, by looking at the effect of 
single top data on parton flavours more generally. Specifically, we study 
how the relative uncertainties on a range of parton luminosities change upon 
including single top data in a PDF fit. We do this because parton 
luminosities, which describe the partonic content of two colliding protons, 
are in fact the relevant entities for the LHC. Note that we adopt the 
definition of the parton luminosity given in 
Ref.~\cite{Mangano:2016jyj}.

Fig.~\ref{fig:2d-lumi-uncertainties} compares the relative uncertainty on 
the gluon-gluon, gluon-quark, quark-quark, and quark-antiquark parton 
luminosities at $\sqrt{s}=13$~TeV (whereby a sum over all quark or antiquark
flavours is performed), and Fig.~\ref{fig:2d-lumi-uncertainties-ud} compares
the relative uncertainty on the $u\bar{d}$ luminosity,
to which single top data are expected to be particularly sensitive.
The uncertainty is plotted in the invariant mass $M_X$ and the rapidity $y$ of 
the final state. The plots on the left-hand side show the relative 
uncertainties on the parton luminosities for the Optimal fit. The results on 
the right-hand side show the relative difference in uncertainties between the 
Baseline Fit (Fit 1) and the Optimal fit. The relative difference is found by 
taking the uncertainty for the Baseline fit away from the uncertainty for the 
Optimal fit, and normalising to the former. 
We see that for each combination of partons, there is a reduction in 
uncertainties for regions where the luminosities are already reasonably well 
constrained. This is most obvious for the luminosities involving quarks, which 
is as expected given that we saw a more significant decrease in PDF 
uncertainties for the up quark and the down quark compared
to the gluon (see Fig.~\ref{fig:pdf_plots}).

\section{Conclusions} 
\label{conclusions}

In this paper we systematically studied the impact of $t$-channel single 
top-quark and top-antiquark production data from the LHC on the determination 
of the proton PDFs. Specifically, we considered total cross sections, 
top-antitop cross section ratios, and differential distributions measured by
the ATLAS and CMS experiments at centre-of-mass energies of 7, 8 and 13 TeV.
We presented a critical appraisal of the data, studying in particular how 
their description is affected by the theoretical details that enter the 
computation of the corresponding observables: QCD corrections, here 
systematically extended up to NNLO for the first time; EW corrections, possibly 
mixed with QCD corrections, up to NLO; the flavour scheme, either 5FS or 4FS; 
and the value of the bottom-quark matching point. We performed a series of 
fits to the data within the NNPDF3.1 framework at NNLO, demonstrated the 
constraining power of the data, and their correlation with the rest of the 
dataset.

We found that the data is overall fairly well described with NNLO QCD in the
5FS. The effect of NNLO QCD corrections is typically of the order of few
percent on the total cross sections and on the distributions differential
in the rapidity of the top quark and top antiquark; they can be larger (up
to 10\%) for the distributions differential in the transverse momentum of 
the top quark and top antiquark, especially for the highest $p_T$ bins.
In comparison, NLO EW and mixed EW$\times$QCD corrections account for a 
sub-percent effect, which is currently negligible in comparison to the data
and the PDF uncertainty. We also found that the 4FS, which might in principle 
be more adequate than the 5FS to describe single top data as it takes into 
account bottom-quark mass corrections, typically leads to lower predictions
that are in worse agreement with the data, unless one chooses a smaller 
factorisation scale. Furthermore, single top total cross sections at NLO
decrease by about 5-7\% (10-12\%) if the bottom-quark matching point $\mu_b$ in 
the 5FS is raised to twice (five times) the mass of the bottom quark, 
in such a way that they become closer to the 4FS results. The effect would be 
significantly smaller if the 5FS cross section had been computed at NNLO,
as shown in Ref.~\cite{Bertone:2017djs}. Further studies aiming to determine
the value of $\mu_b$ that minimises the effect of $m_b$ power corrections in a 
5FS, or aiming to compute the single top distributions in a matched scheme in
order to explicitly include such $m_b$ power corrections in the theoretical 
predictions, will become necessary to describe single top data 
within a percent accuracy, once their experimental precision is improved.

Despite our careful investigation of all of the theoretical assumptions 
detailed above, we observed two significant inconsistencies between data and 
theory, for which we are unable to envision a satisfactory explanation. 
The first concerns the ratio of single top-quark to top-antiquark total
cross section from ATLAS at 8 TeV; the second concerns the distributions (both 
absolute and normalised) differential in the top-antiquark transverse 
momentum. In the first case, we noted that the analogous measurement from CMS,
which differs from the ATLAS one by approximately $\sqrt{2}\sigma$, is instead
very well described; in the latter case, we noted that the analogous 
measurement from ATLAS at 8 TeV (albeit experimental correlations are not 
available) is also very well described.

We then assessed the sensitivity of PDFs to the single top data under scrutiny
by performing a series of fits whereby we supplemented the NNPDF3.1 dataset
(from which we removed jet data that lacks NNLO QCD $C$-factors)
with suitable combinations of single top data. We found that the data-theory 
inconsistencies observed before the fit are not resolved. Because we did not 
observe any apparent tension with the rest of the dataset, even when this was 
reduced to a maximally consistent dataset made up entirely of HERA data, we concluded
that transverse momentum differential distributions should not be included in a 
fit. In light of these results, the optimal combination of single top data that
we recommend to include in a fit is made of the following measurements:
the available ratios of single top-quark to top-antiquark measurements
(from ATLAS at 7, 8 and 13 TeV and from CMS at 8 and 13 TeV), the total cross 
section if the former is not available (for CMS at 7 TeV), and the 
top-quark and top-antiquark rapidity distributions (for ATLAS at 7 TeV).
Concerning the latter, we prefer the normalised distribution. While this choice
neglects unavailable experimental correlations with the total cross sections,
it minimises unaccounted theoretical uncertainties (such as missing higher 
order QCD and EW uncertainties, the dependence on the flavour scheme or on the 
bottom matching threshold). 

Finally, we demonstrated that this optimal combination of data effectively 
constrains the light quark and the gluon PDFs by reducing their relative 
uncertainty by a fraction of a percent in the region $10^{-3}\leq x \leq 0.5$
(with a reduction of up to 25\% on the ratio $u/d$ around $x\sim 0.1$).
As expected, for the $u$ and $d$ such an effect is driven by the EW nature of 
the process, while for the gluon it is driven by the perturbative generation of 
the initiating $b$-quark. The information provided by single top
data is complementary to that provided by other processes in the NNPDF3.1
dataset, both EW-induced, such as $W$-boson production or CC DIS, and 
gluon-sensitive, such as jet or top-pair production. Single top production will
therefore be a useful ingredient in the next generation of PDF 
determinations.

\appendix
\section{The ATLAS 8 TeV differential distributions}
\label{app:ATLAS8TeVdiff}

In this Appendix we illustrate the ATLAS data of Ref.~\cite{Aaboud:2017pdi},
that is, single top-quark and top-antiquark differential distributions 
measured at a centre-of-mass energy of 8 TeV. As discussed in 
Sect.~\ref{sec:exp_data}, these are not included in the fits presented in this 
work because no information on how experimental uncertainties are correlated is
provided. Nevertheless, we believe that a data-theory comparison, along the
lines of what was presented in Sect.~\ref{sec:data-theory-comparison}, is
instructive, in particular to investigate whether the discrepancies observed in 
the ATLAS 7 TeV data persist or not.

\begin{figure}[!t]
\centering
\includegraphics[angle=270,scale=0.252]{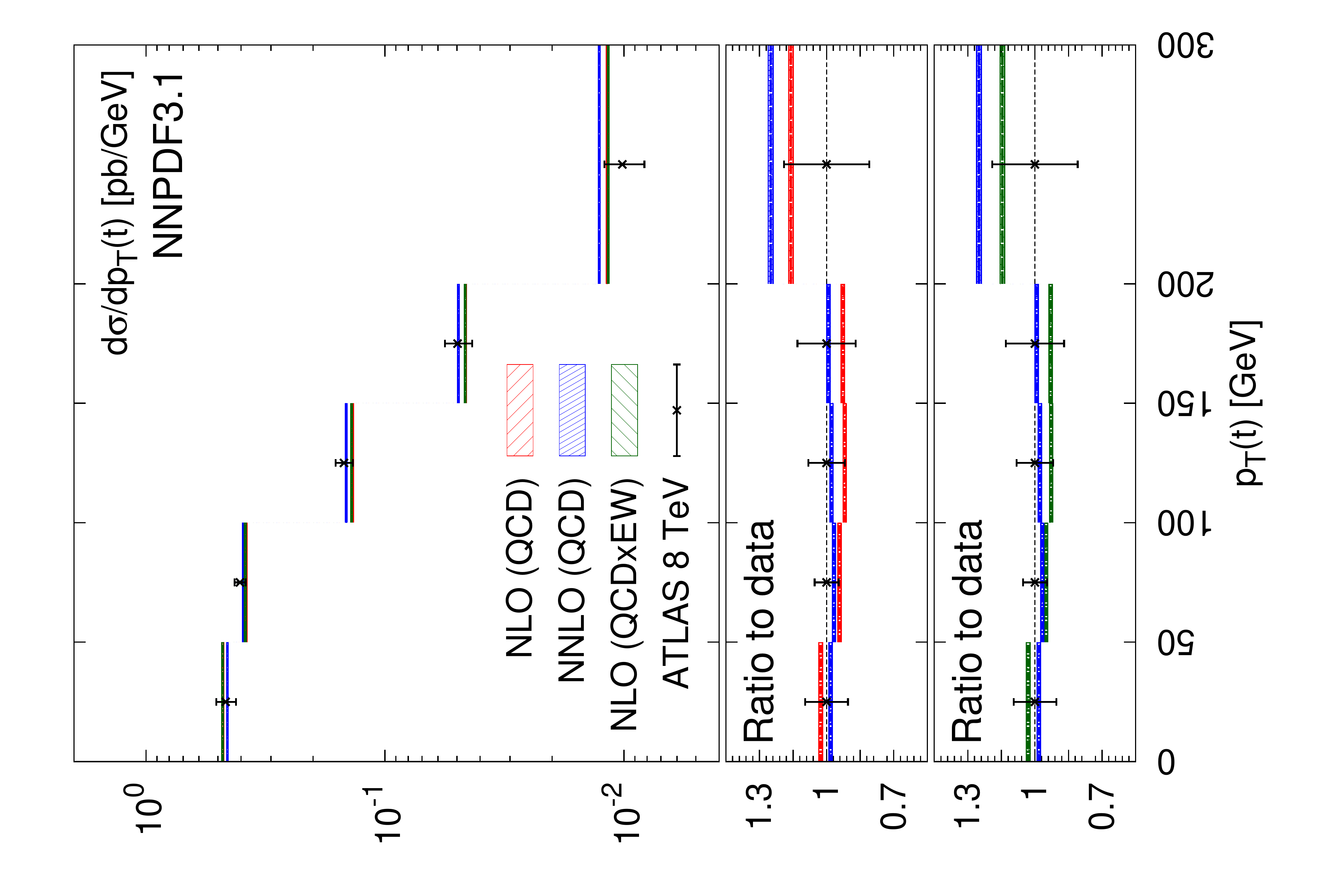}
\includegraphics[angle=270,scale=0.252]{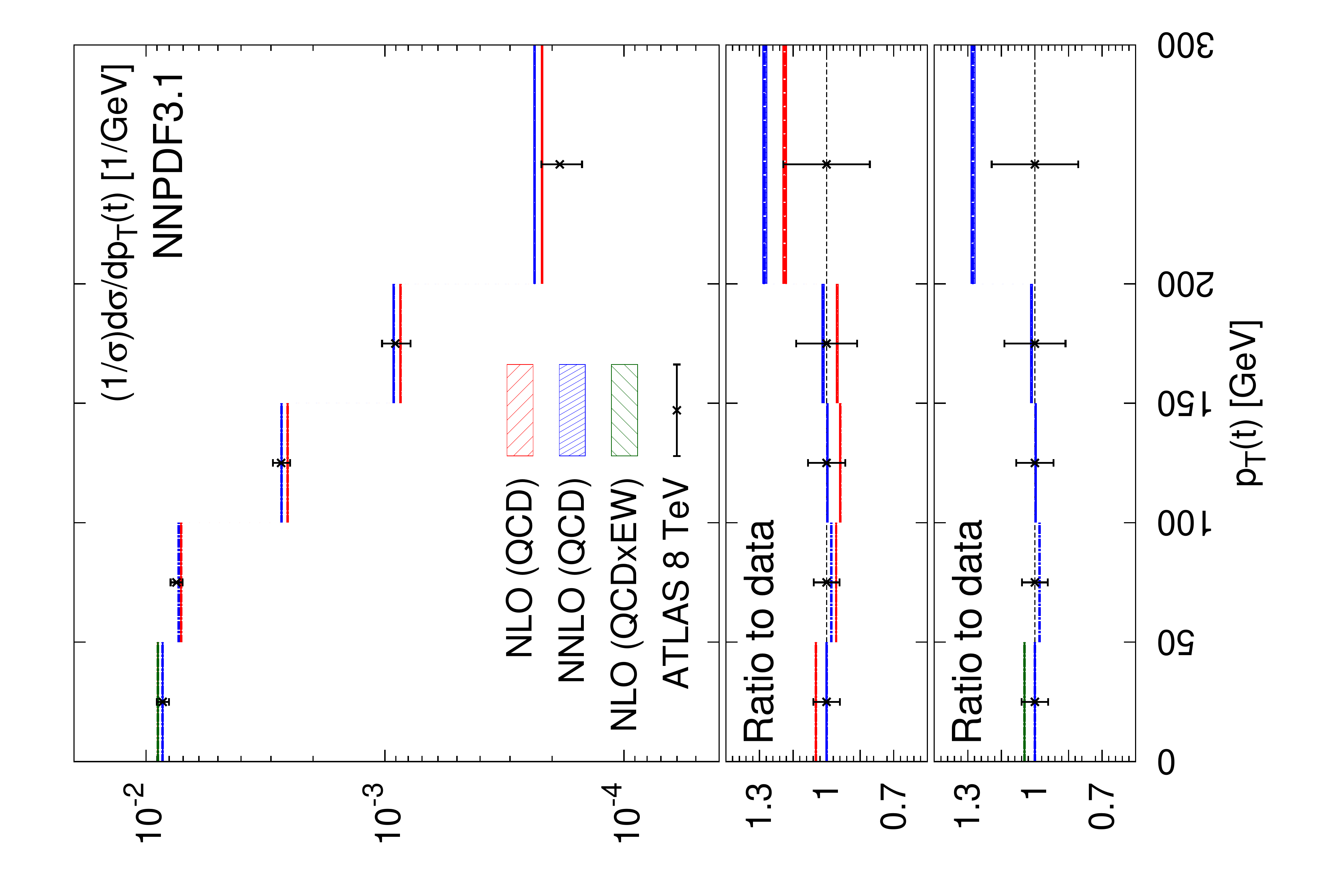}\\
\includegraphics[angle=270,scale=0.252]{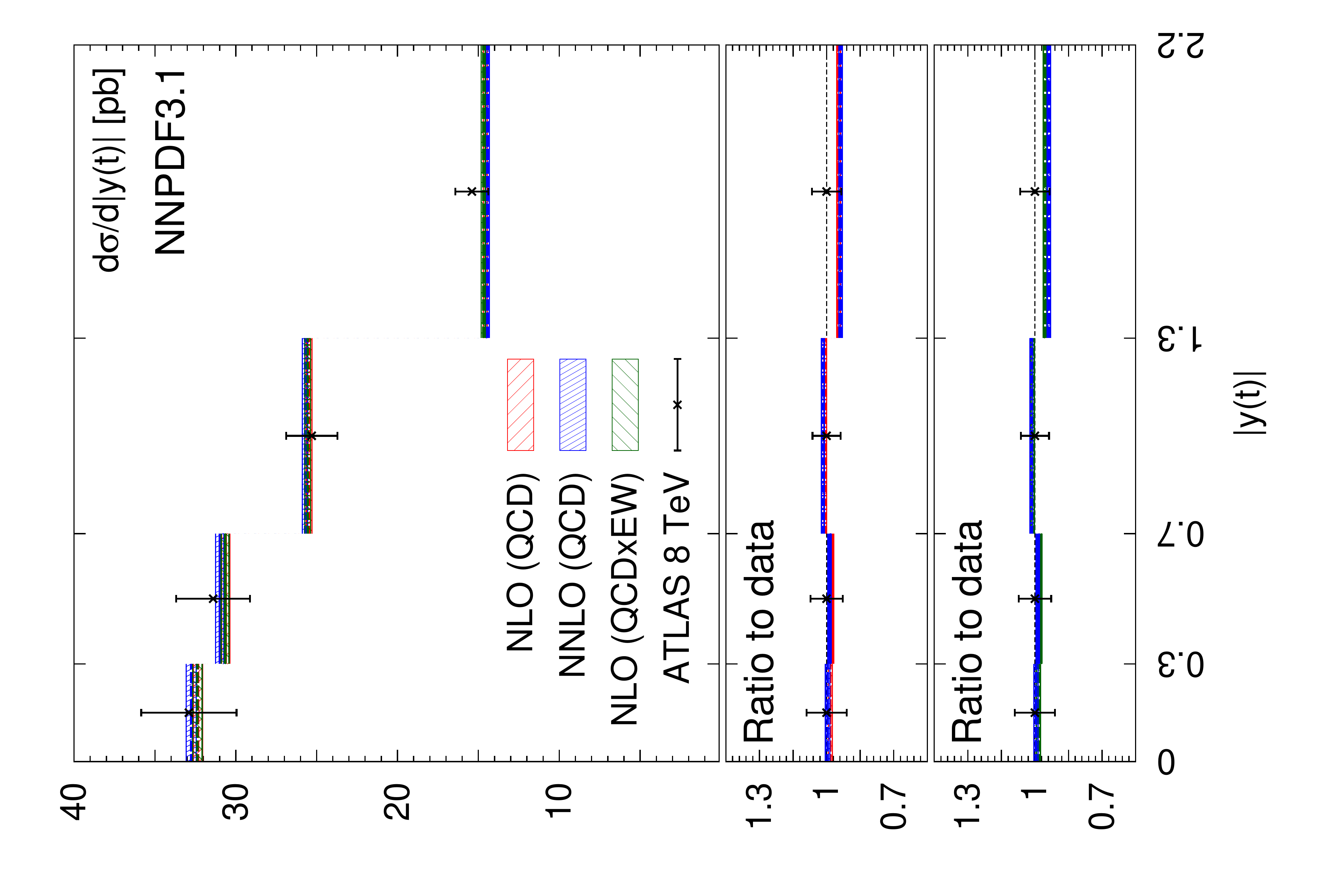}
\includegraphics[angle=270,scale=0.252]{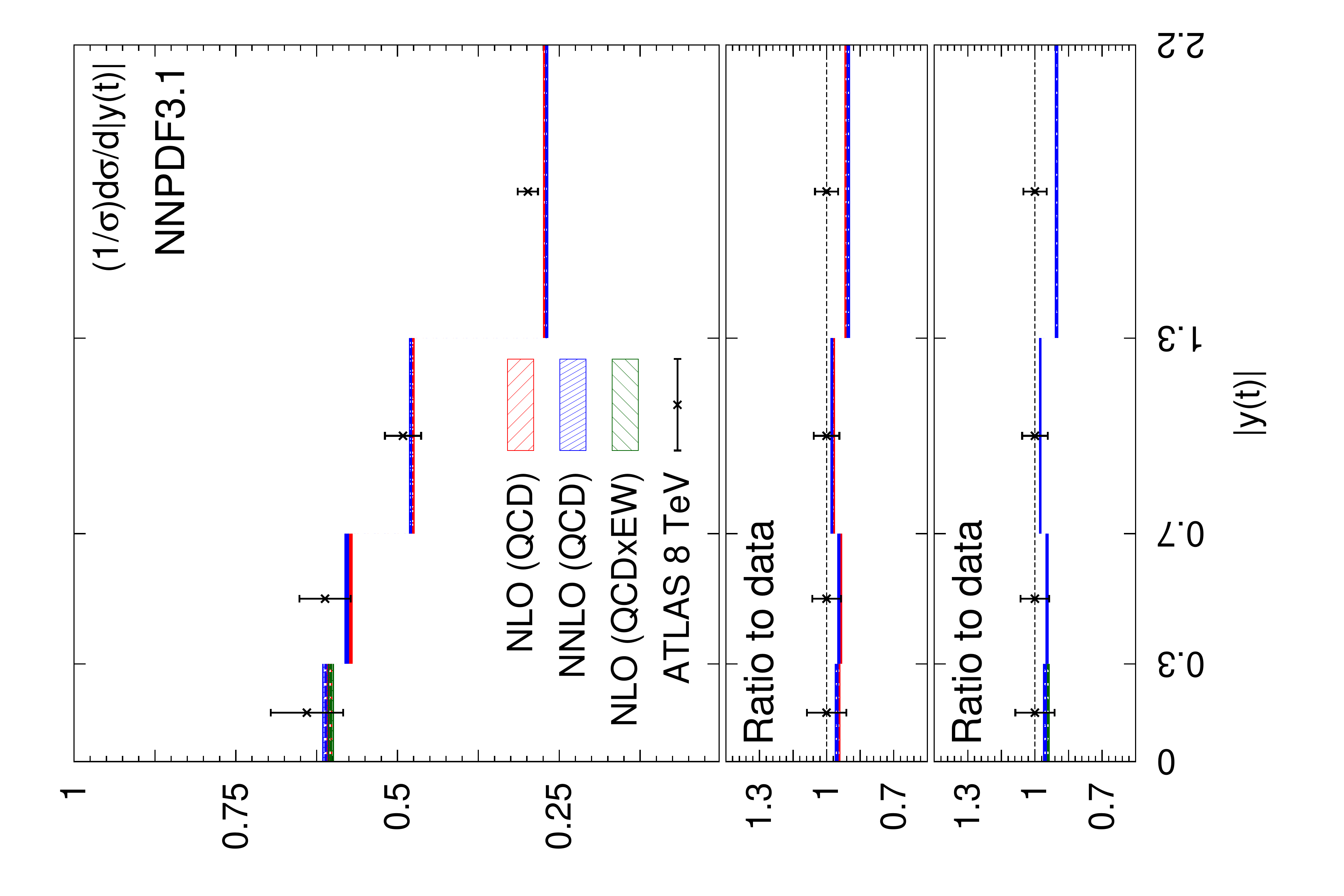}\\
\caption{Same as Fig.~\ref{fig:data_theory_top}, but for the ATLAS data of 
         Ref.~\cite{Aaboud:2017pdi}.}
\label{fig:data_theory_top_8}
\end{figure}

\begin{figure}[!t]
\centering
\includegraphics[angle=270,scale=0.252]{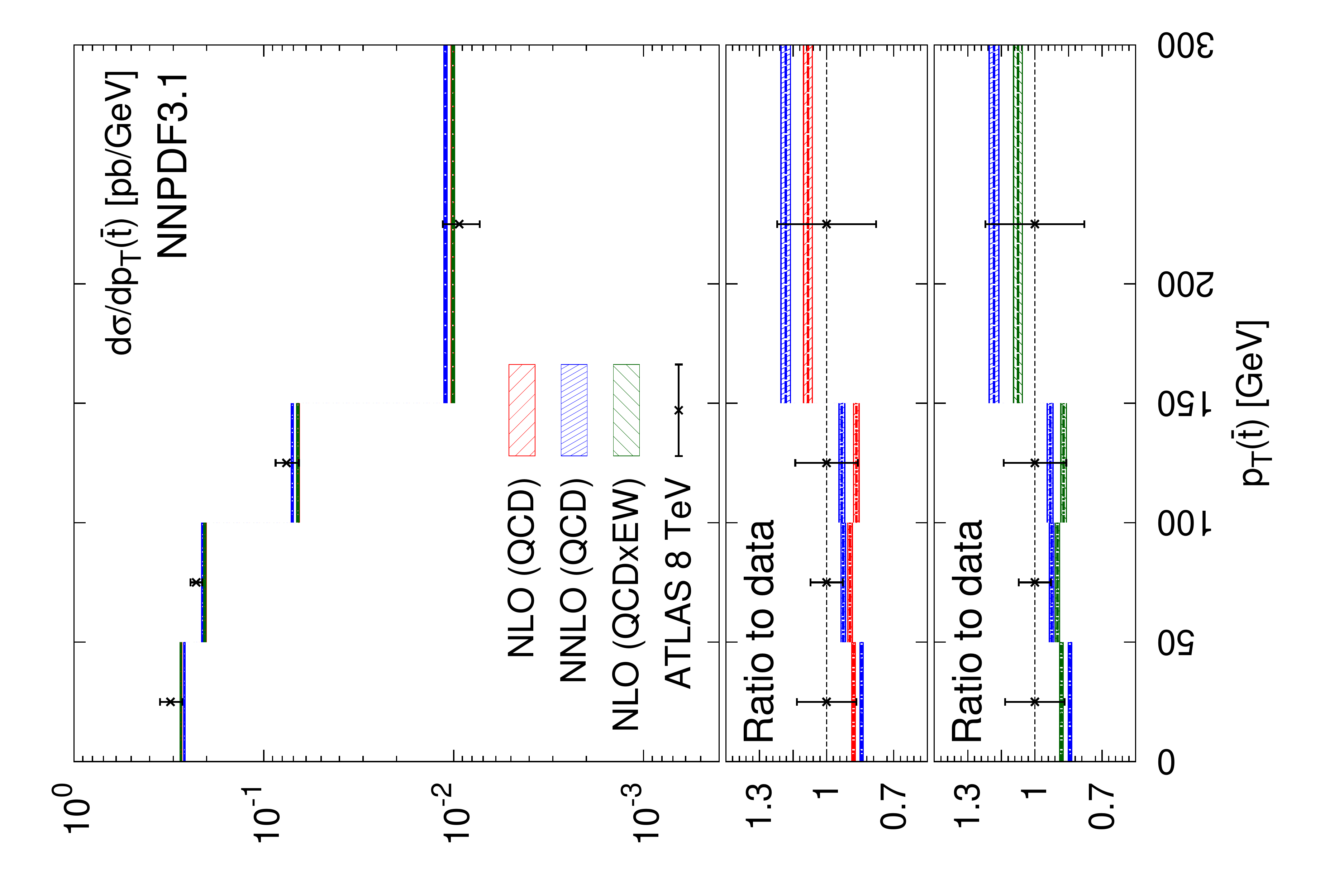}
\includegraphics[angle=270,scale=0.252]{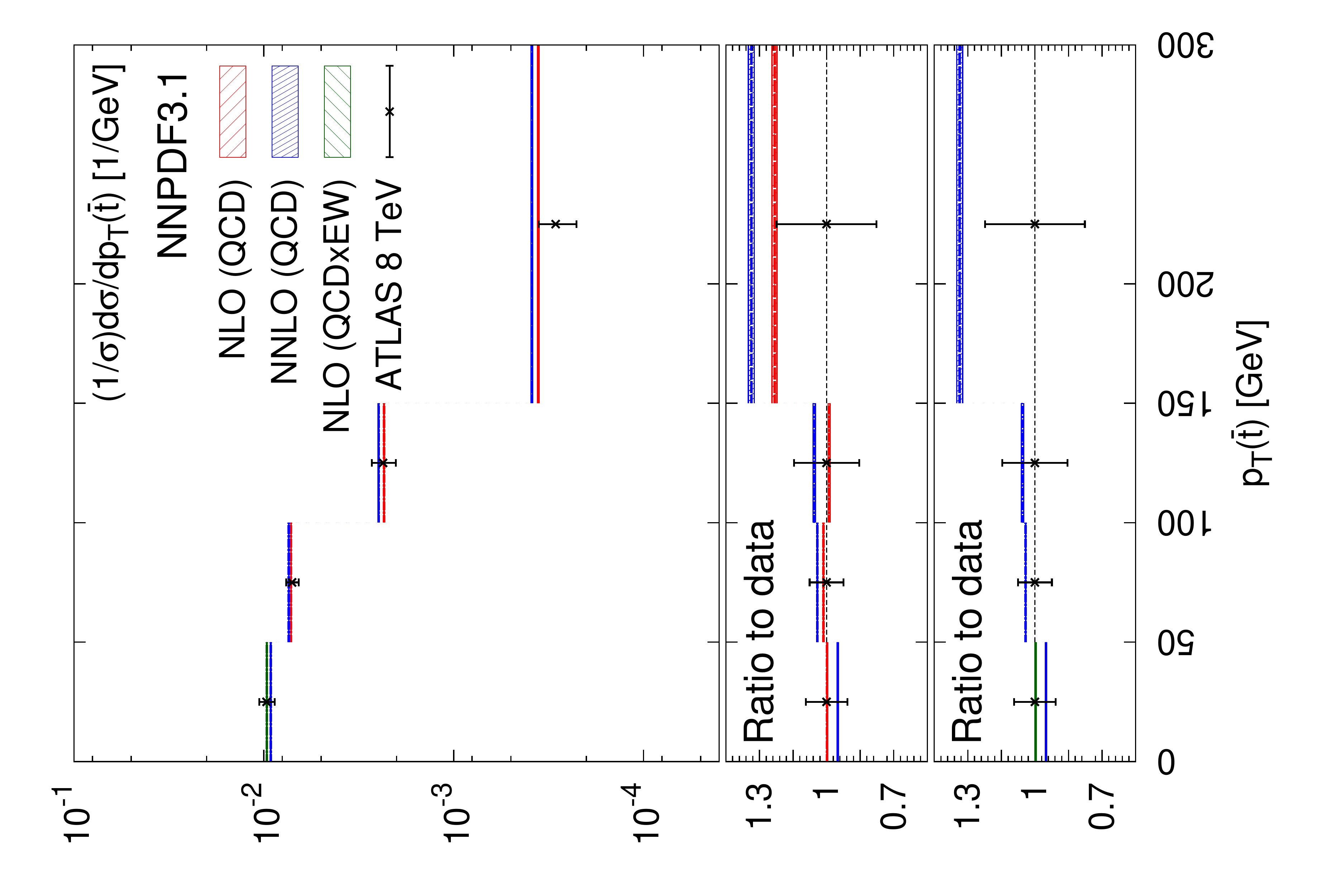}\\
\includegraphics[angle=270,scale=0.252]{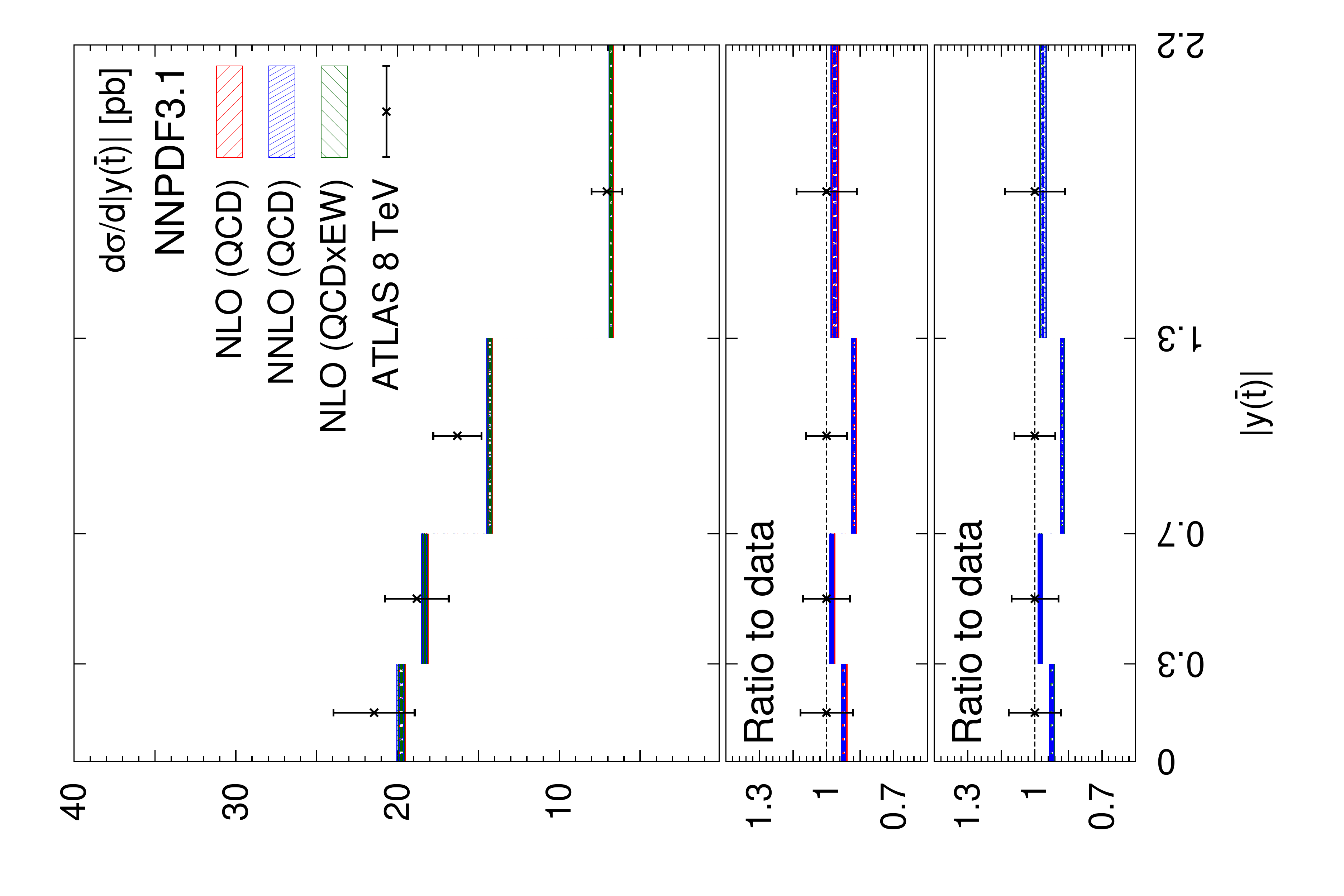}
\includegraphics[angle=270,scale=0.252]{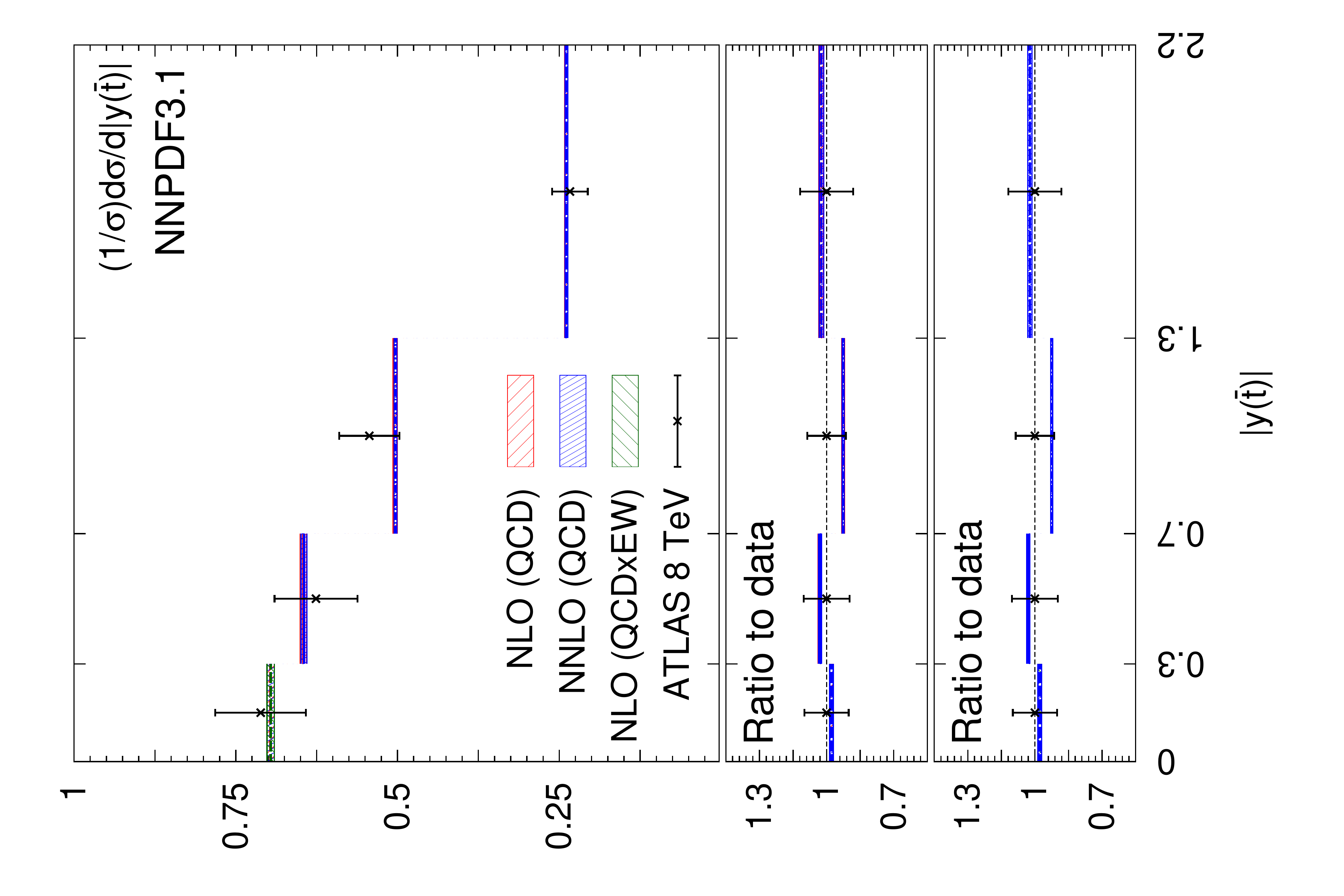}\\
\caption{Same as Fig.~\ref{fig:data_theory_atop}, but for the ATLAS data of 
         Ref.~\cite{Aaboud:2017pdi}.}
\label{fig:data_theory_atop_8}
\end{figure}

The data is compared to the corresponding QCD (at NLO and NNLO), and 
QCD$\times$EW (at NLO) theoretical predictions for both absolute and 
normalised  differential cross sections of top-quark production in
Fig.~\ref{fig:data_theory_top_8} and of top-antiquark production in
Fig.~\ref{fig:data_theory_atop_8}. Theoretical predictions are computed 
with the same computational settings, physical parameters and input PDFs as
delineated in Sects.~\ref{sec:QCD}--\ref{sec:EW}. The error bands correspond to
the 68\% CL PDF uncertainty. The features of the data are summarised in 
Table~\ref{tab:chi2_8}, where we display, for each distribution, its kinematic 
range, the number of data points, $N_{\rm dat}$, and the values of the $\chi^2$
obtained at NLO and NNLO in the 5FS (pure QCD). As in 
Sect.~\ref{sec:data-theory-comparison}, the last bin of the normalised 
distributions is removed from the computation of the $\chi^2$, but not from 
the data-theory comparisons displayed in 
Figs.~\ref{fig:data_theory_top_8}-\ref{fig:data_theory_atop_8}.

\begin{table}[!t]
  \centering
  \scriptsize
\begin{tabular}{cccll}
\toprule       
Distribution & Kinematic range
           & $N_{\rm dat}$  
           & $\chi^{2\ {\rm (5FS)}}_{\rm NLO}$ 
           & $\chi^{2\ {\rm (5FS)}}_{\rm NNLO}$\\
\midrule
$d\sigma/dp_T(t)$ & $0<p_T(t)<300$ GeV                
           &  5 & 0.71 & 0.46 \\
$d\sigma/dp_T(\bar{t})$  & $0<p_T(\bar{t})<300$ GeV             
           &  4 & 0.90 & 1.36 \\
$d\sigma/d |y(t)|$ & $0<|y(t)|<2.2$             
           &  4 & 0.17 & 0.29 \\
$d\sigma/d|y(\bar{t})|$ & $0<|y(\bar{t})|<2.2$         
           &  4 & 0.40 & 0.37 \\
$(1/\sigma)d\sigma/dp_T(t)$ & $0<p_T(t)<300$ GeV        
           &  4 & 0.80 & 0.05 \\
$(1/\sigma)d\sigma/dp_T(\bar{t})$ & $0<p_T(\bar{t})<300$ GeV 
           &  3 & 0.02 & 0.51 \\
$(1/\sigma)d\sigma/d|y(t)|$ &  $0<|y(t)|<2.2$       
           &  3 & 0.24 & 0.19 \\
$(1/\sigma)d\sigma/d|y(\bar{t})|$ & $0<|y(\bar{t})|<2.2$
           &  3 & 0.33 & 0.34 \\
\bottomrule
\end{tabular}

  \caption{The values of the $\chi^2$, Eq.~\eqref{eq:chi2}, computed for all 
  of the datasets presented in Ref.~\cite{Aaboud:2017pdi}. We indicate the 
  distribution, its kinematic range, and the corresponding number of data 
  points, $N_{\rm dat}$. The bin edges for the transverse momentum distributions
  are 0, 50, 100, 150, 200 and 300 GeV, while for the rapidity distributions 
  they are 0, 0.3, 0.7, 1.3 and 2.2.
  Theoretical predictions are computed at either NLO or NNLO accuracy in the 
  5FS in pure QCD; the input PDFs are taken from the 
  {\tt NNPDF31\_nlo\_as\_0118} and {\tt NNPDF31\_nnlo\_as\_0118} sets, 
  respectively. Experimental correlations are not provided, therefore they
  are not included in the $\chi^2$ computation.}
\label{tab:chi2_8}
\end{table}

As is apparent from Table~\ref{tab:chi2_8} and
Figs.~\ref{fig:data_theory_top_8}--\ref{fig:data_theory_atop_8},
the description of the data provided by NNPDF3.1 is qualitatively and 
quantitatively excellent for all the differential distributions. 
The impact of EW corrections is negligible, while the impact of NNLO QCD
corrections is moderate. These lead to a general improvement of the $\chi^2$,
except for the distributions differential in the transverse momentum of the 
top antiquark, and in the absolute rapidity of the top quark. This seems to 
be a consequence of the fact that the NNLO prediction for both the largest 
$p_T(\bar{t})$ and $|y(t)|$ bins is farther from the data than its NLO
counterpart. All of the $\chi^2$, however, remain perfectly acceptable; any
further conclusion must be reassessed when experimental correlations
become available.

The overall excellent description of the data includes, most notably, the 
distributions differential in the transverse momentum of the
top antiquark, for which we found large data-theory tensions in the analogous 
ATLAS measurement at a centre-of-mass energy of 7 TeV (see 
Sect.~\ref{sec:data-theory-comparison}). 
We interpret this result as further evidence of the fact that the discrepancy 
observed in Sect.~\ref{sec:data-theory-comparison} is likely not to have a
theoretical origin. We believe that this specific 
measurement deserves further investigation, possibly on the original 
experimental analysis, and should therefore not be included in a PDF fit.

\section*{Acknowledgements}
The Authors would like to thank Jun Gao for providing them with the NNLO
computation of Refs.~\cite{Berger:2016oht,Berger:2017zof}, Marco Zaro for 
support with {\amc}, Zahari Kassabov for his help with the {\sc ReportEngine} 
software, Lucian Harland-Lang for a critical reading of the manuscript,
Reinhard Schwienhorst and Alberto Orso Maria Iorio for 
clarifications on the ATLAS and CMS experimental data, respectively, and the 
members of the NNPDF Collaboration, in particular Stefano Forte, Zahari 
Kassabov, Juan Rojo and Luca Rottoli, for useful comments on the manuscript. 
E.R.N. is supported by the European Commission through the
Marie Sk\l odowska-Curie Action ParDHonS\_FFs.TMDs (grant number 752748).
M.U. is funded by the Royal Society grant DH/150088 and supported by the 
STFC consolidated grant ST/L000681/1 and the Royal Society grant RGF/EA/180148.
C.V. is supported by the STFC grant ST/R504671/1.

\renewcommand{\em}{}
\bibliographystyle{UTPstyle}
\bibliography{paper}

\end{document}